\documentclass[namedreferences]{SolarPhysics}
\pdfoutput=1
\usepackage[optionalrh]{spr-sola-addons} 
\usepackage{graphicx}                    
\usepackage{color}                       
\usepackage{url}                         

 \newcommand{\blg}[1]{{#1}}

\newcommand{\todash}{\,--\,}
\newcommand{\degree}{\mbox{\ensuremath{^\circ}}}

\newcommand{\etal}{{\it et al.}}
\newcommand{\eg}{{\it e.g.\ }}

\newcommand{\ion}[2]{#1\,{\sc #2}} 

\long\def\symbolfootnote[#1]#2{\begingroup%
\def\thefootnote{\fnsymbol{footnote}}\footnote[#1]{#2}\endgroup} 
\begin{document}

\begin{article}

\begin{opening}

\title{Modeling the Subsurface Structure of Sunspots\symbolfootnote[1]{\bf Invited Review}}

%
\author{H.~\surname{Moradi}$^{1}$\sep
        C.~\surname{Baldner}$^{2}$\sep
        A.C.~\surname{Birch}$^{3}$\sep
        D.C.~\surname{Braun}$^{3}$\sep
        R.H.~\surname{Cameron}$^{1}$\sep
        T.L.~\surname{Duvall~Jr.}$^{4}$\sep
        L.~\surname{Gizon}$^{1}$\sep
        D.~\surname{Haber}$^{5}$\sep
        S.M.~\surname{Hanasoge}$^{1}$\sep
        B.W.~\surname{Hindman}$^{5}$\sep
        J.~\surname{Jackiewicz}$^{6}$\sep
        E.~\surname{Khomenko}$^{7}$\sep
        R.~\surname{Komm}$^{8}$\sep
        P.~\surname{Rajaguru}$^{9}$\sep
        M.~\surname{Rempel}$^{10}$\sep
        M.~\surname{Roth}$^{11}$\sep
        R.~\surname{Schlichenmaier}$^{11}$\sep
        H.J.~\surname{Schunker}$^{1}$\sep
        H.C.~\surname{Spruit}$^{12}$\sep
        K.G.~\surname{Strassmeier}$^{13}$\sep
        M.J.~\surname{Thompson}$^{10,14}$\sep
        S.~\surname{Zharkov}$^{14,15}$
       }


\runningauthor{H. Moradi \etal}
\runningtitle{Modeling the Subsurface Structure of Sunspots}

%
  \institute{$^{1}$ Max-Planck-Institut f{\"u}r Sonnensystemforschung, 37191 Katlenburg-Lindau, Germany \\email: \url{moradi@mps.mpg.de} \\
             	$^{2}$ Department of Astronomy, Yale University, P.O. Box 208101, New Haven, CT 06520, USA \\
             	$^{3}$ Colorado Research Associates, A Division of NorthWest Research Associates, Inc., 3380 Mitchell Lane, Boulder, CO 80301-5410, USA \\
             	$^{4}$ Laboratory for Solar Physics, NASA/Goddard Space Flight Center, Greenbelt, MD 20771, USA \\
	         $^{5}$ JILA, University of Colorado, Boulder, CO 80309-440 \\
             	$^{6}$ New Mexico State University, Astronomy Department, P.O. Box 30001, MSC4500, Las Cruces, NM 88003, USA \\
             	$^{7}$  Instituto de Astrofisica de Canarias, 38205, C/Via L\'actea, s/n, Tenerife, Spain \\
             	$^{8}$ National Solar Observatory, Tucson, AZ 85719, USA \\
             	$^{9}$ Indian Institute of Astrophysics, Bangalore, India  \\
             	$^{10}$ HAO/NCAR, P.O. Box 3000, Boulder, CO 80307, USA \\
             	$^{11}$ Kiepenheuer-Institut f\"ur Sonnenphysik, Sch\"oneckstr. 6, 79104 Freiburg, Germany \\
             	$^{12}$ Max-Planck-Institut f\"ur Astrophysik, Karl-Schwarzschild-Str. 1, 85748 Garching, Germany\\
		$^{13}$ Astrophysikalisches Institut Potsdam, An der Sternwarte 16, 14482 Potsdam, Germany \\
             	$^{14}$ School of Mathematics and Statistics, University of Sheffield, Houndsfield Road, Sheffield, S3 7RH, UK \\
	          $^{15}$ Mullard Space Science Laboratory, University College London, UK \\
             }

\begin{abstract}
While sunspots are easily observed at the solar surface, determining their subsurface structure is not trivial. There are two main hypotheses for the subsurface structure of sunspots: the monolithic model and the cluster model. Local helioseismology is the only means by which we can investigate subphotospheric structure. However, as current linear inversion techniques do not yet allow helioseismology to probe the internal structure with sufficient confidence to distinguish between the monolith and cluster models, the development of physically realistic sunspot models are a priority for helioseismologists. This is because they are not only important indicators of the variety of physical effects that may influence helioseismic inferences in active regions, but they also enable detailed assessments of the validity of helioseismic interpretations through numerical forward modeling. In this paper, we provide a critical review of the existing sunspot models and an overview of numerical methods employed to model wave propagation through model sunspots. We then carry out an helioseismic analysis of the sunspot in Active Region 9787 and address the serious inconsistencies uncovered by \citeauthor{gizonetal2009}~(\citeyear{gizonetal2009,gizonetal2009a}). We find that this sunspot is most probably associated with a shallow, positive wave-speed perturbation (unlike the traditional two-layer model) and that travel-time measurements are consistent with a horizontal outflow in the surrounding moat.
\end{abstract}

%

\end{opening}
\newpage
\tableofcontents
\newpage
%
\section{Introduction}\label{s:intro}
\subsection{Why Sunspots are Interesting}
Sunspots are the surface manifestations of intense magnetic flux 
concentrations that have intersected with
the solar surface. As such, they represent one of the major connections of
the internal magnetic field of the Sun with its wider environments
and are the main sites of solar activity phenomena.
They are at the center of many ongoing challenges in the study of the Sun, as
the structure and evolution of sunspots, individually and collectively,
are still not fully understood.

Sunspots tend to appear at well defined latitudes, which vary with the 11-year solar cycle,
as summarized in the so-called butterfly diagram.
Any theory on the mechanism of the solar global dynamo has to be able
to explain this collective behavior.
Understanding dynamo processes is of utmost importance, as they are believed
to play crucial roles in many astrophysical phenomena,
and sunspots are the best known candidates to provide us important clues
on how they operate.
\begin{figure}[ht]
\centering
\includegraphics[width=\textwidth]{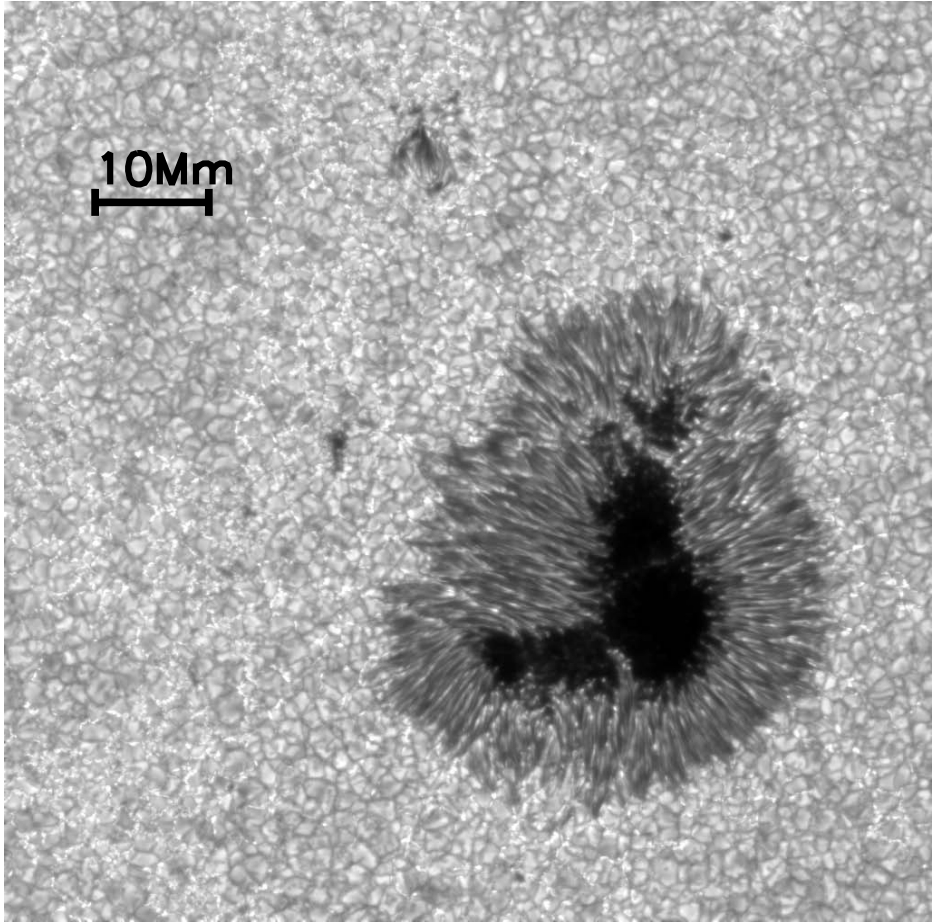}
\caption{A \textit{Hinode} G-band image of the well-developed sunspot AR 10953,
acquired at 10:10:16UT, 2 May 2007 (the spatial scale is
indicated by the horizontal bar in the upper left). There is
a light-bridge close to the northern (upper) tip, and another
is beginning to form near the southern edge. }
\label{fig:hinode}
\end{figure}

While the sunspots are easily observed at the surface, determining their subsurface
structure is not at all trivial. There are two main hypotheses for
the structure of the subsurface magnetic configuration of the spot:
the monolithic model (\textit{e.g.} \citeauthor{cowling1946}~\citeyear{cowling1946,cowling1957,cowling1976}) and the jellyfish/cluster/spaghetti model (\textit{e.g.} \citeauthor{parker1975}~\citeyear{parker1975,parker1979}; \opencite{spruit1981}; \opencite{zwaan1981}). Determining the parameters of these tubes, that is, typical size, field strength \textit{\textit{etc.}}, will help reveal details of the operation
of the solar dynamo and how magnetic field is transported up through
the convection zone.

Our understanding of sunspot structure has been somewhat hampered in the past
by a lack of high-resolution observations.
However, recent advances in adaptive optics, image selection and reconstruction
at ground-based telescopes, and the advent of high-resolution space observations with \textit{Hinode} (see Figure \ref{fig:hinode}), have all led to a wealth of
detailed information about the fine structure of sunspot umbrae and penumbrae (see \textit{e.g.} \opencite{scharmeretal2002}; \opencite{tw2008}).
On the other hand, the subsurface structure of sunspots is still poorly understood.

Local helioseismology (\textit{e.g.} time--distance helioseismology, helioseismic holography, acoustic imaging, ring-diagram analysis,
see \opencite{gb2005} and references therein) is the only means by which we can investigate subphotospheric structure.
However, interpretations of data have been somewhat ambiguous and inconsistent
for applications of local helioseismic methods in solar active regions.
Furthermore, current linear inversion techniques do not yet allow helioseismology to
probe the internal structure of sunspots with sufficient accuracy to
distinguish between monolith and cluster models. But progress has been made in
addressing some of these inconsistencies (\textit{e.g.} \opencite{bb2008}; \opencite{birchetal2009}; \opencite{gizonetal2009}; \opencite{gizonetal2009a})
and significant advances have also been made in the simulation of helioseismic wave-propagation in magnetized plasmas
(\textit{e.g.} \opencite{cgd2007}; \opencite{pk2007a}; \opencite{cgd2008}; \opencite{hanasoge2008}; \opencite{mhc2008}; \opencite{khomenkoetal2009}; \opencite{pk2009}; \opencite{sheylagetal2009}; \opencite{cameronetal2010}).

\subsection{The Need for Sunspot Models in Helioseismology}
Sunspot models are important indicators of the variety of physical effects that
may influence helioseismic inferences in active regions as well as to enable
assessments of the validity of helioseismic inversions. Currently it
is also important, in the absence of adequate nonlinear inversion
techniques, to have models that may be close to the truth as starting points for linear inversions.
The associated danger of course is an over-reliance on a small range of models
that may limit our imagination of what structures may exist and which may
bias the helioseismic inversion results.
\begin{figure} 
\centering
\includegraphics[angle=90,width=\textwidth]{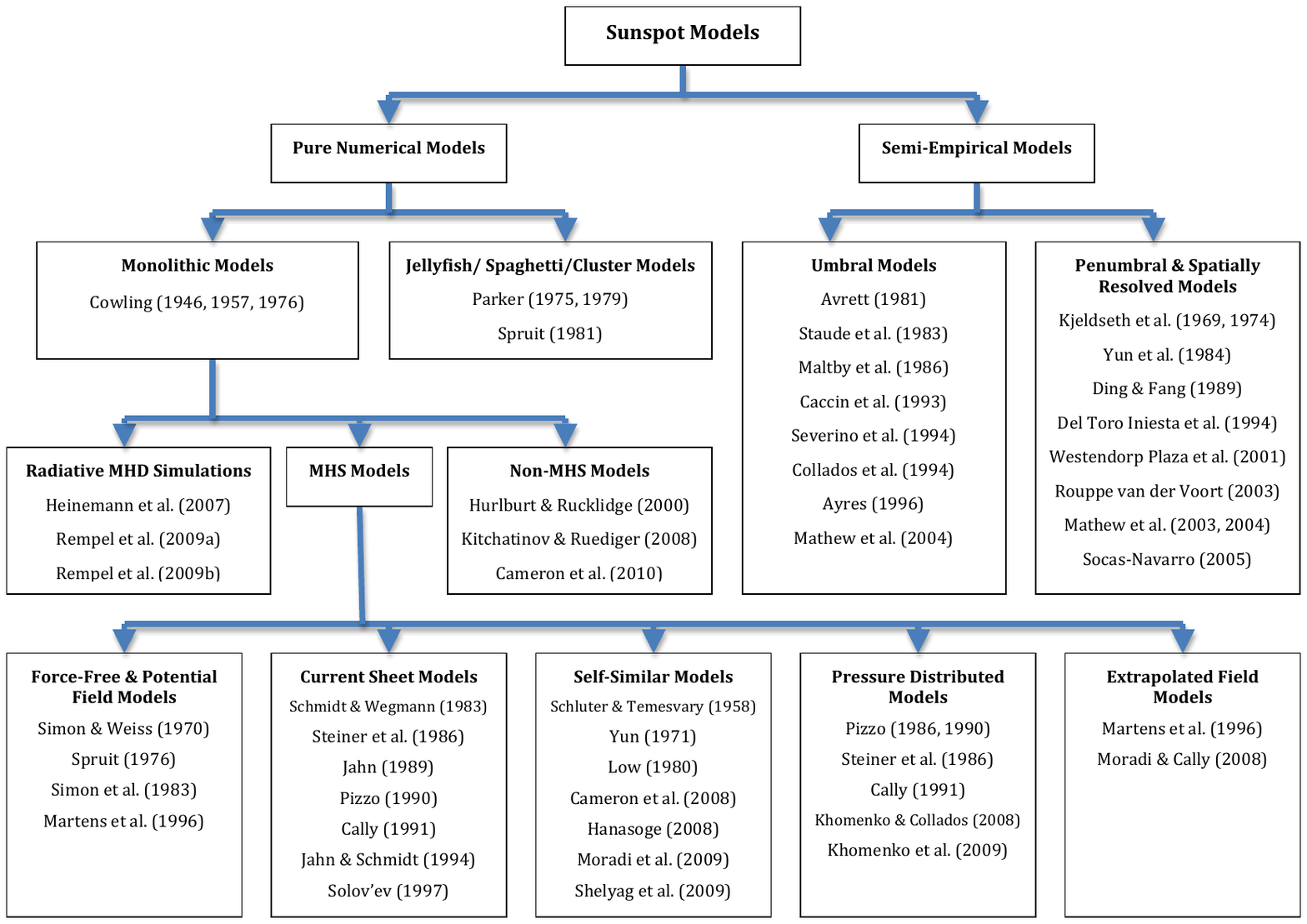}
\caption{An overview diagram representing the different classes of existing sunspot models. A number of references for each class are provided as a guide.  }
\label{fig:overview}
\end{figure}

As illustrated in Figure~\ref{fig:overview}, there are several existing classes of sunspot
models -- radiative magnetohydrodynamical (MHD) numerical simulations, semi-empirical models,
monolithic models, cluster/spaghetti models, self-similar models, potential
field models, current sheet models, pressure distributed models, \textit{etc.} --
which are more or less suited to local helioseismology. 

The availability of powerful computers has only recently made it possible
to produce numerical simulations of sunspots using realistic physics
(radiative MHD equations; see Section\,\ref{rempel}). These simulations are
more than adequate for studying surface dynamics and fine structure,
conducting experiments regarding the subsurface structure, and for
providing very useful artificial oscillation data that can be used
to test helioseismic inversion methods. Although this is a huge step
forward, this approach is unfortunately not suitable for computational
helioseismology (where detailed parametric studies need to be conducted
using existing numerical wave propagation codes) for a number of reasons:
\textit{i}) The subsurface structure of the ``realistic'' simulations depends on
the lower boundary and initial conditions, neither of which is necessarily
realistic. The surface properties obtained are more or less correct
because the timescales for the evolution at the surface are essentially
determined by the radiative loss term (\textit{i.e} convection on the Sun is
driven from the surface). The timescale for the subsurface structure
of the spot is huge, and is dependent on both the bottom the boundary
and initial conditions. \blg{Nevertheless, work is currently underway to
establish whether models with such artificially restrictive lower boundary
conditions are able produce results that are compatible with helioseismic
measurements.} \textit{ii}) The cumbersome computational requirements
of such \textit{ab-initio} simulations. As a guide, simulating the
two-hour evolution of a pair of sunspots in \inlinecite{rempeletal2009}
takes a number of weeks on extremely powerful supercomputers. \textit{iii})
Such simulations still do not address the question of the nature of the
deep structure of sunspots, \blg{as while one may ask whether a sunspot
model with the magnetic field clamped 10 Mm below the photosphere is,
or is not, compatible with helioseismic measurements, probing down to
these depths however will first require that we adequately model (and
remove) the effects of the surface layers.}

Examining a number of different sunspot models is not only essential to test the robustness of helioseismic inferences, but it also an indispensable need in computational heliosiesmology where effective inference by matching wave-field simulations to observations relies on the availability of sunspot models. For example, sunspot models with very different subsurface configurations (\eg cluster or monolith, connected or disconnected \textit{etc}.) can be used to verify what helioseismic signatures would be observed for a sunspot with that particular subsurface structure. It is also essential to be able to continually tune the surface parameters of the model to actual observations in order to match the wave-field signatures observed. So the aim here is not to have the best sunspot model, but to have a group of models than can be useful for computational helioseismology -- models which can be tuned to match the non-helioseismic observations, and then in turn be used in inversions to determine the subsurface structure. Of course, with multiple models the inversions are unlikely to be unique. Therefore, such parametric studies will provide us with idea of what is, and what is not reliable in the inversions. As well as more or less realistic models, highly simplified models which isolate just a few physical effects can also be useful for elucidating the sensitivity of inference methods in helioseismology to particular effects. Nonetheless, it must be borne in mind if using such models that other physical effects may have seismic signatures that are qualitatively or quantitatively similar.

\subsection{What Basic Properties should be included in the Models?}
Apart from the desired characteristics mentioned above, models of the gross magnetic structure of sunspots for use in computational helioseismology should ideally posses a high degree of flexibility and computational efficiency to allow for extensive parameter studies using existing numerical simulation codes. A number of other essential, observationally derived, characteristics should also be embodied by the models. For example, an accurate prescription for the surface (photospheric) magnetic-field characteristics (of both the umbra and penumbra, \textit{e.g.} field strength, orientation, twist, return flux \textit{etc.}, see Section\,\ref{field} and Section\,\ref{twist}). These should comply with observations and ideally, one should be able to model the sunspot field on extrapolations from observed surface magnetic profiles (\textit{e.g.} vector magnetograms).  One should also be able to choose the profile of thermodynamic parameters (pressure, density, temperature \textit{etc.}, see Section\,\ref {thermo} and Section\,\ref{semodels}) in the umbra and penumbra from either spectro-polarimetric inversions or semi-empirical models of the solar atmosphere. Some important dynamical phenomena (\textit{e.g.} the Evershed flow, moat flow, \textit{etc.}, see Section\,\ref{evershed} and Section\,\ref{mmf}) should also be taken into account, while a realistic and consistent (see \textit{e.g.} Section\,\ref{wd}) description of the Wilson depression is also essential.

\subsection{The Premise of the Article}
The basic premise of this article is to satisfy two complementary goals. The first goal is to present a critical review of the existing physical models for the subsurface structure of sunspots, in the context of local helioseismology and numerical simulations of wave-fields, and magnetic field--wave interaction. As discussed above, physical sunspot models are critically important to assess the validity of the helioseismic inversions.
In addition, numerical simulations of the propagation of solar waves through model sunspots are emerging as a valid and realistic technique to interpret helioseismic data. The success of this approach relies on a very close interaction between sunspot modelers and helioseismologists.

The second goal is to extend the helioseismic analysis undertaken by \citeauthor{gizonetal2009}~(\citeyear{gizonetal2009,gizonetal2009a}) of the sunspot in Active Region (AR) 9787, which was the topic of the Third HELAS (European Helio- and Asteroseismology Network) Local Helioseismology Workshop, held in Berlin on 12\,--\,15 May 2009. This sunspot was observed during the period 20\,--\,28 January 2002 by the SOHO/MDI instrument.  Serious inconsistencies between the different helioseismic methods were uncovered, which cannot be left unanswered.

\section{Surface Observational Constraints}\label{s:obs}
In this section we briefly review some of the main observational characteristics of sunspot formation and evolution. The aim here is to present a very general overview of some of the pertinent issues related to sunspot observations that need to be considered when developing a realistic sunspot model. Comprehensive reviews by \inlinecite{solanki2003}, \inlinecite{tw2004}, \inlinecite{tw2004a}, \inlinecite{schlichenmaier2009}, and the books by \citeauthor{tw2008}~(\citeyear{twbook1992,tw2008}) (and references therein), have excellent extended discussions on both the observational and theoretical aspects of sunspots and active regions.

\subsection{Sunspot Formation and Evolution}
\subsubsection{Flux Emergence}\label{emergence}
Individual bundles of magnetic flux are believed to rise from deep in the convection zone and break
through the surface of the Sun. As they approach the surface, each flux bundle is shredded into many separate strands which, upon emergence, are quickly concentrated into small, intense (kG strength) magnetic flux bundles (or
elements) by the vigorous convection occurring in the thin superadiabatic layer at
the top of the convection zone. These small flux elements then accumulate at the boundaries
between granules or supergranules, and some of them coalesce to form small pores \cite{km1996,ls1998}. Some pores and flux elements in turn coalesce to form sunspots.

A number of studies have examined the buoyant rise of magnetic flux sitting initially just
below the surface and rising into a non-magnetized atmosphere (\textit{e.g.} \opencite{ml2001}; \opencite{ml2003}; \opencite{manchesteretal2004}; \opencite{csm2007}). A number of studies modeling the rise of thin flux tubes through the convection zone have also shown that the tubes must have a significant amount of twist in order to maintain their integrity and not fragment in the face of hydrodynamic forces, and indeed observations show that magnetic flux usually emerges at the surface already in a significantly twisted state (\textit{e.g.} \opencite{riethmulleretal2008}).

\subsubsection{Sunspot Formation and Decay}
When the flux does emerge, then it is often in the form of pore structures of the order of a Mm or so in size \cite{vrabec1974,zwaan1978,zwaan1992,mcintosh1981}. Pores have continuum intensities ranging from 80\% down to 20\% of the normal photospheric intensity, with maximum magnetic-field strengths of 1500\,--\,2000 G.
If a growing pore reaches a sufficient size (a diameter of about 3500 km, but sometimes as much as 7000 km) or a sufficient total magnetic flux of order $10^{20}$ Mx \cite{ls1998}, and if the magnetic field reaches an inclination from the vertical that is greater than about $\gamma=35^{\circ}$  \cite{mp1997}, then it forms a penumbra at its periphery and becomes a fully fledged sunspot. The formation of a penumbra is a rapid event, occurring in less than 20--30 minutes, and the characteristic sunspot magnetic-field configuration and Evershed flow are both established within this same short time period \cite{ls1998,yangetal2003}. Furthermore, the fact that the largest pores are observed to be bigger than the smallest sunspots also provides evidence that the pore--sunspot transition is associated with hysteresis (Bray and Loughhead 1964; Rucklidge, Schmidt, and Weiss 1995). 

The time scale for the formation of a large sunspot is between a few hours and several days. A sunspot can span a lifetime of months, but more typically of weeks \cite{solanki2003}. However, this life expectancy is considerably shorter than the magnetic diffusion time, $t_D=l^2/\eta$ (where $l$ is the width of the current sheet and $\eta=1/\sigma$ is the magnetic diffusivity), across a solar active region where estimates for $t_D$ range from hundreds, to thousands of years (see \opencite{pf2000}). This reduced lifetime suggests that a convective instability sets in that enhances the decay process, via fragmentation. Another possible process is the action of turbulent diffusion, owing to the nonlinear dependence of the diffusivity on the strength of the magnetic field \cite{petrovay+moreno1997}. An overview on sunspot decay was presented by \inlinecite{martinez2002}.

\subsubsection{Surface Evolution}
Observational evidence indicates a significant change in the dynamical properties of sunspots and active regions from ``active'' to ``passive'' evolution shortly after their emergence \cite{schussler1987,st1999,sr2005}. Initially, the emerging magnetic flux displays the characteristics of a rising, fragmented flux tube and evolves according to its internal large-scale dynamics (\textit{e.g.} \opencite{mcintosh1981}; \opencite{strousetal1996}). Within a few days after emergence, the proper motion of the sunspots with respect to the surrounding plasma begins to decay. Larger magnetic structures also start to fragment into small-scale flux concentrations, which are largely dominated by the local near-surface flows (granulation, supergranulation, differential rotation, meridional circulation). The magnetic flux is then passively advected by these velocity fields, gradually dispersing over a large area. This process is efficiently described by well-established surface flux-transport mechanisms which model the evolution of the magnetic field at the solar surface (\textit{e.g.} \opencite{wangetal1989}; \opencite{vanballegooijenetal1998}; \opencite{schrijver2001}; \opencite{baumannetal2004}).

\subsection{Sunspot Surface Structure}\label{surface}
\subsubsection{Field Strength and Orientation}\label{field}
A sunspot's maximum magnetic-field strength tends to increase approximately linearly
with sunspot diameter, from around 2000 G for the smallest, to 4000 G or more for the largest sunspot
(\textit{e.g.} \opencite{rj1960}; \opencite{bz1982}; \opencite{kr1992}; \opencite{colladosetal1994}; \opencite{livingston2002}; \opencite{livingstonetal2006}). The field strength drops steadily towards the sunspot's periphery, becoming 700\,--\,1000 G at the edge of the visible sunspot (\textit{e.g.} \opencite{mathewetal2004}).

The field strength decreases with height within the visible outline of the spot. At photospheric levels in
the umbra, line-of-sight field strength decreases of $\approx 1$\,--\,$3$ G\,km$^{-1}$ are observed \cite{balthasar+schmidt1993,schmidt+balthasar1994}, but when averaged over a height range of 2000 km or more, the field strength reduces to $0.3$\,--\,$0.6$~G\,km$^{-1}$ \cite{solanki2003}. The strongest field within a sunspot is usually associated with the darkest part of the umbra and is generally close to vertical. At the visible outer sunspot boundary it is inclined by 70\,--\,80$^{\circ}$ to the vertical \cite{bellotetal2003,mathewetal2004} if one applies inversions that assume a constant field inclination in the atmosphere. Assuming more than one magnetic component or gradients along the line-of-sight, high-resolution polarimetric studies present evidence for magnetic flux that returns to the surface in the outer penumbra (\textit{e.g.} \opencite{wpetal2001}; \opencite{bbc2004}; \opencite{borreroetal2004}; \opencite{langhansetal2005}; \opencite{borreroetal2006}; \opencite{ichimotoetal2007}; \opencite{ichimoto2009}; \opencite{beck2008}; \opencite{jb2008}), \textit{i.e.}, there is evidence that a fraction of the sunspot magnetic field does not extend into the chromosphere, but submerges beneath the photosphere in the outer penumbra. 

\subsubsection{Field Twist}\label{twist}
As we have already mentioned, it appears that at least a small amount of twist is needed in a rising flux tube in order for it to resist fragmentation and preserve its identity as it rises through the convection zone, hence systematic twists of sunspot fields are relevant to the emergence and stability of sunspot magnetic fields. Observation also indicate that the magnetic field of regular sunspots can be twisted, with an azimuthal twist $\phi \approx 10^\circ$\,--\,$35^{\circ}$ \cite{hwc1977,gh1981,ls1990,slm1994,wpetal2001}.

However, the more recent observations of \inlinecite{mathewetal2003} indicate that for regular isolated sunspots, the global azimuthal twist of the field does not significantly exceed $20^{\circ}$. Moreover, \inlinecite{yun1971} and \inlinecite{of1983} have included the effects of an azimuthal field twist in their (self-similar) sunspot models and find that the introduction of a moderately twisted field, compatible with observations, contributes little to the force balance in spots and only slightly changes the main characteristics of their sunspot models (\textit{e.g.} the mixing length parameter, effective temperature, Wilson depression, and the central field strength remain practically the same). This is not surprising, in the view of the fact that the measured $B_{\psi}$ from observations is small compared to $B_z$ and $B_r$ over most of the sunspot region.

\subsubsection{Umbral Dots}
Umbral dots, bright dot-like feature observed inside an umbra, are found in almost all sunspots and also in pores \cite{sobotka1997,sobotka2002}. They are observed to cover only 3\,--\,10\% of the umbral area but contribute 10\,--\,20\% of the total umbral brightness \cite{sbv1993}. Their distribution is not uniform: they can occur in clusters and alignments, and no large dots are found in dark nuclei \cite{rimmele1997}.

On average, umbral dots are 500\,--\,1000 K cooler than the photosphere outside a spot, but about 1000 K hotter than the coolest parts of the umbra itself \cite{kitaietal2007}. The magnetic field in umbral dots appears to be weaker than that in the umbral background \cite{sobotka1997}. \inlinecite{snetal2004} found differences of several hundred gauss and deduced that the fields were more inclined to the vertical, by about $10^{\circ}$ . Furthermore, a small upward velocity of 30\,--\,50 m s$^{-1}$ and 200 m s$^{-1}$ has been reported in umbral dots relative to the umbral background by \inlinecite{rimmele1997} and \inlinecite{sn2002}.

 \inlinecite{sv2006} have presented realistic numerical simulations of umbral magnetoconvection in the context of the monolithic model, by assuming an initially uniform vertical magnetic field. Their model reproduces all of the principal observed features of umbral dots, including their dark lanes \cite{rimmele2008}. Their results provide support for the monolithic model, demonstrating that umbral dots can arise naturally as a consequence of magnetoconvection in a space-filling vertical magnetic field. The more recent numerical simulations of \inlinecite{heinemannetal2007} and \inlinecite{rsk2009} also confirm this.

\subsubsection{Evershed Flow}\label{evershed}
The Evershed flow is observed as an outward directed (horizontal) flow observed in the photospheric layers of
penumbrae. The inverse Evershed flow is an inward directed flow in chromospheric layers. A number of well-established observational properties of the Evershed flow are:
\begin{itemize}
\item The averaged flow velocity increases from the inner to the outer penumbra. From observed line asymmetries it is concluded that the flow is located in the deep photosphere \cite{maltby1964,sbt2004}.
\item  Observed velocities of the flow typically exceed 6 km s$^{-1}$ \cite{rvdv2002,bellotetal2003} in the outer penumbra, but can also exceed 10 km s$^{-1}$ in localized patches (\textit{e.g.} \opencite{bbc2004}).
\item The direction of the Evershed flow is close to horizontal. Azimuthal averages reveal that the flow angle varies from 70\degree~(average flow points upwards) in the inner penumbra to some 100\degree~(flow points downward) in the outer penumbra \cite{schl+schmidt2000}.
\item The Evershed flow is magnetized. This is obvious from Stokes $V$ profiles with more than two lobes. These additional lobes are Doppler shifted (\textit{e.g.} \opencite{schl+collados2002}).
\end{itemize}

Two models have been proposed to explain a number of observational properties: the ``siphon-flow'' model as proposed by \inlinecite{mt1997}, and the ``moving-tube'' model by \inlinecite{sjs1998}. \inlinecite{mt1997} elaborated on the idea of \inlinecite{ms1968} that the flow is driven by a gas pressure difference between the footpoints of a thin magnetic flux tube in magnetohydrostatic (MHS) equilibrium. On the other hand, \inlinecite{sjs1998} developed a dynamical 2D model of a thin magnetic flux tube that acts as a convective element in a superadiabatic and magnetized penumbral atmosphere. In their model, the convective rise of the thin flux tube to the surface initiates a local pressure gradient build up, leading to a gas flow along the tube. Penumbral grains are then identified as the hot upflow locations where the gas reaches the (optical) surface.

Numerical simulations of radiative magnetoconvection in inclined magnetic fields (\textit{e.g.} \opencite{heinemannetal2007}; \opencite{snh2008}; \opencite{rsk2009}; \opencite{rempeletal2009}) are only beginning to reproduce the structure of the outer penumbra, with its horizontal and returning magnetic fields and fast Evershed flows along arched channels. They have already succeeded in reproducing single elongated filaments with lengths of up to a few Mm which resemble in many ways what is observed as thin light bridges and penumbral filaments of the inner penumbra. In their simulations, they find that the progression of the filament heads toward the umbra during their formation phase is not caused by the inward motion of a narrow flux tube, but rather due to the expansion of the sheet-like upflow plumes along the filament.

\subsubsection{Moat Flow and Moving Magnetic Features}\label{mmf}
The moat flow is an outflow that initiates immediately after the formation of a penumbra. Moats are typically 10 to 20 Mm wide, with the outer radius of the moat appearing to scale with the size of the enclosed sunspot, being about twice the radius of the spot itself \cite{bl1988}. The moat flow velocity is about 0.5 to 1 km s$^{-1}$, and can be seen by proper motions of granules as well as by Doppler shift measurements \cite{balthasar+etal1996}. The flow usually persists over the duration of the spot's life, while the area and magnetic flux of the sunspot decrease at a roughly constant rate. As the moat flow evolves, it pushes the magnetic flux to its periphery, leaving the moat largely free of magnetic field except for small magnetic features (known as moving magnetic features) that move outward across the moat at speeds of about 1 km s$^{-1}$ \cite{sheeley1969,sheeley1972,vrabec1971,vrabec1974,hh1973,bl1988,wz1992,ywg2001,sb2008}.

The moat flow is only present in the presence of a penumbra.  Pores that have no penumbra also lack the moat flow. Observations of irregular sunspots by \inlinecite{vdetal2007} indicate that the moat flow exists only on sunspots sides where a penumbra has formed. On sunspot sides where the umbra and the granulation are adjacent, no moat flow is detected. Moreover, in such irregular configurations moat flows are only observed as radial extensions of penumbral filaments, but not perpendicular to the filament.

The moat flow has also been detected through local helioseismology, using $f$-mode time--distance helioseismology. Using azimuthally-averaged MDI data. \inlinecite{gdl2000} find an outflow extending well beyond the sunspot boundary (up to 30 Mm) which reaches a peak of 1 km s$^{-1}$ just outside the
penumbra. This flow is consistent with the moat flow. In another recent study, \citeauthor{gizonetal2009}~(\citeyear{gizonetal2009,gizonetal2009a}) used $f$ to $p_4$ ridge-filtered time--distance travel times to produce linear inversion for flows around AR 9787. These inversions showed an azimuthally-averaged horizontal outflow in the first 4 Mm beneath the surface, reaching an amplitude of 230 m s$^{-1}$ at a depth of 2.6~Mm and radial distance of some 5 Mm outside the outer spot boundary. The inversion results were in line with observations of the moat flow in AR 9787 presented by \citeauthor{gizonetal2009}~(\citeyear{gizonetal2009,gizonetal2009a}), the strength and extent of which was characterized by measuring the observed motion of the moving magnetic features (MMFs) using a local correlation tracking method. Their measurements indicated a peak amplitude of 230 m s$^{-1}$ for the moat flow, extending out to about 45 Mm from the spot center. 

As the moat flow is unmagnetized and has velocities that are much smaller than the Evershed flow, the link between the two is not obvious. One possibility is that the gas pressure that builds up beneath the penumbra could drive the moat flow (beneath the penumbra the magnetopause is largely inclined). \inlinecite{sainz+bellot2008} detect bipolar MMFs within the penumbra which migrate outwards into and throughout the moat. This is consistent with an scenario proposed by \inlinecite{schl2002}: a magneto-convective overshoot instability in an Evershed flow channel leads to a bipolar MMF that travels outwards along the magnetic canopy.

\subsection{Sunspot Thermodynamics in the Photosphere}\label{thermo}
There are a number of semi-empirical and observational models as reference, consisting of both one- and two-component models, for the umbra and for the penumbra (see Section\,\ref{semodels} for a more detailed discussion). The basic assumption underlying almost all single component models is that it is possible to describe all umbrae (or at least those above a certain size) by a single thermal model.

Thus, a very important question in this context is whether sunspot brightness (temperature) or magnetic field strength actually varies with the size of the sunspot. Theoretical models of sunspots based on MHS equilibrium and inhibition of convective heat transport (\textit{e.g.} \opencite{deinzer1965}; \opencite{yun1970}) typically obtain lower temperatures for stronger magnetic fields. However, observations by \inlinecite{rs1970} and \inlinecite{am1981} indicated no dependence of umbral intensity on umbral size for umbral
diameters greater than about $8^{\prime\prime}$. 
On the other hand, the observations of \inlinecite{kr1992}
showed a nearly linear decrease in umbral brightness with umbral radius for six sunspots. A similar result was obtained for seven spots at visible wavelengths by \inlinecite{mpv1993} and \inlinecite{colladosetal1994}. This was also confirmed in a more recent study of continuum images of more than 160 sunspots taken during solar Cycle 23 by \inlinecite{mathewetal2007}. Indeed, \inlinecite{ng2004} use this dependence to predict the peak field strength of a sunspot from its brightness to an accuracy of about 100 G. Hence, even though the relatively homogeneous umbral nuclei cannot be described by a single universal atmosphere, these models may nonetheless be used as boundary conditions in theoretical global sunspot models.

Another related question is whether the umbra-photosphere brightness ratio of large sunspots varies over the solar cycle. \citeauthor{am1978}~(\citeyear{am1978,am1981}) find that sunspots were darkest at the beginning of sunspot Cycle 20 and that spots appearing later in the cycle were progressively brighter (with a nearly linear dependence on the phase of the cycle), up until the new Cycle 21 spots appeared which were again darkest. Subsequent observations showed the same behavior occurred in Cycle 21 \cite{maltbyetal1986}. \inlinecite{pl2006} found similar behavior during Cycle 23. However, the results of \inlinecite{mathewetal2007} find no significant change in umbral brightness over Cycle 23.

\subsection{The Wilson Depression}\label{wd}
The reduced opacity in a sunspot, and the consequent depression of the $\tau=1$ level,
arise mainly from two effects: \textit{i}) the reduced temperature in the spot atmosphere leads to a decrease in the H$^{-}$ bound-free opacity, and \textit{ii}) the radial force balance (including magnetic pressure and curvature forces) demands a lower gas pressure within the spot, further reducing the net opacity.

A purely observational determination of the Wilson depression is complicated by evolutionary changes in the shape of the spot. Estimates from observations range from 400\,--\,1500 km for mature spots \cite{bl1964,gz1972,bw1983}. \inlinecite{mpv1993} assume a linear relation between magnetic pressure and temperature in a spot (as indicated by observations), in which case the radial force balance yields a simple relationship between the net magnetic curvature force and the Wilson depression. They computed a Wilson depression of 400\,--\,800 km in the umbra of their observed sunspot (the range being due to different assumed values of the curvature forces). \inlinecite{swl1993} and \inlinecite{mathewetal2004} used this approach to determine the variation of the Wilson depression with radius across a sunspot. They found values of some 100 km in the penumbra and some 400 km in the umbra, with a fairly sharp transition at the umbra-penumbra boundary. In a more recent study, \inlinecite{watsonetal2009} compared observations
of sunspot longitude distribution and Monte Carlo simulations of sunspot appearance using  different models for spot growth rate, growth time and depth of Wilson depression, deducing a mean depth for the umbral $\tau=1$ layer of 500\,--\,1500 km.

\section{The Deep Structure of Sunspots}\label{spruit}
Though the small-scale structure seen at the surface of a sunspot is not exactly in equilibrium, the spot itself survives on much longer time scales than would be expected if it were a superficial structure. It lives for much longer than the time scale on which it would evolve if it were not in a stable equilibrium, which is the time for the Alfv\'en speed to cross the size of the spot (on the order of an hour).

The magnetic field of a spot cannot be just a surface phenomenon, however, since magnetic-field lines have no ends. The extension of the spot's field lines above the solar surface can be observed in the chromosphere and corona, but they must continue below the surface as well. In contrast with a scalar field like pressure, the magnetic field of a sunspot cannot be kept in equilibrium simply by pressure balance at the surface: the tension in the magnetic-field lines continuing below the surface exerts forces as well. The question of spot equilibrium thus involves deeper layers, down to wherever the field lines continue. This is the well-known ``anchoring'' problem of sunspots (see \opencite{parker1979}): at which depth, and by which agent is the sunspot flux bundle kept together?

\subsection{The Anchoring Problem}
As an answer to this problem, \inlinecite{parker1979} postulates the existence of a horizontal flow at some depth below the surface, converging on and flowing through the magnetic flux bundle of the spot. The drag force of this flow on the field would prevent the bundle from fragmenting. There are some obstacles to this idea. A horizontal flow is observed around spots (the moat flow), but it is of the opposite sign to the proposed inflow. There is no theory for the cause of the proposed flow of \inlinecite{parker1979}. In fact, the ``heat flux blocking'' by the sunspot would cause a flow of opposite sign, and this is actually observed in the form of the moat flow. It is also questionable if the proposed flow would actually be sufficient to keep the flux bundle together, in view of the (interchange) instabilities to be expected in this picture \cite{schuessler1984}. What has kept Parker's proposal alive, however, is the helioseismic inference \cite{duvalletal1996} of a huge downflow flow of 2 km s$^{-1}$ of the sign proposed by Parker. It is a puzzling observation, which if confirmed, would require new theoretical ideas. However, more recent helioseismic inferences derived from inversions of subsurface flows around the sunspot in AR 9787 (see Section\,\ref{sec:haber}) do not confirm the existence of this downflow.

A second proposal (already implicit in \opencite{cowling1953}, see also \opencite{babcock1963}; \opencite{leighton1969}; \opencite{sr1983}) is that the flux bundle of a spot actually continues all the way to the base of the convection zone, to the layer of toroidal field from which the active region erupted. At this depth, a field strength of $\approx 80$ kG is inferred from flux tube emerging calculations (\textit{e.g.} \opencite{dc1993}; \opencite{cms1995}). The Alfv\'en travel time for a change at the base to propagate to the surface (more accurately: the propagation time of a transverse tube wave) is then about five days. This is the time scale on which the spot would change if there were nothing to maintain conditions at the base. The idea is thus that a spot, while in equilibrium at the surface, would actually be a transient structure in the layers from which its magnetic field originates. This proposal agrees roughly with the lifetime of most small spots. It is too short, however, to explain the anchoring of stable long-lived spots.

\subsection{The Issue of Flux Emergence}
The anchoring problem of long-lived spots is thus still somewhat open. Quite clear, however, is the picture of flux emergence: the process by which a loop of flux rises from a horizontal layer of magnetic flux at the base of the convection zone to form an active region, as discussed briefly in Section\,\ref{emergence}. Simplified 1D calculations in ``thin tube'' approximation \cite{dc1993,ffm1994,cms1995} point to a field strength of about $10^5$ G at the base of the convection zone. At this strength agreement is reached with three independent key properties of active regions: \textit{i}) the time scale of emergence (a few days), \textit{ii}) the heliographic latitude range of emergence, and \textit{iii}) the tilt of active region axes (\textit{e.g.} \opencite{csm1998} and references therein). This is also the field strength at which the horizontal field is first expected to become unstable to the development of bends in the field lines, creating loops rising to the surface \cite{schuessleretal1994}.

However, the picture presented by rising flux tube simulations is somewhat incomplete, since the last stages of the flux emergence process in the near photospheric layers are typically excluded. It has been pointed out by \inlinecite{sr2005} that the typical size of active regions correspond to a wavenumber of $m=$ 10 to 60, while buoyant flux tubes typically prefer $m=$ 1 or 2 instabilities. From this mismatch in scales one would expect an unrealistic drift of sunspots apart from each other much further than actually observed due to magnetic tension forces. Possible solutions could include a ``dynamical disconnection'' as suggested by \inlinecite{sr2005} or a flux emergence process that prefers larger wave numbers from the beginning. Observational evidence for this process has recently been provided by \inlinecite{sks2009}.

Overall, the convergence of these lines of evidence supports the basic correctness of the flux emergence picture of active region formation that has long been intuitively evident from observations \cite{cowling1953}. It has the important implication that the energy density of the magnetic field at its source is more than two orders of magnitude larger than energy density in convective flows; at such a strength the field is nearly unaffected by convective flows. The same is the case at the surface of a sunspot. This is incompatible with the cornerstone of traditional mean-field dynamo models based on the effect of convective turbulence acting on magnetic fields. The picture sketched above and the mean field convective dynamo model thus cannot both be correct. This fact is remarkably politely ignored in discussions on theories of the solar cycle.

\subsection{Anchoring Where?}
Some details of the evolution of an active region give additional clues about the anchoring of sunspots. After their formation, long-lived spots wander about a bit in longitude and latitude (\textit{e.g.} like AR 9787, see \citeauthor{gizonetal2009},~\citeyear{gizonetal2009,gizonetal2009a}), settling to their final position over the course of a few days (\textit{e.g.} \opencite{mcg1990}). This is similar to the inferred Alfv\'en travel time from the base of the convection zone and the surface. The observed settling process thus agrees with the notion of anchoring in deep layers. The ``settling time scale'', the Alfv\'en travel time from the anchor to the surface, agrees with the field strength at the base of the convection zone inferred from the other properties of the emergence process mentioned above.

While anchoring at the base of the convection zone thus agrees with a range of observational indications, this does not in itself prove that other locations within the convection zone are excluded. Such locations are, however, somewhat unlikely. Apart from the boundary with the stably stratified interior it is hard to find a plausible location for anchoring a magnetic field in a convecting stellar envelope, which is itself unstable to fluid displacements. The anchor of a long-lived spot needs to act over a time longer than the convective turnover time of the envelope as a whole (which would be on the order of a month, in the classical mixing length view).

\section{Sunspot Models}\label{models}

\subsection{Semi-Empirical Models of the Sunspot Atmosphere}\label{semodels}
Semi-empirical models of umbral or penumbral atmospheres give the
variation of thermodynamic variables and magnetic-field vector
with optical depth based on empirical data
and theoretical considerations of mechanical equilibrium and
radiative transfer. Semi-empirical models can be divided into two
groups based on the observational material and methods used for
their derivation.

One group of such models are spatially unresolved
one-dimensional models based on multi-wavelength observations of
the continuum, weak and strong spectral lines. The general
procedure in constructing a model atmosphere usually involves
first determining a temperature-optical depth relation, as a best
fit to the empirical data, and then to determine the gas and
electron pressures by integrating the equation of hydrostatic
equilibrium. In general, a number of assumptions/generalizations
are made when calculating this type of semi-empirical models:
\textit{i}) For the lower photospheric layers, measurements of
the center-to-limb variation of continuum intensity at several
spectral intervals and the profiles of the weak spectral lines (or
the wings of strong lines) are used together with the LTE
assumption. While for the high chromospheric layers strong
spectral lines are used like \ion{Ca}{ii} H and K and IR lines,
H$\alpha$, \ion{Na}{i} D, \textit{etc.} and the analysis is carried in NLTE; \textit{ii}) Hydrostatic equilibrium is assumed. The question of the
magnetic-field distribution is not addressed.

Most models consist of a single-component, meant to represent a
horizontal average over the umbra or penumbra, while a number of models consist of
two-components, designed to treat bright and dark regions
separately in observations that do not fully resolve them
spatially.

Another group of models uses a relatively new strategy based on
the inversion of the radiative transfer equation of polarized
light (\textit{e.g.} \opencite{rd1992}). The input
material for these models are spectro-polarimetric or spectral
observations (typically high-resolution). Strict mathematical
methods are used to find the best fit to the spectral line
profiles and only few spectral lines are required to produce a
model atmosphere. This method provides sunspots models (umbra,
penumbra and the surrounding plage) spatially resolved to the
extent allowed by the observations. The following assumptions are
typical:
\begin{itemize}
\item The inversion of radiative transfer equation is done under LTE
conditions for photospheric lines. NLTE inversion results have
been reported for the chromospheric ion{Ca}{ii} infrared triplet
\cite{sn2005}. All variables are derived
as a function of optical depth.
\item The gas stratification is obtained at heights of formation
of spectral lines, at most $\approx$120 km below, and up to
$\approx$300\,--\,400 km above the continuum level. The data about the
chromospheric layers are less frequent (see \textit{e.g.} \opencite{sn2007}).
\item The models include magnetic-field vector and line of sight
velocity component with or without gradients.
\item For practical reasons, the atmosphere is assumed to be in MHS
equilibrium at each spatial location.
\item The Wilson depression is not an inherent part of the models,
but instead must be determined from additional considerations
(\textit{e.g.} lateral pressure balance).
\end{itemize}
Such semi-empirical models can include one, two or more magnetic
components describing the variations of thermodynamic parameters,
velocities and magnetic-field vector at each individual spatial
location, supposing these variations are not completely resolved
by observations. The shape of the polarization profiles in the
penumbra suggest that they result from at least two different
magnetic components, one of them carrying the Evershed flow
\cite{br2003}. Note that the magnetic-field strength and its
inclination in the penumbra are correlated with the brightness of
the penumbral filaments \cite{bs1969,schmidtetal1992,titleetal1992,br2003,solanki2003},
thus the components with different field strength also typically
have different temperatures.

All these models are useful in constraining certain physical
processes that determine the structure of a real sunspot
atmosphere, and also in providing a background model for studies
of element abundances, wave propagation, and other behavior in a
sunspot.

\subsubsection{Umbral Models}
Below is a summary of some of the more often used umbral models.
The first three belong to the first group of models described
above and only provide variations of thermodynamical quantities:
\begin{itemize}
\item \inlinecite{avrett1981}: Provided a combined model of the umbral
photosphere, chromosphere and transition zone, by combining the
low photospheric model of \inlinecite{am1981} with the upper photospheric and chromospheric parts
of \inlinecite{ls1982} model and the transition region model of \inlinecite{nicolasetal1981}.
\item \inlinecite{staudeetal1983}: Derived a comprehensive umbral model
covering the full height range from the photosphere to the corona.
Their model was based on a large set of observations (radio, optical, EUV,
X-ray). The photospheric structure was taken largely from \inlinecite{sw1975}, while its lower
chromospheric stratification is similar to that of \inlinecite{tgs1977}. They also introduced two components (hot and cold) in higher layers.
\item \inlinecite{maltbyetal1986}:  Came up with an improvement of the \inlinecite{avrett1981} model in the deep layers. These
consist of a set of three models (each for a different phase of the
solar cycle). Modifications of the \inlinecite{maltbyetal1986} models have been proposed by \inlinecite{cgs1993}, \inlinecite{sgc1994} and \inlinecite{ayres1996}, including constraints from more spectral lines.
\item \inlinecite{colladosetal1994}: Presented models of the umbra of large and
small spots from the inversions of Stokes $I$ and $V$ spectra from
the darkest cores of three different sunspots. The run-with
optical depth of temperature, magnetic-field vector, and velocity
along the line of sight were obtained. Their observations confirmed
that there are noticeable differences in the temperature and field
strength between the umbra of large and small spots.
\end{itemize}

The temperature stratifications of all of the above-mentioned umbral models reveal an umbra which is cooler than the surrounding field-free regions in the photospheric layers, but slightly hotter in the chromosphere, while the transition to coronal temperatures takes place at a lower
height in the umbra (see \opencite{staude1986}; \opencite{solanki1990}; and \opencite{colladosetal1994} for comparison plots).

\subsubsection{Penumbral Models and Spatially Resolved Models}
Below is a brief summary of some semi-empirical average penumbral
models and models obtained from the inversion of the
high-resolution polarized spectra including complete sunspots or
their parts:
\begin{itemize}
\item \inlinecite{mm1960}, \inlinecite{kmm1969}, \inlinecite{ybb1984} and \inlinecite{df1989}: Produced one-component models, with the latter two
being primarily aimed at modeling the average penumbral
chromosphere. \inlinecite{kmm1974} produced a two-component model of similar kind that provided
temperature as function of optical depth in dark and bright
penumbral filaments.
\item \inlinecite{deltoroiniestaetal1994}: Applied an inversion
technique to a series of high-resolution filtergrams, scanning a
magnetically insensitive \ion{Fe}{i} line to study the vertical
temperature and velocity structure over a small penumbral region.
They provided a mean penumbral model atmosphere as a function of optical
depth.
\item \inlinecite{wpetal2001}: Besides deriving the velocity
and temperature profiles, they also derived the magnetic-field
stratification of a sunspot from inversions of observations of
magnetically sensitive \ion{Fe}{i} lines at 630 nm with a spatial
resolution of approximately $1^{\prime\prime}$ and one magnetic component
together with stray light component in each observed pixel.
Similar models were provided by \citeauthor{mathewetal2003}~(\citeyear{mathewetal2003,mathewetal2004}) from the
inversion of infrared \ion{Fe}{i} spectral lines at 1.56 $\mu$m. \inlinecite{mathewetal2004} also calculated maps of Wilson depression and plasma parameter $\beta$.
\item \inlinecite{rvdv2002}: The observed radiation temperatures
in the \ion{Ca}{ii} K wing were used to derive the temperature
stratification of fine-structure elements in the penumbra. Three
atmospheric models were constructed to represent cool,
intermediate, and hot features within the penumbra, with
temperature differences of order 300 K between them.
\item \inlinecite{br2003} and \citeauthor{borreroetal2004}~(\citeyear{borreroetal2004,borreroetal2005,borreroetal2006}): Inverted the polarization profiles from the penumbra and considered the uncombed penumbral model with two magnetic components, where the component with a
more horizontal and weaker field harbors the Evershed flow.
\item \inlinecite{sn2005}: Produced a 3D sunspot
model, up to chromospheric heights, from NLTE inversion of the \ion{Ca}{ii}
infrared triplet polarized spectra.
\item \inlinecite{prm2008}: By assuming that the fine structure of the penumbra
is spatially resolved in high-resolution \textit{Hinode}/SOT data, they invert the data with a single component
model. The advantage of their modeling is a very detailed and self-consistent translation from optical depth to geometrical height
scale taking into account terms of the Lorentz force and using a
generic algorithm. The Wilson depression was derived this way and
the model was checked to be divergence-free and in equilibrium in
the horizontal and vertical directions.
\end{itemize}

Unlike quiet Sun models, sunspot models should not be used
indiscriminately to all cases and can only be taken as
representative examples. As was discussed in Section\,\ref{thermo}, a number of observations have shown that the
thermal stratification of sunspots depends sensitively on its
magnetic-field strength, sunspot size, and possibly, on the solar cycle as well.

\subsection{MHS Models}\label{ss:mhs}
The simplest static models of a pore or a sunspot ignore azimuthal variations and
treat it as an axisymmetric, poloidal magnetic field (with no azimuthal component) confined to a homogeneous flux tube of circular cross-section. 
The equation (in cgs units) describing the MHS equilibrium of the flux tube is:
\begin{equation}\label{eq:fb}
-\mathbf{\nabla} p + \rho \mathbf{g} + \frac{1}{c}(\mathbf{J}\times\mathbf{B})=0,
\end{equation}
where $p$ is the gas pressure, $\rho$ is the gas density, $\mathbf{g}$ is the acceleration due to gravity, $c$ is the speed of light, and $\bf{B}$ is the magnetic-field vector. The electric current density is given by Ampere's Law:
\begin{equation}
\mathbf{J}=\frac{c}{4\pi}(\mathbf{\nabla}\times\mathbf{B}).
\end{equation}

In reality, magnetic flux concentrations are devoid of symmetry (like most sunspots). However, occasional long-lived spots, sufficiently separated from other large flux concentrations, tend to be round and to have regular, ring-like penumbrae. Therefore, an isolated, axially symmetric spot with a unipolar magnetic field is somewhat justified. Furthermore, the lifetime of sunspots is much longer than any dynamical timescale in the solar photosphere. Some sunspots persist for several rotations, essentially unchanged, whereas perturbations that propagate with the Alfv\'en speed would need about an hour to cross a large sunspot at its photosphere.

There are also a number of other assumptions/generalizations to keep in mind when considering MHS models: \textit{i}) The MHS models are generally limited to two dimensions, the vertical and radial directions; \textit{ii}) Dynamic phenomena (important for shaping small-scale magnetic and thermal structure) and all fluctuations related to convective motions are usually ignored; \textit{iii}) Most MHS models (unrealistically) tend to treat the force balance in isolation of the energy balance; \textit{iv}) Unless the energy equation is solved together with the force balance equation, either the magnetic field, temperature, or gas pressure must be specified, as well as an equation of state; finally, \textit{v}) For models in which the force balance and energy equations are consistently solved throughout the sunspot the convective transport is treated by applying the mixing length formalism and the radiative transport by the diffusion approximation (\textit{e.g.} \opencite{chitre1963}; \opencite{deinzer1965}; \opencite{cs1967}; \opencite{yun1970}; \opencite{jahn1989}).

\subsubsection{Force-Free and Potential Field Models}
These are the simplest static model where the magnetic field within the sunspot flux tube is assumed
to be force-free, resulting in the atmosphere being horizontally stratified with variations only
in the vertical direction.
A number of conditions must be satisfied in order to create such a model: \textit{i}) the force-free field must satisfy the condition:
\begin{equation}
(\nabla \times \mathbf{B})\times \mathbf{B}=0,
\end{equation}
and \textit{ii}) the assumed axial-symmetry also requires that the field satisfy the current-free
condition $\nabla\times \bf{B}$\,=\,0, and hence $\bf{B}$ is a potential field $\bf{B}=-\nabla\phi$ (where $\phi$ satisfies Laplace's equation, $\nabla^2\phi=0$).

Potential field models can thus be effortlessly derived by taking solutions of Laplace's equations (\textit{e.g.} using dipole or Bessel function potentials). Simple models of pores were constructed \inlinecite{sw1970}, using Bessel function solutions of Laplace's equation. Subsequent advances of this model were made  by \inlinecite{spruit1976} and by \inlinecite{swn1983}, who represented the field in a pore by a potential field such that, at the photosphere, $B_z$ was uniform over a disc with a prescribed radius and zero outside it.

One obvious advantage of this method is that direct solution of equation (\ref{eq:fb}) is avoided, however the solution now involves the solution of a non-linear boundary value and free-surface problem (\textit{i.e.}, determining the location of the current sheet, as potential fields fill the whole atmosphere unless bounded by a current sheet), which is nontrivial.

\subsubsection{Current Sheet Models}
A more satisfactory potential field model can be derived by constructing a field contained within a flux tube such that the difference between the internal [$p_\mathrm{i}$] and external [$p_\mathrm{e}$] gas pressures is balanced by the Lorentz force in a thin current sheet bounding the flux tube, such that
\begin{equation}
p_i+\frac{B^2}{8\pi}=p_e.
\end{equation}
This bounding current sheet is referred to as the ``magnetopause''.

Approximate solutions have been found and applied to sunspots and pores by \inlinecite{sw1970} and \inlinecite{swn1983}. \inlinecite{wegmann1981} produced the first general solution to the free boundary problem and \inlinecite{sw1983} were the first to apply this technique to sunspots, successfully modeling pores. More advanced current sheet models have succeeded in producing relatively realistic sunspot models, including the penumbra. These models are briefly summarized below:
\begin{itemize}
\item \inlinecite{jahn1989}: Provided an extension of the \inlinecite{sw1983} model by introducing body/volume currents (distributed in the outer parts of a flux tube, below the photosphere of the penumbra), in addition to a current sheet at the sunspot-quiet Sun boundary. The body currents contribute to the lateral force balance and affect the pressure stratification, so that the gas in the penumbra is hotter, thus layers of equal gas pressure assume higher levels than in an umbra where the field is current free.  A fit to the magnetic and photometric profiles (taken from \opencite{bs1969}) provided a distribution of the electric currents in their model.
\item \inlinecite{js1994}: Introduced two current sheets, one at the outer boundary of the flux tube (the magnetopause) and the other at the interface between the penumbra and the umbra. They did this in order to obtain a more realistic thermal structure of the sunspot with distinctively different umbral and penumbral thermal mechanisms. They also assumed that the umbra is thermally insulated from the penumbra, while some of the energy radiated from the penumbra itself is supplied by convective processes that transfer energy across the magnetopause.
\item \inlinecite{pizzo1990}: Extended the earlier work of \opencite{pizzo1986} (see Section\,\ref{fullmhs}), using multi-grid techniques to calculate the magnetic structure of a sunspot bounded by a current sheet.
\end{itemize}

In general, current sheet models are consistent with the concept that sunspots represent discrete, erupted, magnetic entities \cite{solanki2003}. Observational support for the current-sheet description of spots is taken from the sharp transition between the umbra and quiet photosphere in pores and from the relatively uniform photometric appearance of most umbrae \cite{gz1972}. The fact that the magnetic field is so large at the white-light boundary of the sunspot also strongly suggests that sunspots are bounded by a current sheet \cite{ss1993}. However, as \inlinecite{solanki2003} points out, the rugged nature of the sunspot boundary in white-light images means that the current sheet is not as well defined as one might picture on the basis of simple flux-tube models. Further evidence for a current sheet surrounding sunspots comes from observations suggesting that the field inside sunspots is close to potential. This suggests that the currents bounding the strong field must be mainly located in a relatively thin sheet at the magnetopause (\textit{e.g.} \opencite{ls1990}).

\subsubsection{Self-Similar Fields}\label{selfsimilar}
There is considerable arbitrariness in assigning the distribution of the volume current and the corresponding radial structure in the sunspot atmosphere. As first shown by \inlinecite{st1958} (and later extended by others, \textit{e.g.} \opencite{chitre1963}; \opencite{jakimiec1965}; \opencite{jz1966}; \opencite{cs1967}), the problem can be greatly simplified by assuming a self-similar profile for the magnetic field.

In cylindrical coordinates, the (untwisted) $B_z$ and $B_r$ components of the magnetic field take the form:
\begin{equation}\label{eq:bz}
B_z(r,z)=f(\zeta) B_0(z),
\end{equation}
\begin{equation}\label{eq:br}
B_r(r,z)=-\frac{r}{2}f(\zeta)\frac{\mathrm{d}B_0(z)}{\mathrm{d}z},
\end{equation}
where $r$ and $z$ refer to the radial and vertical coordinates respectively, $B_0(z)$ is the field strength at the flux-tube axis, and $\zeta=r\sqrt{B_0(z)}$. The shape of the function $f(\zeta)$ may be freely chosen (usually a Gaussian). For a non-constant $f(\zeta)$, inserting Equations (\ref{eq:bz}) and (\ref{eq:br}) into (\ref{eq:fb}) reduces the equation of MHS equilibrium to the following equations:
\begin{equation}\label{eq:ss1}
0=-\frac{\partial p}{\partial r} + \frac{B_z}{4\pi}\left(\frac{\partial B_r}{\partial z}-\frac{\partial B_z}{\partial r}\right),
\end{equation}
\begin{equation}
0=-\frac{\partial p}{\partial z} - \frac{B_r}{4\pi}\left(\frac{\partial B_r}{\partial z}-\frac{\partial B_z}{\partial r}\right) -\rho g.
\end{equation}
Integrating Equation (\ref{eq:ss1}) over $r$ from 0 to infinity for constant $z$ leads to the following  expression for $\Delta p(z)=p_e(z)-p_i(z)$, the difference in gas pressure between the external [$p(\infty,z)$] and internal [$p(0,z)$] regions:
\begin{equation}
\Delta p(z)=-\frac{1}{8\pi}\left(\frac{\Phi}{2\pi} y \frac{\mathrm{d}^2y}{\mathrm{d}z^2}-y^4\right),
\end{equation}
where $\Phi$ denotes the total magnetic flux and $y=\sqrt{B_0(z)}$. Thus this method has the advantage in that it simplifies the mathematical treatment of MHS equilibrium by reducing the partial differential equation to a second-order ordinary differential equation for the field strength at axis of the spot by specifying the cross-sectional shape of the magnetic field distribution within an axisymmetric flux tube.

A general assumption made is that the distribution of magnetic flux on horizontal planes is geometrically similar at each depth. Furthermore, in contrast to current sheet models, self-similarity allows for a continuous variation of field strength and gas pressure across the spot. The field falls off smoothly from the central axis value to zero at large radial distances (hence, no clear definable ``inside'' or ``outside'' of the spot in this description, therefore essentially ignoring the fact that sunspots have sharp edges). Thus, there is the computational convenience as treatment of the discontinuity associated with a current sheet is avoided. However, the similarity law enforces a somewhat arbitrary distribution of electric current, resulting in the appearance of a bright ring in the emergent intensity. Those currents determine (to some extent) the horizontal temperature variations, which is in general need not comply with the observed photometric profile \cite{jahn1992}.

A further disadvantage of a purely self-similar sunspot model is that negative pressures and densities are often obtained in the photospheric and upper atmospheric layers of the flux tube, due to the fact that the hydrodynamic pressure and density are decreasing exponentially with height, while the magnetic field does not quite decrease at the same rate. There have been some proposed workarounds to this problem (see \textit{e.g.} \opencite{hanasoge2008}), however such methods are not ideal as they tend to substantially alter the governing differential equations because they require the inclusion of terms in the ideal MHD equations which are not physical.

Furthermore, since the Lorentz force also drops with height, the magnetic field essentially becomes an unbounded, force-free field within a few pressure scale heights above the photosphere, resulting in a field configuration not too different from a potential field. This fact implies that self-similar models are essentially not force-free in the regions where they should be. Since the shape of the magnetic field at the surface is sensitive to this, one has here a model that breaks down in an essential way in just those regions where diagnostics are best.

Nonetheless, due to their simplicity, similarity expansions have been utilized to generate the field configuration for a number of studies, some of which are summarized below:
\begin{itemize}
\item \inlinecite{deinzer1965}: Generalized the \inlinecite{st1958} model, where the similarity law for the magnetic field is coupled with the thermodynamic structure along the axis of a spot, as described by mixing length theory.
\item \inlinecite{yun1970}: Improved the \inlinecite{deinzer1965} model by the introduction of an ``effective surface monopole'', which controlled the inclination of the particular field lines identified with the outer edge a spot at the surface. Hence, the upper boundary condition takes into account the fact that the gas pressure difference at the photospheric level is not negligible (as assumed by \opencite{deinzer1965}). Lower boundary conditions were also modified, as the effects of partial ionization on the relation between the internal and external pressures and temperatures were included.
\item \inlinecite{yun1971} and \inlinecite{of1983}: Demonstrated that the introduction of a moderately twisted field ($\approx 17^{\circ}$ near the surface, compatible with observations) contributes little to the force balance in spots and only slightly changes the main characteristics of the model (as already mentioned in Section\,~\ref{twist}).
\item  \inlinecite{lf1979}: Imposed an Evershed-type radial velocity distribution in the upper region of the spot atmosphere, in order to get a satisfactory continuum intensity profile across the sunspot.  However, relatively large values of the Evershed flow (\textit{i.e.}, close to 10 km s$^{-1}$) were required to obtain a satisfactory temperature profile.
\item \inlinecite{low1980}: Prescribed a method for generating exact solutions of MHS equilibrium describing a cylindrically symmetric magnetic flux tube oriented vertically in a stratified medium. Given the geometric shape of the field lines, compact formulas were presented for the direct calculation of all the possible distributions of pressure, density, temperature and magnetic-field strength compatible with these field lines under the condition of static equilibrium. A particular solution was obtained by this method for a medium sized sunspot whose magnetic field obeys the similarity law of \inlinecite{st1958}.
\item \inlinecite{osherovich1982}: Extended self similar models to include field lines in the outer part of the sunspot that return to the solar surface just outside the visible sunspot (return flux). The emerging flux constitutes the penumbra. The predicted continuum intensity of return flux models was not much closer to the observations than the standard self-similar models.
\item \inlinecite{fso1982}: Applied the return-flux model to a spot with the observational data of \inlinecite{ls1982}  for pressure, maximum field strength and size. The force balance equation was solved to obtain self-consistent magnetic field, pressure, and temperature distributions. The resulting distributions appeared to yield improved representations of umbral--penumbra and penumbra--quiet Sun boundaries compared to regular (\textit{e.g.} \opencite{st1958}) self-similar models. However, it appears that one needs to introduce an Evershed flow to eliminate the apparent umbral bright ring in the continuum.  Similar work on return-flux sunspot models was undertaken by \inlinecite{ol1983}, \inlinecite{og1989} and \inlinecite{ls1996}.
\item \inlinecite{solovev1997}: Extended the self-similar sunspot models by introducing a current sheet at the sunspot boundary.
\item \inlinecite{myp1998}: Included a description of the energy balance and the observed horizontal variation of the Wilson depression when determining the shape function from the observed radial dependence of the magnetic field
\item \inlinecite{cgd2008}, \inlinecite{hanasoge2008}, \inlinecite{mhc2008}, and \inlinecite{sheylagetal2009}: All employed simple self-similar toy sunspot models in conducting numerical simulations of helioseismic wave propagation through magnetized plasmas.
\end{itemize}
As we shall see in the proceeding sections however, constructing more physically realistic sunspot models is practical, especially with the MHD codes (Section\,\ref{forward_models}) that are currently used for the wave-propagation problem.

\subsubsection{Solution of Full MHS Force Balance}\label{fullmhs}
These methods involve solving for the magnetic field on the basis of full MHD equilibrium, with and without a current sheet. Pressure is specified throughout the numerical domain (usually as a function of depth and taken from semi-empirical models), partly depending on the distribution of field lines (hydrostatic equilibrium acts along each field line).

 \inlinecite{pizzo1986} utilizes the description of \inlinecite{low1975}, who proposed transforming the thermodynamic parameters (the pressure and temperature of the gas) into functions of the magnetic vector potential and depth. In this form, a functional form is prescribed for the gas pressure, but not for the magnetic field as in the similarity models, and the equilibrium is solved as a classical non-linear boundary value problem.

By transforming the pressure and density into functions of the field line constant $u$ (used in both \opencite{low1980} and \opencite{pizzo1986}, it essentially determines the shape of the field lines) and height $z$, and requiring hydrostatic equilibrium along the field lines, Equation \ref{eq:fb} can be reduced to a single scalar equations describing the magnetostatic equilibrium of an axisymmetric, poloidal field:
\begin{equation}\label{eq:low}
\frac{\partial^2 u}{\partial r^2}-\frac{1}{r}\frac{\partial u}{\partial r}+\frac{\partial^2 u}{\partial z^2}=-4\pi r^2\frac{\partial P (u, z)}{\partial u},
\end{equation}
where $P$ is a function related to the gas pressure distribution. \inlinecite{low1975} provides an approximation for the distribution of gas pressure along the magnetic-field lines in a vertical, axisymmetric flux tube in magnetostatic equilibrium,
\begin{equation}
P(u,z)= P_0(u)\exp\left[-\int^z_0 \frac{\mathrm{d}z'}{h(u,z')}\right],
\end{equation}
where $P_0(u)$ is the gas pressure along the lower boundary, $h(u,z)$ is the isothermal scale height ($h=\mathcal{R}T/\mu g$, where $\mathcal{R}$ denotes the ideal gas constant, $T$ the temperature, $\mu$ the mean molecular weight, and $g$ is the acceleration due to gravity) for a plasma obeying the ideal gas law ($p=\rho\mathcal{R}T/\mu$). The $u=$~constant curves describe the field lines of the system. The vertical and radial field components may then be expressed in terms of $u$:
\begin{equation}
B_z=\frac{1}{r}\frac{\partial u}{\partial r}
\end{equation}
and
\begin{equation}
B_r=-\frac{1}{r}\frac{\partial u}{\partial z}.
\end{equation}
The range of validity is determined by the representative pressure distributions along the axis and in the field-free atmosphere.

The gas pressure difference between the quiet photosphere and the axis of the spot is needed for the computation. \inlinecite{pizzo1986} takes these values from the semi-empirical models of the umbral photosphere derived by \inlinecite{avrett1981}. \inlinecite{pizzo1986} then develops a method for the iterative numerical solution of Equation (\ref{eq:low}), essentially a second order, non-linear, elliptic partial differential equation, which can be easily solved using standard numerical techniques in the case of fixed boundary conditions.

An advantage of this model is that the Wilson depression and net internal-external pressure difference can be adjusted by vertical translation of the absolute height scales of the two reference atmospheres. However, the configuration considered by \inlinecite{pizzo1986} has its base placed ~120 km below the visible surface of the umbra, which corresponds to $z=0$ in the \inlinecite{avrett1981} model, hence his models do not address the question of the spot structure in deeper layers. The model also assumes a Gaussian profile of the magnetic field across the base (\textit{i.e.} self-similar), thus ignoring the existence of a discontinuous transition from the magnetized to field free plasma (\textit{i.e.}, a current sheet). However, \inlinecite{pizzo1990} later extends his method by incorporating the free-surface problem in the solution of the equation of magnetostatic equilibrium for a flux tube surrounded by an infinitely thin current sheet, utilizing a body-fitted mesh generation and multi-grid relaxation techniques for solving Equation (\ref{eq:low}).

\inlinecite{sps1986} also developed a method for the iterative numerical solution of Equation (\ref{eq:low}), including a boundary current sheet and also field twist in their treatment, while \inlinecite{cally1991} adopted a full multi-grid method to tackle the free boundary problem by formulating it in terms of inverse or flux coordinates, in which the magnetic-field lines become coordinate lines. This results in the energy equation reducing to ordinary differential equations along field lines when the radiation is optically thin. Also, if steady plasma flow is allowed, Alfv\'en's theorem guarantees that there can be no cross-field component of velocity, \textit{i.e.}, that the fluid flows along field lines.

 \inlinecite{kc2006} used the \inlinecite{pizzo1986} sunspot model in numerical simulations of magnetoacoustic wave propagation, and more recently \inlinecite{kc2008} produced a new set of models consisting of concatenation of self-similar models in the deep layers, where the gas pressure dominates over magnetic pressure, with models in which the pressure distribution is prescribed on the axis. In the deep photospheric layers, a self-similar solution for the magnetic field is calculated following the method of \inlinecite{low1980}, while the pressure and density distributions with height and radius are found from analytical expressions. A potential solution is then generated above some arbitrary height using the method of \inlinecite{pizzo1986} -- the bottom boundary of this model coincides with the top boundary of the deep photospheric/self-similar model (approximately $z=-1$~Mm, can be adjusted however to limit the upper boundary of the self-similar model). This solution is then used as an initial guess in the integration of the complete force balance equation along the magnetic-field lines, \textit{i.e.}, Equation (\ref{eq:low}).

The analytical description of the pressure distribution along magnetic-field lines is taken from \inlinecite{pizzo1986} and \inlinecite{low1975}. The Model S \cite{cdetal1996} pressure distribution for the field-free atmosphere is smoothly joined to the VAL-C \cite{val1981} model of the solar chromosphere. At the axis, the \inlinecite{avrett1981} umbral core model is used in the upper layers, while the linear inversion model of \inlinecite{kds2000} is used for deeper layers (down to a depth of 1~Mm, thus ignoring the ``hot'' layer), which takes the Wilson depression to be at 450 km. A smooth transition between the models is then calculated for the gas pressure and scale height distributions, in the same manner as \inlinecite{pizzo1986}, and by changing the parameters of the solution, a set of models with the desired properties can be produced. However, as with the \inlinecite{pizzo1986} and all other pressure-distributed sunspot models, the vertical extent of these models is severely limited by available semi-empirical data on the sunspot axis.

\subsubsection{Extrapolated Field Models}
\inlinecite{martensetal1996} present a force-free constant-$\alpha$ model for the magnetic field in and above a fluted sunspot. They demonstrate that magnetograms for round sunspots can essentially be matched by a series of Bessel functions. Their model parameters are chosen to reproduce the high resolution observations (magnetograms) of \inlinecite{titleetal1993} at the 1-m Swedish Solar Observatory at La Palma, and an analytical expression is obtained for the 3D magnetic field emanating from the sunspot's umbra and penumbra. The model accurately reproduces the azimuthal variation in inclination angle, as well as the mean constancy of the magnetic-field strength, and the appearance of a highly corrugated neutral line on the limb side of off-center sunspots.

Another model based on matching mnagnetograms with analytical expressions is the axisymmetric sunspot model of \inlinecite{mc2008}, which consists of a non-potential, untwisted, 3D MHS sunspot model constrained to fit observed surface magnetic-field profiles. The preferred surface field configuration of the sunspot model was derived from constrained polynomial fits to the observed scatter plots of the radial ($B_r$) and vertical ($B_z$) surface magnetic-field profiles of AR 9026 on 5 June 2000 -- a fairly symmetrical sunspot near disk-center, ideal for helioseismic analysis -- obtained from IVM (Imaging Vector Magnetograph) vector magnetograms. The surface field is therefore quite realistic, which is important because there is evidence (\textit{e.g.} \opencite{sc2006}) that magnetic effects in helioseismology are dominated by the top few hundred kilometres. The fits of $B_r$ and $B_z$ are then used to derive an analytical form for the flux function. Instead of a current sheet along the boundary, the authors formulate an analytical form for the outermost field line to allow field strength to smoothly drop to zero. However, in a similar vain to other sunspot models that derive their thermodynamic properties solely from a prescribed magnetic-field configuration (\textit{e.g.} self-similar models, Section\,\ref{selfsimilar}), the axial pressure and density of these models posses a high degree of sensitivity near the upper boundary, with a tendency to become negative in the photospheric layers and beyond.

\subsection{Non-MHS Models}
\subsubsection{Semi-Emprical Sunspot Models}\label{cameronss}
Very close to the solar surface, the properties of sunspots
can be inferred from spectro-polarimetric measurements. These
layers are the most important to model since it is exactly
in this region that the effects on the wave cannot be treated
assuming only weak perturbations.
\inlinecite{cameronetal2010} have emphasized
the desirability of incorporating this information into helioseismic
wave-propagation simulations by constructing simplified, axisymmetric sunspot models which are designed to capture the effects of the first few hundred km. 
\begin{figure}[ht]
\centering
\begin{tabular}{cc}
\includegraphics[trim= 10mm 0mm 70mm 0mm, clip,width=0.45\textwidth]{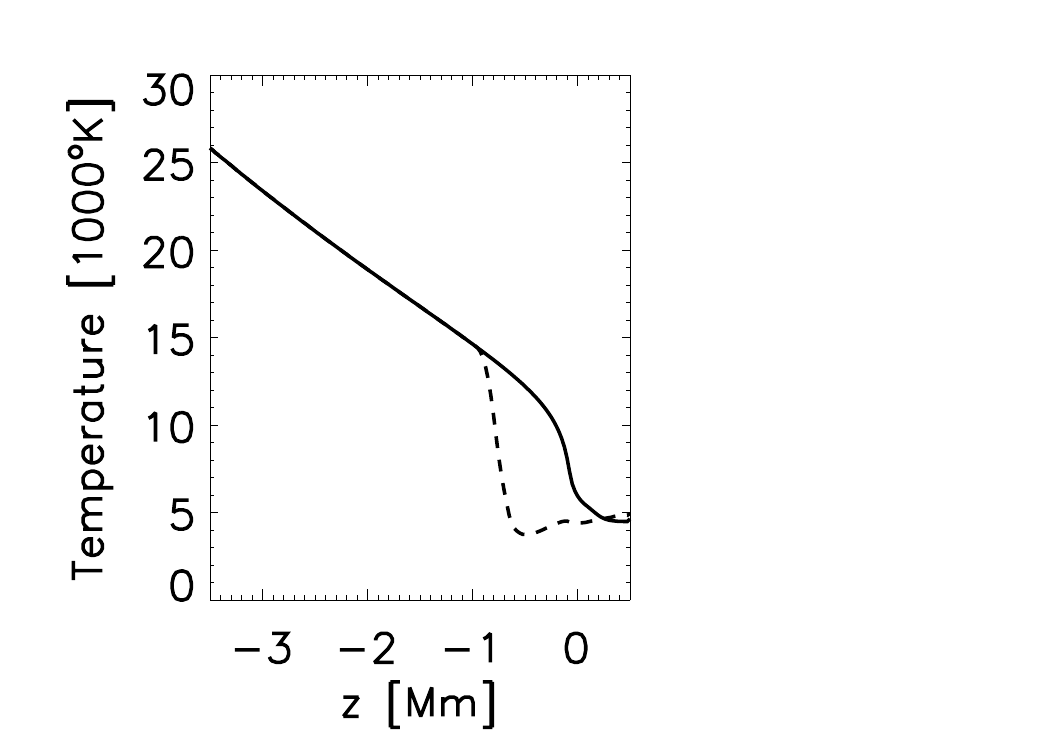} &
\includegraphics[trim= 10mm 0mm 70mm 0mm, clip, width=0.45\textwidth]{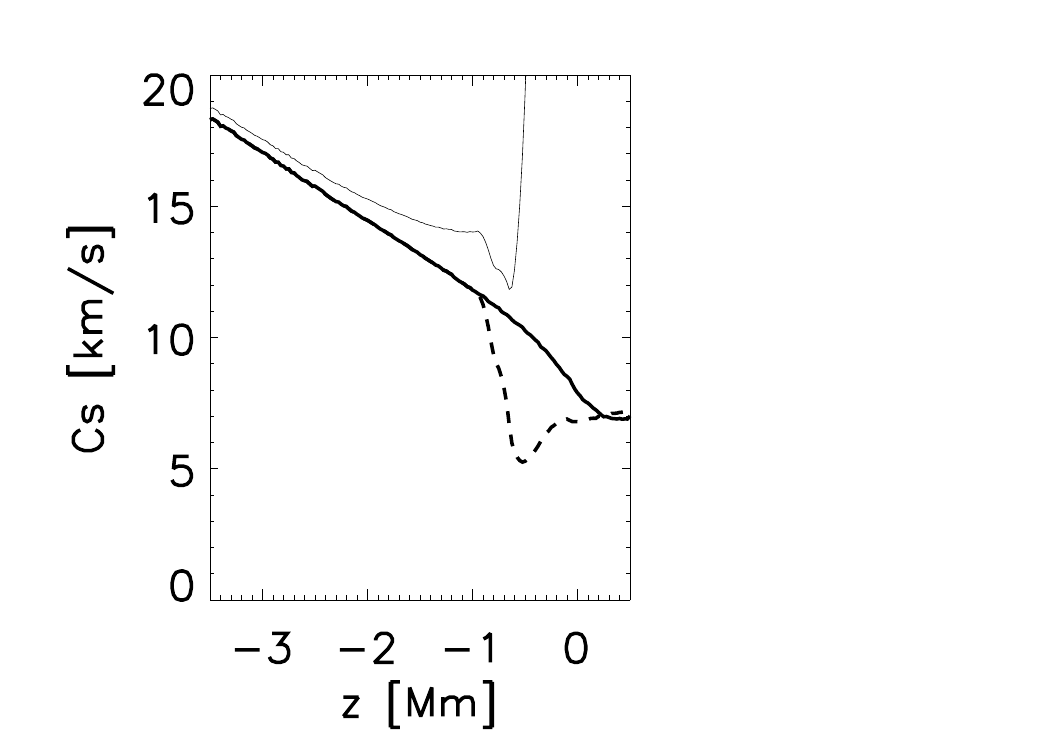}
\caption{Thermodynamic, sound and fast mode speed profiles of the sunspot model of AR 9787 produced by Cameron \etal~(2010). Left panel: Vertical temperature profile of reference stratification (solid)
and near the center of the umbra (dashed). Right panel:  Vertical sound speed profile of reference stratification (solid)
and near the center of the umbra (dashed). The dotted line shows the fast mode
speed in the center of the umbra.}
\label{fig:cameronss}
\end{tabular}
\end{figure}

Since full spectro-polarimetric inversions are available for very few spots (\textit{e.g.} \opencite{mathewetal2004}), \inlinecite{cameronetal2010} utilize a combination of existing semi-empirical models: the umbral model of
\inlinecite{maltbyetal1986} and the penumbral model of \inlinecite{df1989}, which cover the range of heights from about 500~km below the quiet-Sun $\tau=1$ level to the lower chromosphere.
The OPAL equation of state tables are used to infer sound-speed,
temperature, pressure and density profiles. Since the
aim is to only model the near-surface region, the atmospheric
models are smoothly matched to the Model S \cite{cdetal1996} quiet-Sun atmosphere
below $\tau_{5000}=0$ and 400~km  above.
\inlinecite{cameronetal2010} combine these 1D models to form a 3D,
axisymmetric model of the sunspot in AR 9787, with the umbral and penumbral radii chosen to match those of the observed sunspot (10 and 20~Mm respectively). The model is then stabilized with respect to convection
by calculating the relative perturbations with respect to the quiet-Sun
atmosphere and imposing these on the stabilized model (see Section\,\ref{background} for details). As good measurements of the magnetic field are unavailable for the sunspot in AR 9787, the model 
assumes that $B_z$ is described by a Gaussian horizontal profile and the
vertical profile is tuned so the field inclination at the umbra--penumbra border is approximately 45\degree. Figure~\ref{fig:cameronss} shows the vertical temperature, sound and fast mode speed profiles in the center of the umbra of the sunspot model.

\subsubsection{Dynamical Models}\label{dynamic}
\inlinecite{kr2007} propose a theoretical sunspot model, with complete dynamics of both magnetic field and flow, which can reproduce the observed bright rings around sunspots (\textit{e.g.} \citeauthor{rastetal1999}~\citeyear{rastetal1999,rastetal2001}). A simplified model was initially used to probe the possibility of reproducing bright rings by heat transport alone. In this model the mean flow velocity is put to zero, the spot-like structure of the magnetic field is prescribed and stationary, and the diffusion equation for the entropy is solved. The authors find the brightness of the region occupied by magnetic field decreases due to magnetic quenching of the thermal diffusivity, however the resulting surface brightness profile did not show a bright ring.

A more consistent MHD model of a sunspot was then considered using the complete momentum equation together with the induction equation. Amplitudes of the initial uniform field of several hundred Gauss were considered. The surface flow generated is convergent near the spot but divergent
at larger distances. The amplitude of the flow modeled was approximately 800 m s$^{-1}$. Both brightness and field strength (central value of about 2700 G) were modeled to be almost uniform in the central parts of the sunspot and changing rapidly with radius beyond. The radial heat-flux profiles from their simulations show consistent bright rings around the spots, appearing to be somewhat brighter than observed rings. To probe the contribution of the flow to the bright rings, the field is switched off. The bright rings do not disappear, which leads the authors to conclude that both the flow and the reduced diffusivity quenching contribute to the resulting bright rings.

Another set of dynamical MHD sunspot models, consisting of idealized axisymmeric flux tubes (where the magnetic field is matched to a potential field at the upper boundary) in a compressible convecting atmosphere, were presented by \inlinecite{hr2000}. In their models, they find the magnetic flux to be confined by an inward ``collar'' flow at the surface. Further outside, the flow direction appears to reverse and a ÔmoatÕ cell appears. The authors suggest that the collar cell holding sunspots together is hidden beneath their penumbrae, so that only the outflow in the moat cell is visible at the surface. Although  this particular flow pattern (i.e., inflows and down-flows around sunspots, first proposed by \opencite{parker1979}), can in theory help to explain the question of the stability of long-lived sunspots, there are a number of theoretical concerns with this conjecture (already discussed in Section\,\ref{spruit}). Furthermore, the existence of such a flow structure around sunspots is absent from both observations and recent helioseismic inferences (see
\eg Section\,\ref{sec:haber}), as well as realistic radiative MHD simulations of sunspot structure (\opencite{heinemannetal2007}; \opencite{rsk2009}; \opencite{rempeletal2009}; see below for details).

\subsection{Numerical Simulations of Radiative Magnetoconvection}\label{rempel}
Significant progress in our ability to simulate sunspots using realistic
MHD simulations (\textit{i.e.} MHD simulations that include the solar equation of
state and multi-dimensional radiative transfer) was only possible over the past
couple years. This is primarily due to the fact that pursuing radiative MHD
simulations on the scale of sunspots with sufficient resolution for
capturing the essential scales of magneto-convective energy transport requires
fairly large computational domains and accordingly computing power.
Additional to the computational domain size also the physical parameters encountered
in and above the umbral region of a sunspot pose significant numerical
challenges. The combination of several kG magnetic field with the rather
small density scale height leads to a steep increase of the Alfv\'en velocity
above the sunspot umbra reaching values in excess of a few 1000 km s$^{-1}$. Such
high velocities lead to severe time step constraints for explicit codes that
make such a simulation almost impractical unless this constraint is relaxed
by artificially limiting the Lorentz force in low beta regions.
\begin{figure}[ht!]
\centering
\includegraphics[width=\textwidth]{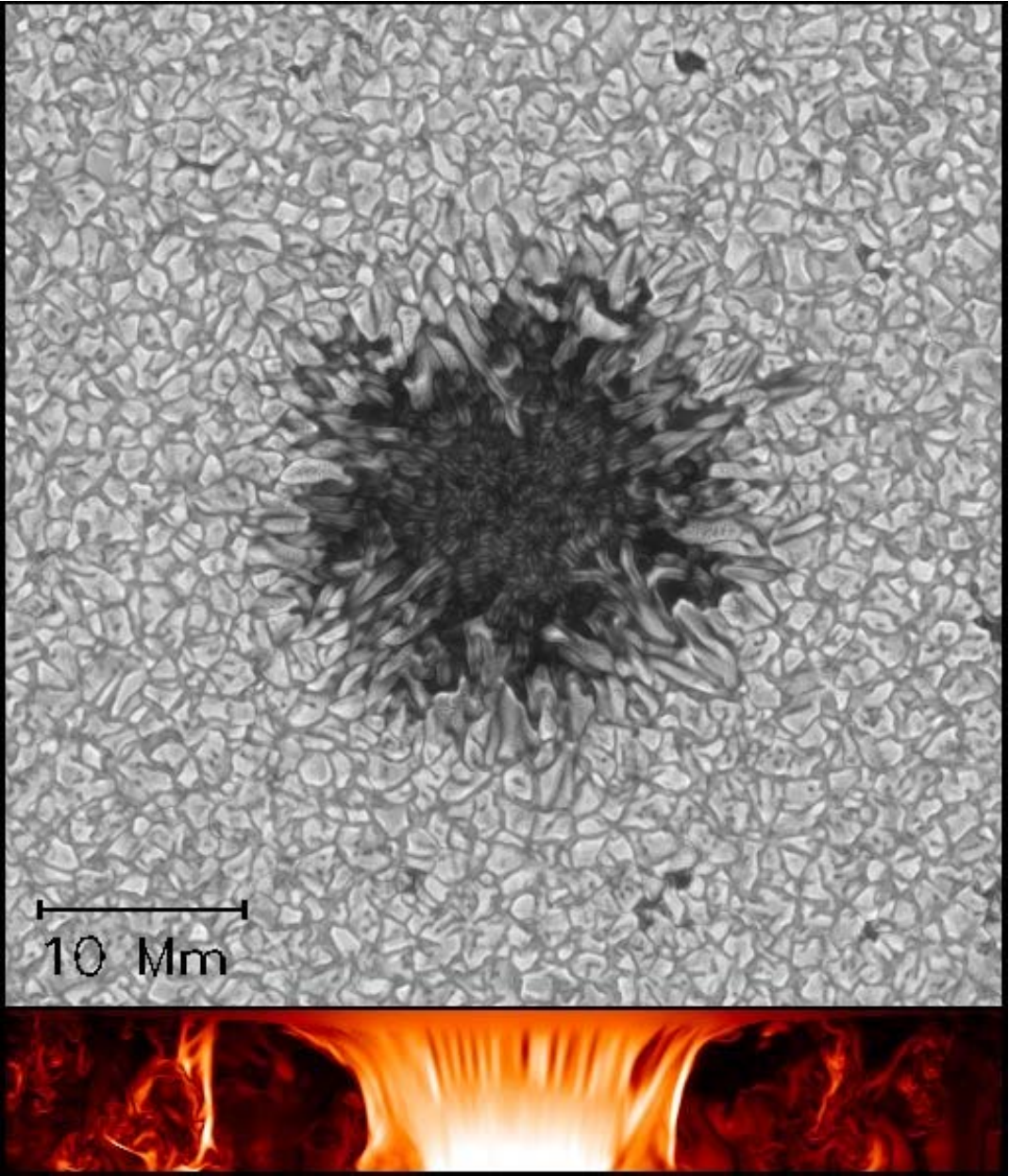}
\caption{Simulated sunspot in a $50\times50\times8$ Mm domain. The top panel displays the bolometric
intensity, the bottom panel field strength on a cut through the center of the
spot.}
\label{fig:fig5}
\end{figure}

3D models with a slender rectangular slab geometry have been
used to study the inner penumbra \cite{heinemannetal2007,rsk2009}. The
results from the MHD simulations show the formation of filamentary structures
resembling those in the inner penumbra of a real sunspot, including bright
filaments containing central dark cores. More recent 3D calculations
of round sunspots, and even pair of opposite polarity spots
(\textit{e.g.} \opencite{rempeletal2009}), also show extended outer penumbrae with
realistic Evershed flows having mean flow velocities of up to 6 km s$^{-1}$.

\begin{figure}[ht]
\centering
\includegraphics[width=\textwidth]{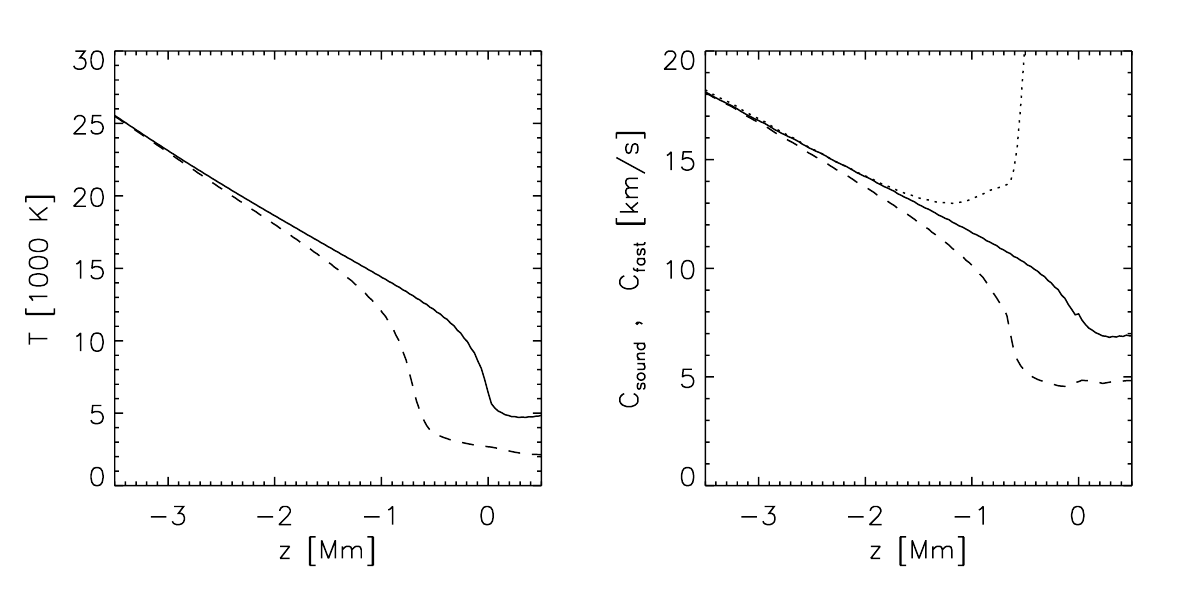}
\caption{Left panel: Vertical temperature profile of the reference stratification (solid)
and near the center of the umbra (dashed) of the sunspot model in Figure~\ref{fig:fig5}. The large temperature perturbation corresponds to a Wilson depression of about 600 km. Right panel:  Vertical sound speed profile of reference stratification (solid)
and near the center of the umbra (dashed). The dotted line shows the fast mode
speed in the center of the umbra.}
\label{fig:fig6}
\end{figure}

Most of these simulations are focused around the sunspot fine structure
(umbral dots and penumbral filaments) and cover only a temporal evolution of
a few hours. Simulations addressing the long term evolution of sunspots (on a
time scale of days) are currently only feasible if the resolution is decreased,
therefore not resolving details of the fine structure as well as penumbral
regions. Figure~\ref{fig:fig5} displays the bolometric intensity and subsurface magnetic-field
strength for a sunspot evolved over a total of 15 hours. Figure~\ref{fig:fig6} compares the
profiles of temperature, sound speed and fast mode speed in the center of the
spot to those of the reference stratification near the edges of the domain.
The temperature profile (left panel) shows a Wilson depression of about
600 km which forms rather quickly during the first few hours of the simulation,
deeper down a slowly progressing cooling front leads to a moderate adjustment
of temperature in layers 1\,--\,2 Mm beneath the $\tau=1$ level in the umbra
corresponding to changes of the speed of sound (right panel) on the order of a
few 100 m s$^{-1}$ to about 1 km s$^{-1}$. The fast mode speed shows a steep increase above
the $\tau=1$ level in the umbra, the maximum speed is here artificially limited to about 60 km s$^{-1}$
to relax the stringent Alfv\'enic time step constraint.

While these simulations provide a realistic description of the thermal and
magnetic structure in the upper most few Mm of a sunspot, they cannot address
fundamental questions regarding the subsurface structure and origin of sunspots.
Currently the domains are with 8 Mm still rather shallow (deeper domains up to
16 Mm are in preparation) and the initial field structure is based on a
monolithic model. Furthermore the magnetic field is fixed at the bottom
boundary. Nevertheless these models provide very useful artificial oscillation data
that can be used to test helioseismic inversion methods, since the latter
requires primarily a consistent rather than a fully realistic model.

\section{Diagnostics Potential of Helioseismology}\label{birch}
In the previous section we reviewed models of sunspots; each of these provides a prescription for computing the subsurface structure (\textit{i.e.}, the 3D forms of the sound speed, density, and magnetic field) of a model sunspot over some range of depths. These prescriptions are based on a wide variety of physical assumptions. In general, it is not known which of these assumptions are good approximations (in the sense of accurately describing solar conditions). Helioseismology promises the ability to infer the subsurface structure of sunspots and active regions, and thus to test the physical approximations that are used in building models of these regions.

Helioseismology has shown (see \opencite{gb2005} for a review) that wave propagation is different in sunspots than in the quiet Sun. For example, single-bounce wave travel times are altered by (order of magnitude) about thirty seconds (the details depend strongly on the data analysis filters, the first bounce distance, and the wave frequency, see \textit{e.g.} \opencite{bb2008}).  In Section\,\ref{s:ar9787} we provide a follow up to the detailed helioseismic study undertaken by \citeauthor{gizonetal2009}~(\citeyear{gizonetal2009,gizonetal2009a}) of the active region AR 9787. These measurements clearly demonstrate that helioseismology can measure the changes in wave propagation that are associated with sunspots and active regions.

There have been many conclusions drawn about sunspot and active region structure from local helioseismology
(\textit{e.g.} among others, \opencite{fbc1995}; \opencite{kosovichev1996}; \opencite{jjc1998}; \opencite{csc2000}; \opencite{kds2000}; \opencite{zk2003}; \opencite{couvidatetal2004}; \opencite{bab2004}; \opencite{crouchetal2005}; \opencite{bogartetal2008}). Despite the long history of work on this topic, there is no general agreement on the subsurface structure of active regions. Here we will investigate some of the possible causes of the disagreement shown by \citeauthor{gizonetal2009}~(\citeyear{gizonetal2009,gizonetal2009a}):  the frequency dependence of travel times (Section\,\ref{freqdep}), the effects of phase-speed filters on measurements (Section\,\ref{filters}), and the effects of inversion parameter choices and inclusion of surface terms on ring-diagram structure inversions (Section\,\ref{baldner}).

Another approach to using helioseismology to constrain sunspot models is the direct simulation of
wave propagation through model sunspots and comparison the resulting wave field
with the helioseismic measurements (\textit{e.g.} \opencite{cgd2007}; \opencite{cgd2008}; \opencite{mhc2008}).
This approach bypasses the need for linear forward models (\textit{i.e.}, models for the linear sensitivity of
travel-time shifts or ring-diagram frequency shifts to changes in subsurface structure).  The approximations 
made in the process of formulating linear forward models may be one source of the disagreement shown by \citeauthor{gizonetal2009}~(\citeyear{gizonetal2009,gizonetal2009a}). A limitation of the direct simulation approach is that it is extremely computationally intensive.
The examples shown in Section\,\ref{cameron_comp} demonstrate that numerical simulations are capable of reproducing
many features of the helioseismic measurements (see also \opencite{cgd2008}).  In addition, simulations promise to provide a powerful tool for determining the sensitivity of helioseismic measurements to small changes in the subsurface structure of sunspot models. This type of study is required to be able to make meaningful estimates of the signal-to-noise ratios required for the helioseismic measurements to constrain any particular aspect of sunspot models.

\section{Numerical Forward Modeling of Waves Through Model Sunspots}\label{forward_models}

\subsection{Numerical Methods}\label{ss:methods}
Our best chance at constraining the interior structure of sunspots comes with constructing accurate numerical forward models. There are two major reasons for claiming this: \textit{i}) the departure from the quiet Sun wave propagation physics being so dramatic in a sunspot, it is quite likely that the single scattering Born approximation (\textit{e.g.} \opencite{bkd2004}) entirely fails (\textit{e.g.} \opencite{ghb2006}). It is therefore imperative to solve the wave propagation equations in a fully 3D, highly magnetized environment.  However, the basis of semi-analytical sunspot formalisms is significantly undermined by the sheer mathematical complexity of dealing with the MHD equations (not for the want of trying; \textit{e.g.} \opencite{bogdan1999}), and hence the necessity of using purely numerical techniques. \textit{ii}) Computational solvers are flexible and can be rather easily extended to include more effects, such as flows, changes on wave sources, \textit{etc.}

The difficulties in creating an MHD solver that produces sufficiently accurate results are not to be underestimated. A number of groups are working on this problem \cite{kc2006,cgd2007,pk2007a,hanasoge2008,sheylagetal2009}. The numerical techniques, and validation and verification procedures vary significantly from one code to another. A further complexity is introduced by an Alfv\'en speed that increases exponentially with height, approaching values of several hundred km s$^{-1}$ near the computational upper boundary. Simulating wave propagation in a plasma so highly magnetically dominated is remarkably expensive, especially in 3D. A number of simulations (\textit{e.g.} \opencite{cgd2007}; \opencite{cgd2008}; \opencite{hanasoge2008}; \opencite{mhc2008}; \opencite{rsk2009}) apply an Alfv\'en wave speed limiter to moderate the action of the Lorentz forces, thereby indirectly controlling the plasma-$\beta$. Having described this context, three rather serious issues remain as yet outstanding: \textit{i}) What are the spatio-temporal resolution requirements for simulating waves in an MHD environment? \textit{ii}) How do we treat the upper magnetic boundary? and \textit{iii}) What harm does the Alfv\'en speed limiting term do to the near-and far scattered wave field?

As yet, only parts of these questions have been answered. This may be attributed to a systemic shortsighted forward modeling approach in our intensely observationally driven field: simulate and obtain results as quickly as possible. Evidently, we must attempt to overturn this trend by \textit{i}) ensuring that the numerical methods are high-order and highly precise, \textit{ii}) validating and verifying that the equations are being solved accurately (\textit{i.e.}, testing against a number of known solutions), \textit{iii}) determining and matching the spatio-temporal resolution requirements, \textit{iv}) using stable, physically-motivated well-tested boundary conditions, and finally \textit{v}) testing the approximations used in the calculations do not directly affect the quality of the solution. Of course, attendant questions of computational efficiency must also be addressed, because the forward modeling approach requires exhaustive testing of sunspot models, requiring a large number of calculations.

\subsection{Background Models Stabilized Against Convective Instability}\label{background}
Numerical simulations of seismic wave propagation through random media must begin with an initial background model of the Sun. In helioseismology, it is common practice to begin with Model~S (\textit{e.g.} \opencite{cgd2007}; \opencite{pk2007a}; \opencite{khomenkoetal2009}).  However, most background models include convection and numerical forward models which simulate linear wave propagation are sensitive to this. Therefore, it is essential to remove such convective instabilities from the background in order to be able to successfully simulate linear wave propagation on the time scales ($\approx 8$ hours) required for computational helioseismology, as well as to ensure that convective modes do not swamp the signatures one is interested in analyzing.

There are a number of ways to stabilize background models against convection (\textit{e.g.} \opencite{hanasogeetal2006}; \opencite{cgd2007}; \opencite{pk2007a}; \opencite{sfe2008}), all requiring zero buoyancy so that a vertically displaced packet of gas in adiabatic equilibrium will not continue to rise, \textit{i.e.}, ensuring that the Brunt-V\"ais\"al\"a frequency always remains positive, specifically $N^2/g > 0$ where $g$ is the gravitational acceleration. There are a number of ways to go about this, for example the method employed by \inlinecite{pk2007a} ensures that, when this condition is not met, the value is replaced with zero, or very small values (\textit{e.g.} $3 \times 10^{-5}~\textrm{Mm}^{-1}$). \inlinecite{sfe2008} and \inlinecite{sheylagetal2009} have a slightly different approach, whereby they adjust the pressure and density of Model~S using the equation of state for an ideal gas to retain a constant $\Gamma_1=5/3$, with the additional constraint that the modified sound speed profile does not differ substantially from the original. Both \inlinecite{pk2007a} and \inlinecite{sheylagetal2009} have shown relatively solar-like power spectrums.

Another method has been employed by \inlinecite{cgd2007} and \inlinecite{cgd2008}, and undergone detailed development and testing in \inlinecite{scg2009}, whereby the Model~S background is stabilized against convective instability by ensuring $\partial_z p= \mathrm{max}(c^2\partial_z \rho_0-\epsilon,\partial_z p_0)$, where $\partial_z$ specifies the partial derivative with respect to height, the subscript ``0'' indicates unperturbed Model S values, and $\epsilon=10^{-5}$ g s$^{-2}$ near the surface and zero everywhere else. These changes to the background model alter the eigenmodes of the problem, and this must be taken into account. \inlinecite{scg2009} have attempted to do this by examining the eigenmodes and ensuring that they are solar-like and have successfully demonstrated quite solar-like power spectrums. Other ways to remove the convective instability have not been explored however, and it is not clear which form of stabilization least affects the eigenmodes.

\subsection{Numerical Codes}
In this section we provide a brief summary of the four numerical simulation codes that were discussed and used during the two HELAS workshops on sunspot seismology held in 2007 and 2009. These codes compute the propagation of solar waves through prescribed background models\footnote{We note that other numerical simulation codes also exist (\textit{e.g.} \opencite{pk2007a}; \opencite{hmm2008}).}.

\subsubsection{The IAC MHD Code}
The IAC MHD code, described in the works by \inlinecite{kc2006} and \inlinecite{kcf2008}, solves the non-linear MHD
equations for perturbations, written in the conservative form, using
a fourth-order central difference scheme and advanced in time by a fourth-order Runge--Kutta method. In a similar manner to \inlinecite{sn1998} and \inlinecite{ck2001}, in order to damp high-frequency numerical noise on sub-grid scales, the physical diffusive terms in the equations of momentum and energy are replaced by artificial equivalents. In the induction equation, the magnetic diffusion term is retained, with $\eta$ being replaced by an artificial value.
Depending on the simulation, Perfectly Matched Layers
(PML; \textit{e.g.} \opencite{berenger1994}) are also placed at the boundaries that absorbs the
incoming waves and prevents their spurious reflection and return
back to the physical domain. For the best results, 10 to 15 grid
points are allocated to the PML layer. This code has been used to
study the wave propagation and refraction in a small sunspot \cite{kc2006}; non-linear wave propagation, shock
formation, mode conversion and energy transport in small-scale
flux tubes with internal structure \cite{kcf2008}.

Recently, the code was also used to model the propagation of helioseismic waves below the sub-photospheric structure of sunspots \cite{khomenkoetal2009,kc2009}. The most important results obtained for helioseismic wave propagation below sunspots are the following:
\textit{i}) the fast magneto-acoustic mode represents an analog of quiet
Sun $p$-modes modified by the presence of magnetic field, \textit{ii}) helioseismic waves
 below sunspots are sped up by the magnetic
field by 20\,--\,40 seconds compared to the quiet Sun, \textit{iii}) the magnetic
field produces a strong frequency dependence of the travel times,
\textit{iv}) the eikonal solution gives a qualitatively good approximation
for the numerical solution, and finally, (v) the high-frequency fast mode waves
are refracted in the magnetically dominated layers and inject
additional energy, possibly causing the power increase observed in
acoustic halos surrounding active regions.

\subsubsection{The SLiM Code}
The Semispectral Linear MHD (SLiM) code solves the ideal linearized MHD equations using a spectral expansion in the horizontal directions
and a two-step Lax-Wendroff treatment in the vertical. The code includes two absorbing layers at the top and the bottom of the box. In the top layer the waves are heavily damped and the effect of the Lorentz force is systematically reduced. Likewise the bottom layer damps the waves that propagate downward. The code has been tested against analytic solutions, which are described in detail in \inlinecite{cgd2007}.

In \inlinecite{cgd2008} and \citeauthor{gizonetal2009}~(\citeyear{gizonetal2009,gizonetal2009a}), SLiM was used to study wave propagation through a simplified monolithic model sunspot embedded in a stabilized quiet-Sun model atmosphere (see Section\,\ref{background}). The corresponding wave field computed with SLiM was then compared with MDI observations of $f$- and $p$-mode scattering by the sunspot in AR 9787. The comparisons were quite encouraging as the numerical simulations from SLiM were able to reproduce wave absorption and scattering phase shifts. The code has also recently been used on the sunspot model described in Section\,\ref{cameronss}, as well as in simulations where the magnetic field of the sunspot is essentially ``switched off'', while maintaining the sound speed, pressure and density perturbations of the sunspot model. This type of simulation is made possible by the fact that the background does not need to be in pressure balance. This type of experiment allows one to disentangle the contributions of the Wilson depression and sound speed/thermal changes from mode conversion and other magnetic-field effects.

\subsubsection{The SPARC Code}
The Seismic Propagation through Active Regions and Convection (SPARC; \opencite{hanasoge2010}) code uses techniques developed by \inlinecite{hanasogeetal2006} and \inlinecite{hdc2007} to simulate helioseismic wave propagation in the near-surface layers of the Sun. Waves are stochastically excited by introducing a forcing term in the vertical momentum equation; the forcing function is prescribed such that a solar-like power spectral distribution is obtained. The solution is temporally evolved using a second-order optimized Runge--Kutta integrator \cite{hu1996}. The vertical derivative is resolved using sixth-order compact finite differences with fifth-order accurate boundary conditions \cite{hr2000}. The derivatives in the horizontal directions are computed using FFTs (periodic horizontal boundaries). The upper and lower boundaries are lined with damping sponges in order to enhance wave absorption. The SPARC code has been utilized to study wave propagation through model sunspots (\textit{e.g.} \opencite{hanasoge2008}; \opencite{mhc2008}; \opencite{mh2010}), as well as solar convection \cite{hdd2010}. 

A recent bit of progress with regards the choice of boundary conditions has been the development of a stable, unsplit PML formulation for the stratified linearized ideal MHD equations. Some related formulations have been developed by \inlinecite{pk2007a} and \inlinecite{kc2006}. However, instabilities caused by waves at grazing incidence to the boundary prevent long time integrations. By extending the technique of Convolutional Perfectly Matched Layers (C-PML; \textit{e.g.} \opencite{rg2000}), \inlinecite{hkg2010} have succeeded in devising a stable C-PML formulation.

\subsubsection{The SAC Code}
In \inlinecite{sheylagetal2009}, the propagation and dispersion of acoustic waves in a solar-like 2D sub-photosphere with localized, non-uniform magnetic field concentrations was investigated using the Sheffield Advanced Code (SAC) developed by \inlinecite{sfe2008}. The numerical code is based on a modified version of the Versatile Advection Code (VAC; \opencite{tkb1998}), and employs artificial diffusivity and resistivity in order to stabilize the numerical solutions and relies on variable separation to background and perturbed components to treat gravitationally stratified plasma. The complete MHD equations are solved using a fourth-order central difference scheme for the spatial derivatives, and are advanced in time by implementing a fourth-order Runge--Kutta numerical method.

The standard Model S \cite{cdetal1996} atmosphere is employed as an initial background model, modified so that $\Gamma_1$ is kept constant in such a way as to have the adiabatic sound speed profile closely match the sound speed in Model S. Three different self-similar, non-potential, magnetic-field configurations are employed for the simulations. As the curvature of the magnetic field changes the temperature stratification in the domain, the magnetic configurations chosen  differ by their field strengths and inclination at the visible solar surface. The acoustic source generates a temporarily localized wave packet with the duration of about $600~\mathrm{seconds}$, which has a peak frequency of approximately $3.33~\mathrm{mHz}$. The amplitude of the source is of the order of a few centimeters per second. Such a low amplitude makes sure that convective processes will not be initiated in the otherwise convectively unstable equilibrium, and that the perturbations are kept linear, \textit{i.e.}, they do not change the background strongly.

Three cases, a weak strongly-curved magnetic field with $B_z=120~\mathrm{G}$ at the surface, a strong weakly-curved magnetic field with $B_z=3.5~\mathrm{kG}$, and a strongly-curved strong magnetic field with $B_z=3.5~\mathrm{kG}$, were analyzed by means of time--distance helioseismology. The travel time dependencies show that for the first bounce the main part of effect of magnetic field on the acoustic wave is due to the change of the temperature structure in the sunspot. Nevertheless, the wave mode conversion from purely acoustic to the slow magneto-acoustic wave motion, characterized by an energy leak downwards, is also observed in the cases with strong surface magnetic field.

\subsection{Eikonal Methods}
While numerical simulations of wave propagation have significantly aided our level of understanding of helioseismology, further guidance is still needed in setting up the correct numerical experiments and understanding wave propagation in magnetized plasmas. MHD ray theory \cite{weinberg1962} has traditionally provided a very useful conceptual framework in which to understand wave propagation, even though this is questionable at the surface where the pressure and density scales vary rapidly, since the assumption of slowly varying coefficients may not be justified.

Regardless of these shortcomings, however, ray theory has been used in helioseismology for some time, being one of the several methods that have been applied to asymptotic inversions of helioseismic frequency measurements in the past (\textit{e.g.} \opencite{gough1984}). In general, it has performed well beyond its formal domain of applicability, a prime example being the agreement between the wave mechanical analysis of \inlinecite{cally2005}, the ray theory modeling of \inlinecite{cally2006} and the recent results of \inlinecite{hc2009}, who find very good agreement between generalized ray theory and previously published exact solutions \cite{cally2001,cally2009a}.

\inlinecite{mc2008} recently combined eikonal methods and observational data by constructing a 3D sunspot model based on observed surface magnetic-field profiles to propagate magneto-acoustic rays across the sunspot model for a range of depths to reproduce a skip-distance geometry similar to center-to-annulus cross-correlations used in time--distance helioseismology. In another recently completed work, \inlinecite{mhc2008} compared the results of forward modeling via MHD ray theory and a 3D ideal-MHD solver, concluding that the simulated travel-time shifts were strongly determined by MHD physics and confirmed their strong dependence on the frequency and phase-speed filter parameters used. \inlinecite{khomenkoetal2009} have similarly applied MHD ray theory to validate and analyse the results of their numerical simulations.  In another series of recently published works, \inlinecite{cally2009b} investigates the direct (magnetic) and indirect (thermal) effects of the magnetic field on vertically propagating waves, suggesting that, overall, travel-time perturbations in umbra appear to be predominantly thermal, while in penumbrae they are mostly magnetic.  
\inlinecite{cally2009a} provides strong evidence for significant phase jumps (or discontinuities) associated with fast magneto-acoustic rays that penetrate the $a=c$ level in sunspots. This effect appears to be more pronounced in highly inclined field characteristic of penumbrae.

Most of the above work has centered on the study of properties of individual rays propagating through magnetized atmosphere by solving the ray equations for a point source in non-magnetic solar model. In a recent study, \inlinecite{sheylagetal2009} have shown the importance of considering the full family of rays corresponding to a particular problem's initial conditions, as these define important geometric properties of the excited wave-field such as the wave front, caustics, and phase surfaces.

\section{Update on the Analysis of AR 9787}\label{s:ar9787}
\subsection{Travel Times Comparison: Time--Distance and Helioseismic Holography}
It has been thought by some that travel times computed from
time--distance helioseismology \cite{duvalletal1997} should be very
similar (or maybe identical) to those measured using helioseismic
holography \cite{bl2000}.  However, a detailed comparison of
measured travel times has not been done.  We have done such a
comparison for an 8-hour interval for AR 9787.  For the time--distance case, correlations are calculated between the central point and the quadrants.  For the holography, egression and ingression signals are constructed in a quadrant geometry and correlations are done with the central point.
Normal phase-speed filters were used \cite{couvidatetal2005}.

\begin{figure}[ht]
\centering
\includegraphics[width=0.95\textwidth]{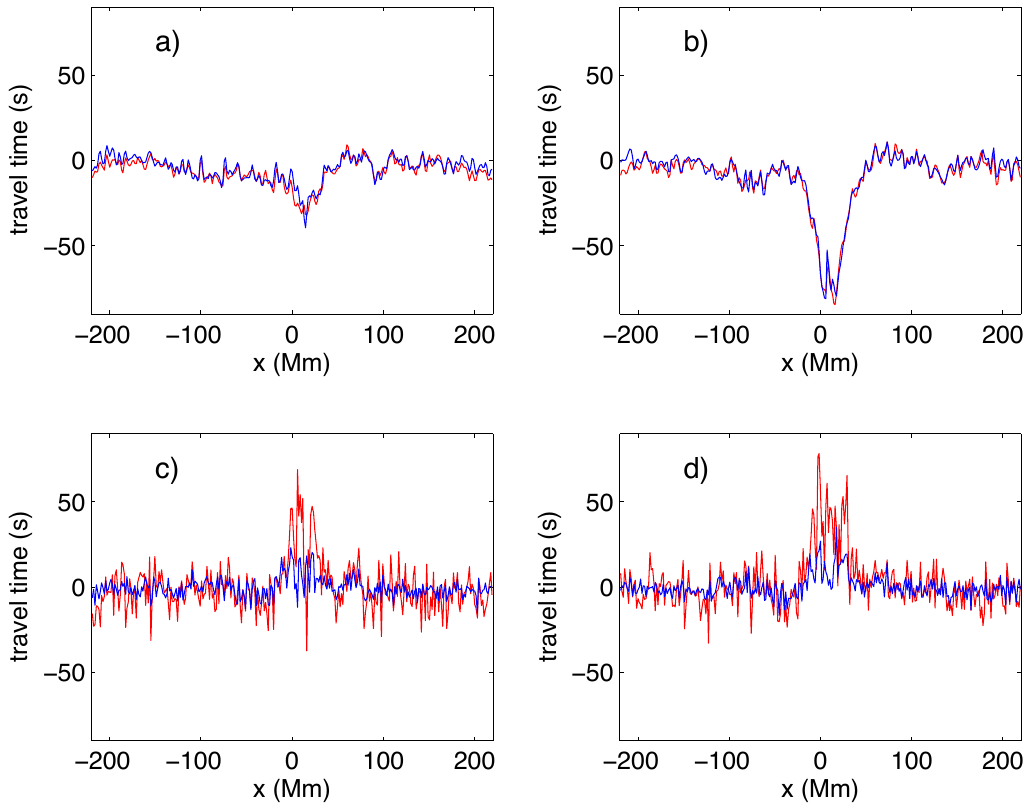}
\caption{
Comparison of travel times computed from helioseismic holography (red) with
time--distance helioseismology (blue). Cuts through the center of the sunspot
in the east-west direction are shown.  a) Inward-going times, distance of $24.35~\rm{Mm}$.
b) Outward-going times, distance of $24.35~\rm{Mm}$.  c) Inward-going times, distance of $6.20~\rm{Mm}$.
d) Outward-going times, distance of $6.20~\rm{Mm}$.
}
\label{fig:duvall}
\end{figure}

Cuts across the travel-time maps for the sunspot are shown in Figure~\ref{fig:duvall} for two distances. Frames
a) and b) show the distance $24.35~\rm{Mm}$ and c) and d) show the
shortest distance $6.20~\rm{Mm}$. The agreement is excellent for the
distance of $24.35~\rm{Mm}$. However, there are still
disagreements for the shortest distances, which we hope to understand
soon.

\subsection{Frequency Dependence of One-Way Travel Times}\label{freqdep}
The dependence of mean and difference travel times on the central frequency of the wave packets measured over a sunspot region was shown and interpreted by \citeauthor{bb2006}~(\citeyear{bb2006,bb2008}), as an indication of perturbations largely confined to a region not deeper than a few Mm. The largest frequency variations in travel times were seen for the small travel distances $\Delta$, which are of the size (diameter) of the spot or smaller. In this section, we conduct a similar study for the sunspot in AR 9787. This type of study is important as, in principle, it should help us to constrain models of the subsurface structure of sunspots.
\begin{figure}[ht]
\centering
\includegraphics[trim= 25mm 130mm 25mm 30mm, clip, width=1.0\textwidth]{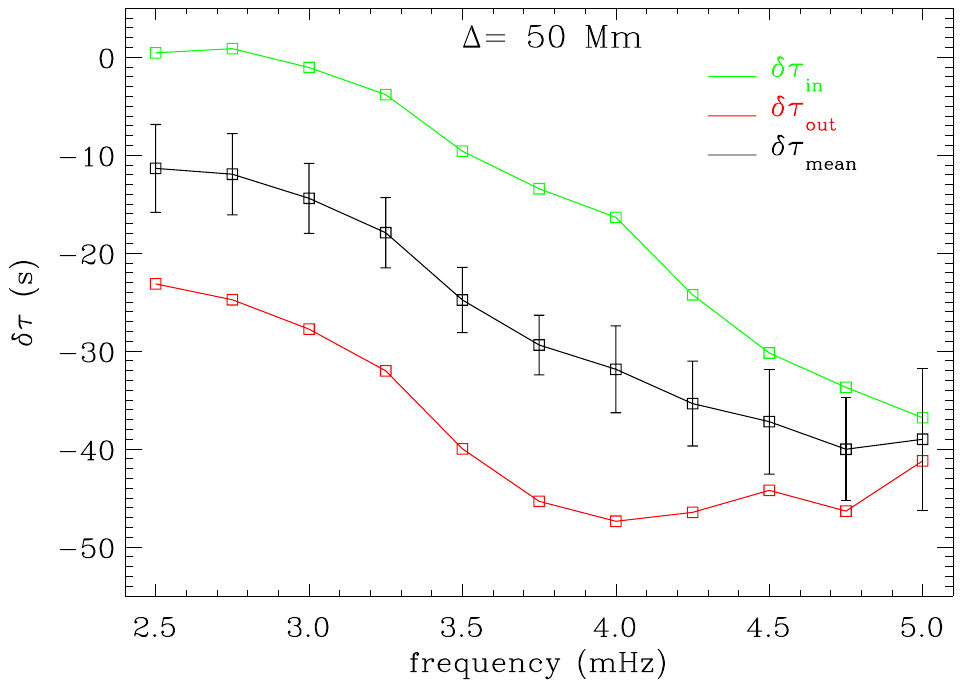}
\caption{Frequency dependence of surface-focus travel times (shown as one-way and mean umbral averages of perturbations) for $\Delta$ = 50 Mm, for the sunspot in AR 9787. For clarity, only error bars for mean travel times are shown. The error bars represent standard deviations of travel time perturbations over the umbral pixels.}
\label{fig:raj}
\end{figure}

Apart from a phase speed filter centered around a phase speed $v_{\rm ph}$= 54 km s$^{-1}$ with
a width (FWHM) of about 40\% of $v_{\rm ph}$, we also use Gaussian frequency filters, of 1-mHz width, centered at every 0.25 mHz interval between 2.5 and 5.0 mHz to study the frequency dependencies of the travel times. In Figure \ref{fig:raj} we show the umbral averages of perturbations associated with in-going and out-going wave travel times (measured using center-to-annulus surface focus geometry), $\rm{\delta\tau_{in}}$ and $\rm{\delta\tau_{out}}$, as well as mean travel-time perturbations, ${\rm\delta\tau_{mean}}$, against frequency for $\Delta$ = 50 Mm. There appears to be a strong frequency dependence associated with both $\rm{\delta\tau_{in}}$ and $\rm{\delta\tau_{out}}$ measured across the umbra of the sunspot in AR 9787. This result appears to be in good qualitative agreement with the measurements of \inlinecite{bb2008}, who look at frequency dependence of the travel-time difference, albeit for another sunspot\footnote{In particular, see top right panel of Figure 5 in \inlinecite{bb2008}, which contains the travel-time difference results for Filter I (distances of 48\,--\,55 Mm).}.

The underlying reason for such a strong frequency dependence of travel times at large distances is not entirely clear however. The interaction of helioseismic waves with sunspots can be influenced by a number of effects: the vertical extent of the sunspot and the Wilson depression, $p$-mode absorption (e.g., \opencite{bdl1987}; \opencite{braun1995}; \opencite{ccb2003}), the contribution from subsurface flows, radiative transfer effects \cite{rajaguruetal2006} and the impact of power reduction and source suppression in sunspots \cite{rajaguruetal2007,hanasogeetal2007,chouetal2009}. More detailed analyses and modeling of the sunspot in AR 9787 (see \textit{e.g.} Section\,\ref{cameron_comp})  will be required to isolate these effects and identify the cause(s) of the observed frequency dependence of travel times.

\subsection{Effects of Filtering on Travel Times}\label{filters}
In this section we show a simple toy model for the sensitivity of time--distance travel times to changes in the power spectrum of solar oscillations. We will use this simple model to develop a qualitative understanding of the ridge and inter-ridge measurements \cite{tz2008} of AR 9787 \cite{gizonetal2009}.

\subsubsection{A Model Power Spectrum}\label{sec.birch_model_power}
Before studying the impact of data analysis filters on time--distance measurements, it is necessary to
have a model for the power spectrum of solar oscillations. Here we describe a simple model obtained from fitting the azimuthal average (over the angle of the wavevector $\bf{k}$) of a 24~hr power spectrum from full-disk MDI data from a quiet Sun region. We then carry out a least squares fit  to the azimuthally averaged power spectrum with a model spectrum of the form,
\begin{equation}
P(k,\omega) =  |G(k,\omega)  |^2  + B(k,\omega) \; ,
\end{equation}
where $k=\| \bf{k} \|$ is the horizontal wavenumber, $\omega$ is the temporal frequency, $B$ is the background power,  and the function $G$ is given by
\begin{equation}
G(k,\omega) = \sum_{n=0}^{n_{\rm max}} a_n(k) \left\{ \frac{1}{  \left[ \omega-\omega_n(k)\right] \omega_n(k)}   - \frac{1}{  \left[ \omega+\omega^*_n(k)\right]  \omega^*_n(k)} \right\}  \; .
\end{equation}
This form of $G$ is motivated by Equations~(47)~and~(48) of \inlinecite{bkd2004} in the case where the imaginary parts of the mode eigenfunctions are small.  The free parameters in the fit are the complex mode frequencies $\omega_n(k)$, the mode amplitudes $a_n(k)$, and the parameters describing a background power spectrum $B$ that is linear in $\omega$ at each value of $k$.   Here $n$ is the radial order and $n_{\rm max}$ is the maximum radial order used in the normal mode summation. In the examples shown here, $n_{\rm max}=2$.  Throughout this toy model we will work in plane-parallel geometry; this is appropriate as we will be considering distances and wavelengths that are small compared to the solar radius.

\subsubsection{Sensitivity of Travel-Time Shifts to Change in Mode Frequencies}
For problems where the statistics of the wavefield are horizontally translation invariant and the expected (limit) power spectrum is azimuthally symmetric, the expectation value of the time--distance cross-covariance $C(\Delta,t)$ can be obtained from the limit power spectrum $P(k,\omega)$ as
\begin{equation}
C(\Delta,t) =2\pi\int_{-\infty}^{\infty}\mathrm{d}\omega\int_0^\infty k\mathrm{d} k \; J_0(k\Delta) {\mathcal F}^2(k,\omega) P(k,\omega) e^{-\mathrm{i}\omega t}\;
\label{eq.birch_C_from_P}
\end{equation}
where $J_0$ is the zero-order Bessel function, and ${\mathcal F}$ is the data analysis filter (see \textit{e.g.} \opencite{gb2002}). Using Equation~(\ref{eq.birch_C_from_P}), we can compute the cross-covariance that we would expect for any model of the power spectrum.
\begin{figure}[ht]
\centering
\includegraphics[width=0.66\linewidth]{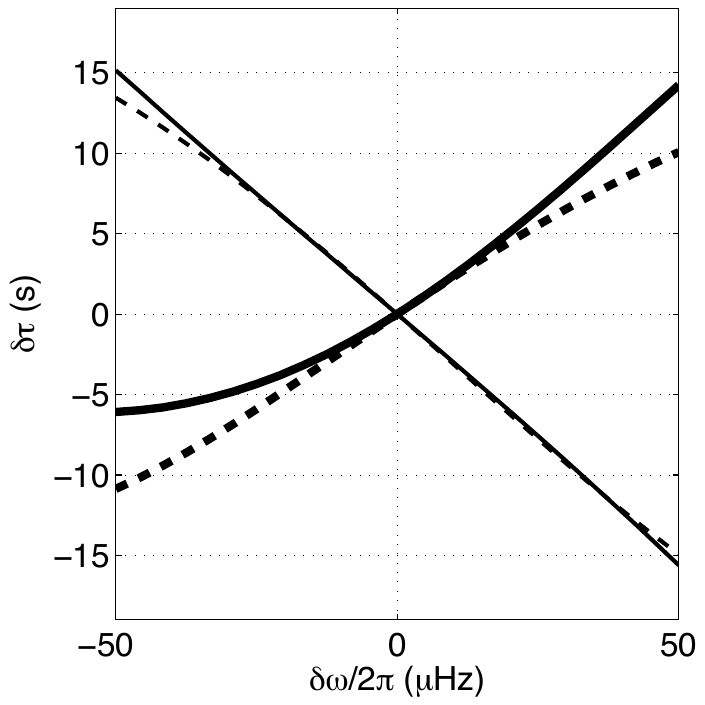}
\caption{Travel-time shifts for the case of the ridge filter (thin lines) and inter-ridge filter (heavy lines) as functions of the change $\delta\omega$ in the real part of the resonance frequency of $n=1$ mode. The solid lines show travel-time shifts obtained from the one-parameter method of Gizon and Birch (2002) and the dashed line show those from the linear definition of Gizon and Birch (2004). In both cases, the fitting window is chosen to be twenty minutes wide and centered at 28 minutes time lag. The travel distance in all cases is $\Delta=16$~Mm. Notice that for the case of the ridge filter, the two definitions of travel-time shift give very similar results. For the case of the inter-ridge filter, the difference between the two definitions increases with the amplitude of the frequency shift. }
\label{fig:birchdt}
\end{figure}

Travel-time shifts can be obtained from the cross-covariance function using a wide variety of techniques (\textit{e.g.} \opencite{duvalletal1997}; \opencite{gb2002}; \opencite{gb2004}).  For the examples shown here we use the linear definition of \inlinecite{gb2004} and also the one-parameter fitting method from \inlinecite{gb2002}.

\inlinecite{tz2008} showed that in sunspots, travel-time shifts measured with ridge filters (filters which isolate power along ridges) and inter-ridge filters (filters which isolate the part of the $k-\omega$ between the ridges) give travel-time shifts (relative to quiet Sun) of opposite sign. They found that the ridge filters yielded decreased phase travel times.  This would seem to imply the waves in sunspots have increased phase speed. The same behavior was observed in the case of AR 9787 by \citeauthor{gizonetal2009}~(\citeyear{gizonetal2009,gizonetal2009a}).
\begin{figure}[ht]
\centering
\includegraphics[width=0.49\linewidth]{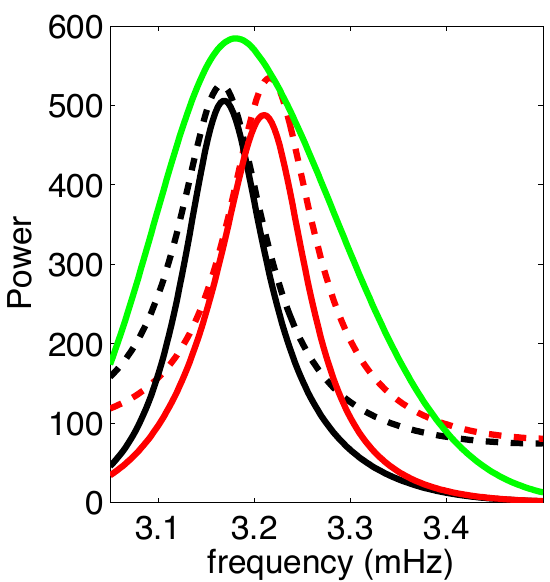}
\includegraphics[width=0.49\linewidth]{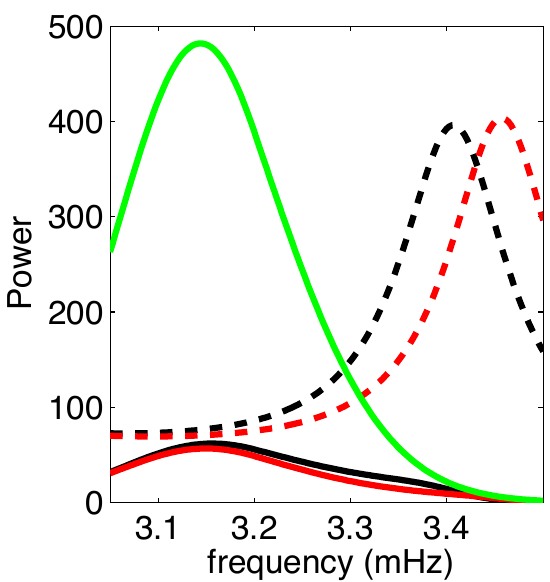}
\caption{Slices through the filtered and unfiltered power spectra for the ridge filter (left) and inter-ridge filter (right).  For the ridge filter case the slice is at $k=1.1$~rad\,Mm$^{-1}$ and for the inter-ridge case the slice is at $k=0.94$~rad\,Mm$^{-1}$ (in both cases the slice is near the $k$ of maximum power). In both panels, the red (black) dotted lines show the unfiltered power spectra with (or without) the frequency shift. The green lines show the filter. The solid red (black) lines show the filtered power spectra with (or without) the frequency shift. In the ridge filter case, the consequence of increasing the mode frequency is that the mean power moves to higher frequency (thus higher phase speed). In the inter-ridge filter case the mean power moves to lower frequency (lower phase speed). }
\label{fig:birchslicepower}
\end{figure}

As a very highly simplified toy model of this situation, we investigate the travel-time shifts caused by increasing the resonant frequencies in the power spectrum.  The reason for choosing increased mode frequencies is to mimic the situation of increased phase speeds (\textit{i.e.}, increased mode frequency at fixed $k$).  The procedure is as follows: \textit{i})~use Equation~(\ref{eq.birch_C_from_P}) to compute a reference cross-correlation from the model power spectrum described above, \textit{ii})~compute a perturbed power spectrum by increasing the model resonance frequencies by a constant value $\delta\omega$, \textit{iii})~compute a perturbed cross-covariance from this perturbed power spectrum, and \textit{iv})~fit the perturbed and reference cross-covariances to obtained the shift in the travel time caused by the change in resonance frequencies.

Figure~\ref{fig:birchdt} shows the results of this procedure for two different choices of the filter ${\mathcal F}$:  the $p_1$ ridge filter and the inter-ridge filter isolating the region between the $f$ and $p_1$ of \inlinecite{tz2008}. In both cases a 1 mHz bandpass filter centered at 3.5~mHz has also been applied \cite{tz2008}. In the results shown here we have only considered the impact of changes in the frequency of the $p_1$ ridge. For the case of the ridge filter, increases in the mode frequencies (\textit{i.e.}, increased phase-speeds for all waves) yield decreased phase times.  This is the expected result.  For the case of the inter-ridge filter, however, we find that increases in wave speeds yield increased travel times. This is an unintuitive result and in this toy model is due to interaction of the inter-ridge filter with the power between the ridges.

Figure~\ref{fig:birchslicepower} shows slices through power spectra for the two cases. The main effect of the perturbations to the mode frequencies is to move the ridges in the unfiltered power spectra. For the inter-ridge filtered case, the effect is more subtle as this filter isolates the part of the spectrum that lies between the ridges. The net result of moving the $p_1$ ridge to higher frequency and applying the inter-ridge filter is to move power to lower frequency (hence lower phase speed) at fixed $k$.

\subsection{Ring-Diagram Structure Inversions}\label{baldner}
The mode frequencies from ring-diagram analysis can be used to determine the
thermodynamic structure of the gas under the tracked region.  The mode parameters
themselves are obtained by fitting a function $P_n(\omega, k_x, k_y)$ to the
3D power spectrum.  
\begin{figure}[htb]
\begin{center}
\includegraphics[trim= 5mm 2mm 15mm 10mm, clip, width=1.0\textwidth]{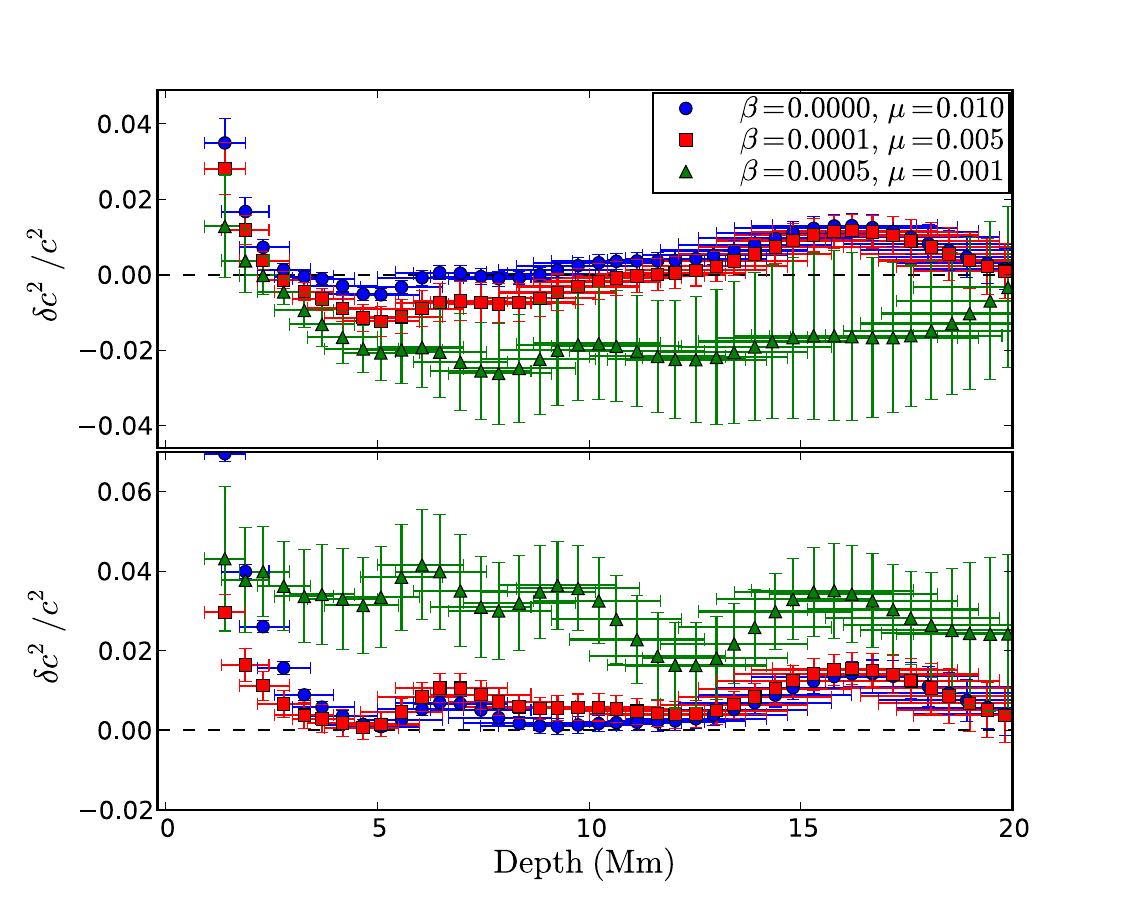}
\caption{The effects of inversion parameter choices on inversion results.  Three
of the four inversion parameters are varied in these inversions -- these inversions were performed with a fixed target width of $W = 0.0055$. The top panel shows inversions
with a surface term removed, and the bottom panels show inversions without
a surface term.  In each case, three different inversions are shown with
different values of the error suppression term $\mu$ and the cross term suppression
term $\beta$.}
\label{fig:inv_parm}
\end{center}
\end{figure}

In this work, we use a fourteen parameter
function defined in \inlinecite{bab2004}. The fit ridges can be interpolated
to wavenumber $k$, and the resulting frequencies $\omega_n(k)$ can be treated
as normal modes using the techniques of global mode analysis.  Typically,
inversions for solar structure are performed by linearizing the stellar
oscillation equations around a reference model. Then, the differences in the
frequency $\omega_i$ of the $i$th mode between the data and reference model, with $i$ representing the pair $(n,k)$, are related to the changes in structure by the following equation:
\begin{equation}\label{inveq}
\frac{\delta \omega_i}{\omega_i} =
\int  K_{c^2}^i(z)  \frac{ \delta c^2}{c^2}(z) \; \mathrm{d}z + \int  K^i_{\rho}(z)  \frac{\delta \rho}{\rho}(z) \;  \mathrm{d}z + \frac{F_\mathrm{surf}(\omega_i)}{Q_i}+\epsilon_i.
\end{equation}
The surface term, $F_\mathrm{surf}(\omega_i)$, is a smoothly varying function of
frequency which accounts for non-adiabatic effects confined to the surface
layers of the Sun, and is normalized by the mode inertia $Q_i$. The observational errors are given by $\epsilon_i$.  
The inversion kernels, $K_{c^2}^i(z)$ and $K_{\rho}^i(z)$, are known functions of the reference model, and give the sensitivity of a
mode to changes at a given depth. In ring-diagram analysis, however, we generally
invert relative to mode frequencies measured in a nearby inactive region of
the Sun.
\begin{figure}[ht]
\begin{center}
\includegraphics[trim= 2mm 2mm 7mm 5mm, clip, width=1.0\textwidth]{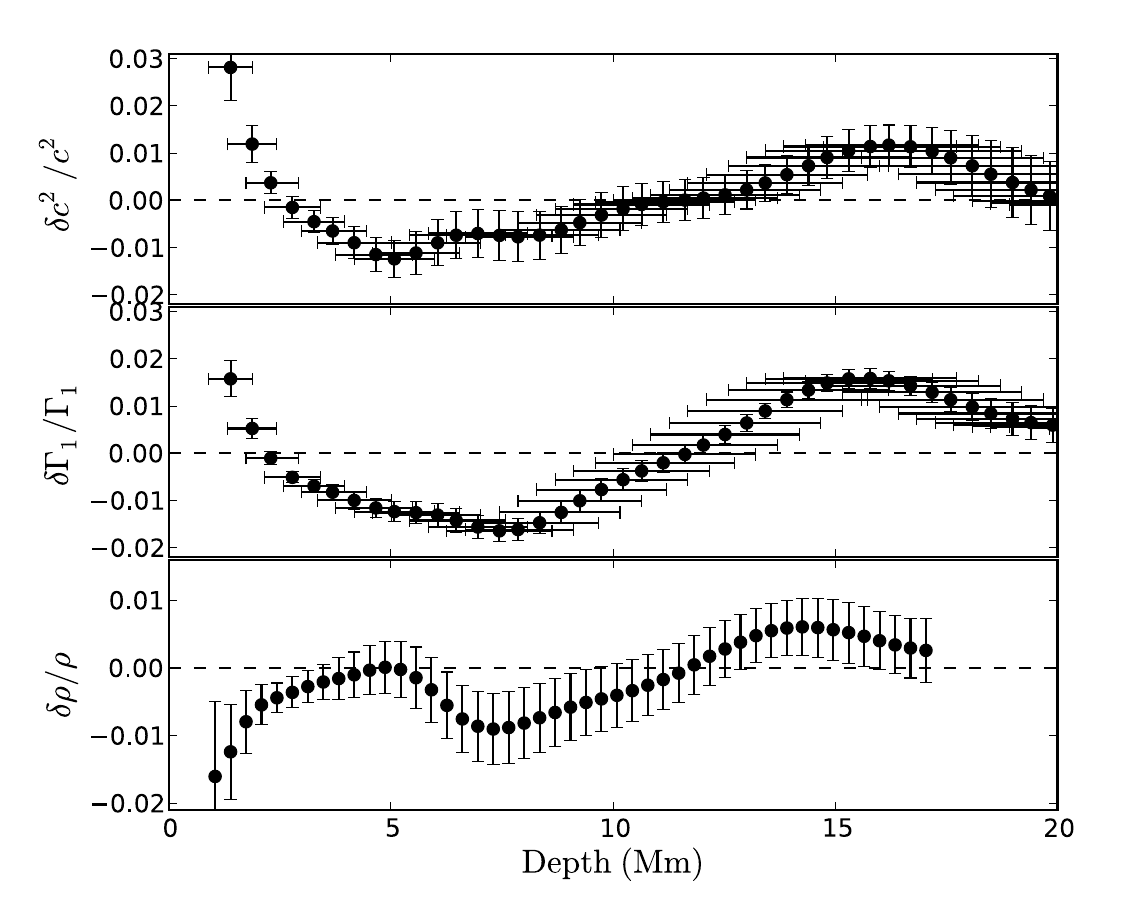}
\caption{Ring diagram inversions for the structure beneath AR 9787.  Inversions
are performed for both sound speed squared $c^2$ (top panel) and adiabatic index
$\Gamma_1$ (middle panel).  The inversions are performed relative to a nearby
quiet region -- the sense of the inversions is active minus quiet.  The results
for $c^2$ and $\Gamma_1$ are used to infer a profile for the density $\rho$ (bottom
panel).}
\label{fig:inv}
\end{center}
\end{figure}

The inversion technique used is called Subtractive Optimally Localized Averages
(SOLA), a technique pioneered in terrestrial seismology \cite{bg1967}. The SOLA method
takes a specified target averaging kernel $\mathcal{T}(z_0,z)$ which is localized around a target height $z_0$, and minimizes
the difference between that target and the actual averaging kernel, along
with the contributions from the cross-term kernel and from the errors (see \textit{e.g.} \opencite{rbc1999}).  
There are two trade-off parameters, $\mu$, which acts to suppress errors,
and $\beta$, which suppresses the cross-term kernel. There are also four free parameters to be chosen in this inversion method: the cross-term suppression term $\beta$, the error suppression term $\mu$, the characteristic width of the target kernel $W$, and $\Lambda$, the number of polynomials used to expand $F_\mathrm{surf}$.  Structure inversions using ring-diagram modes are quite unstable, and very sensitive to the choice of these parameters. Figure \ref{fig:inv_parm} shows sound speed inversions with a variety of different
choices of inversion parameters.  The most dramatic choice to be made is whether
or not to include a surface term in the inversions, and inversions are shown
both with and without a surface term.  A good inversion should have a well-localized
averaging kernel $\mathcal{K}(z_0,z)$ around a target height $z_0$, as well as reasonably small cross-term kernels.

In \citeauthor{gizonetal2009}~(\citeyear{gizonetal2009,gizonetal2009a}), we presented an inversion of the region AR 9787 for
sound speed. In Figure \ref{fig:inv}, we show an inversion of the same region for sound
speed $c^2$ and for adiabatic index $\Gamma_1$. The inverted quantities are 
fairly consistent with the results from other active regions (\textit{e.g.} \opencite{bab2004}; \opencite{bogartetal2008}). There is a depression in sound speed 
between approximately 3 Mm and 10 Mm depth. Below that depth, the perturbation 
becomes positive.

The adiabatic index is depressed below 3~Mm, with an enhancement below
12~Mm. If $c^2$ and $\Gamma_1$ are determined, the other thermodynamic quantities are
also, in principle, known. In Figure \ref{fig:inv}, we also show an inferred density profile.
The density between approximately 5~Mm and 11~Mm appears to be slightly depressed in the
active region compared to the quiet Sun, while the region below that has a lower
density than the quiet Sun. There is a depression in density at the shallow edge
of the inversion, which might be a sign of a Wilson depression of the optical
surface. It should be noted here, however, that results in the shallowest layers
of the Sun are quite uncertain, due both to lack of resolution in the inversions
and to uncertainties in the physics.

\subsection{Moat Flow: Hankel Analysis}
Fourier--Hankel decomposition is a useful analysis tool to study $p$ modes in the vicinity of sunspots. In particular, the method has been used to study the absorption of $p$-mode power by active regions \cite{bdl1988,bogdanetal1993,braun1995,chenetal1996}. In this procedure, the wave signal of the $p$ modes is decomposed into inward and outward propagating modes in an annular region surrounding a sunspot. The annulus is chosen to be small enough to allow describing the spatial part of the solar oscillations with Hankel functions. In the seismic study of AR 9787 undertaken by \citeauthor{gizonetal2009}~(\citeyear{gizonetal2009,gizonetal2009a}), the power absorption of this particular sunspot was demonstrated by Fourier--Hankel decomposition for the $m=0$ $p$ modes.
\begin{figure}[ht]
\centering
\begin{tabular}{cc}
\hspace*{-5mm}
\includegraphics[trim= 14mm 0mm 4mm 0mm, clip, width=0.495\textwidth]{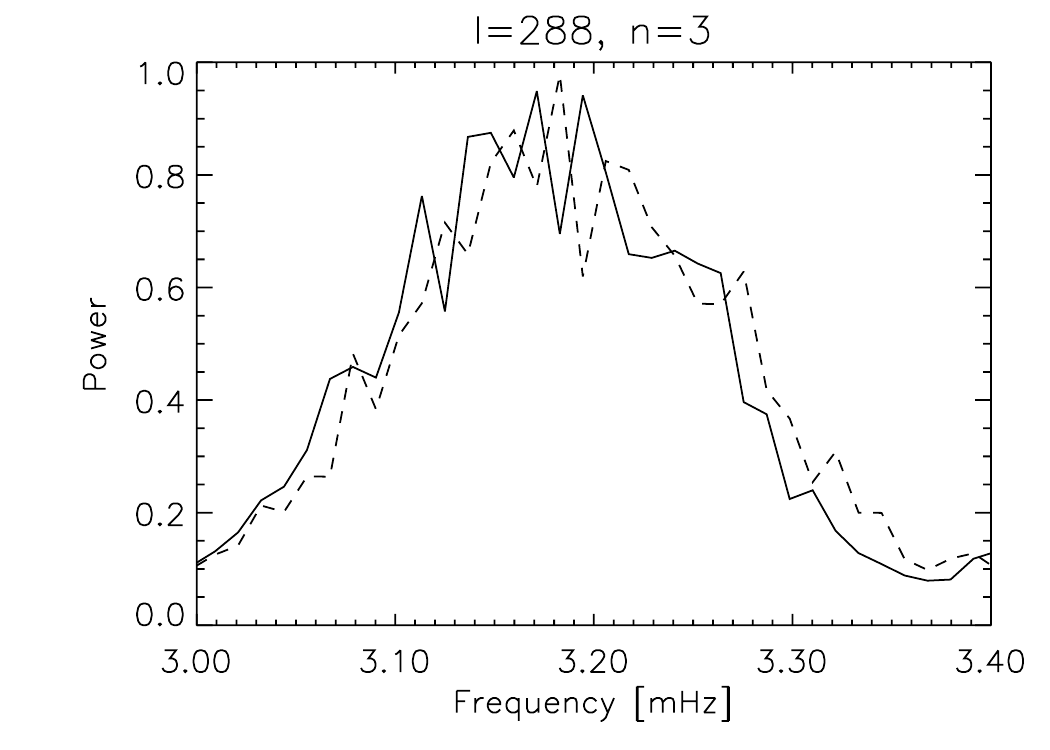}&
\hspace*{-5mm}
\includegraphics[trim= 14mm 0mm 4mm 0mm, clip, width=0.495\textwidth]{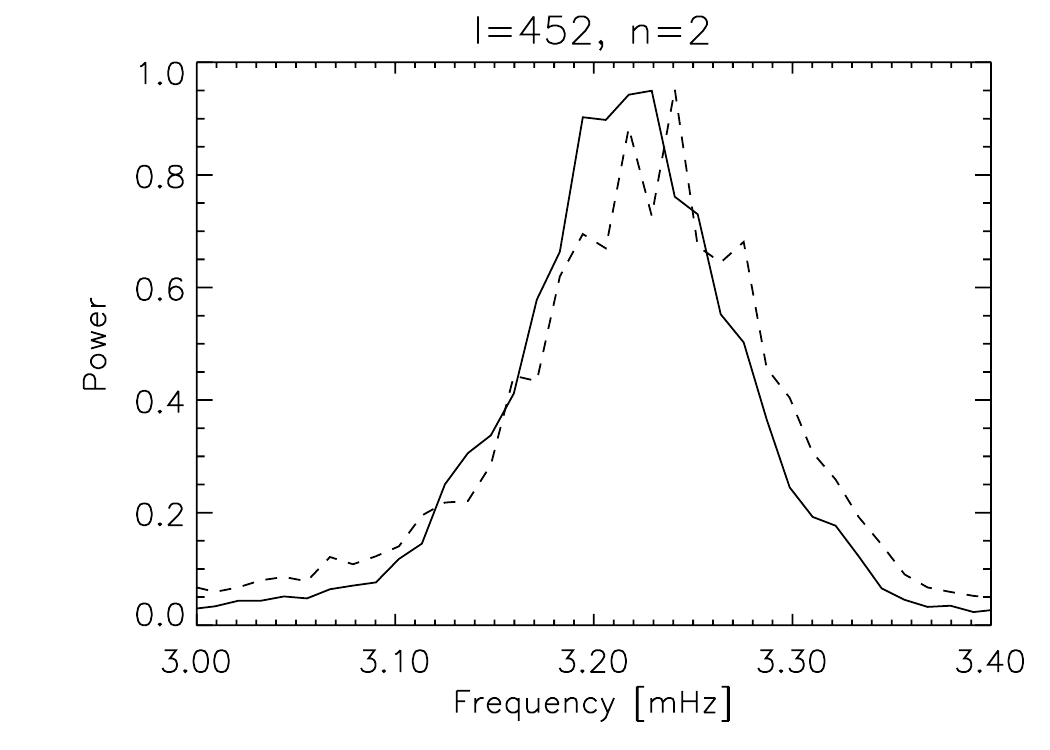}
\end{tabular}
\caption{Power spectra for two inward (solid line) and outward (dotted line) propagating $p$ modes.
Left: $\ell=288,\ n=3$; right: $\ell=452,\ n=2$. The peaks of the outward propagating modes are shifted to higher frequencies.}
\label{fig1}
\end{figure}

Fourier--Hankel decomposition also turns out to be useful for investigating the horizontal outflow associated with active regions. An outflow directed horizontally (the moat flow) would result in Doppler shifts of the $p$ modes traveling into and out of the spot. A radial flow therefore leaves its signature in the power spectra by shifting the $p$-mode ridges.

Here we extend the Fourier--Hankel analysis of AR 9787 using the procedure described by \inlinecite{braun1995}. For this analysis, the annular region around AR 9787 is defined by an inner radius of 30~Mm and an outer radius of 137~Mm measured from the spot center. Power spectra for inward and outward propagating waves were obtained for modes with azimuthal order $m=-5,-4,\dots,5$. Figure~\ref{fig1} displays the resulting spectra for two modes. It is immediately apparent by visual examination of the spectra that the outgoing $p$-mode power is shifted to higher temporal frequencies relative to the ingoing $p$ modes. We note that the power spectra displayed were normalized in order to correct for the power absorption due to the presence of the sunspot. The observed frequency shift is of the order of 10 $\mu$Hz.

\subsection{Ring-Diagram Analysis of Flows}
\subsubsection{Large Scale Flows Around Active Regions}
Dense-pack ring-diagram analyses of MDI and GONG++ data have shown that
subsurface flows associated with active regions have the following
characteristics:
\begin{itemize}
\item Active regions are surrounded by extended inflows and outflows,
depending on depth.
\item Subsurface flows of active regions are twisted.
\item Active regions rotate faster than quiet regions.
\item Flux emergence correlates with the flow divergence signal.
\end{itemize}
The horizontal components of solar subsurface flows are determined over
a range of depths from the surface to about 16~Mm using the dense-pack
ring-diagram analysis \cite{haberetal2002}. Daily flow maps are calculated for
189 dense-pack regions of 15\degree\ diameter with centers spaced by 7.5\degree\ in
latitude and central meridian distance.  The dense-pack technique thus measures flows on
horizontal scales comparable to the size of active regions.

\begin{figure}[ht]
\begin{center}
\includegraphics[trim= 22mm 22mm 30mm 22mm, clip, angle=90,width=1.0\linewidth]{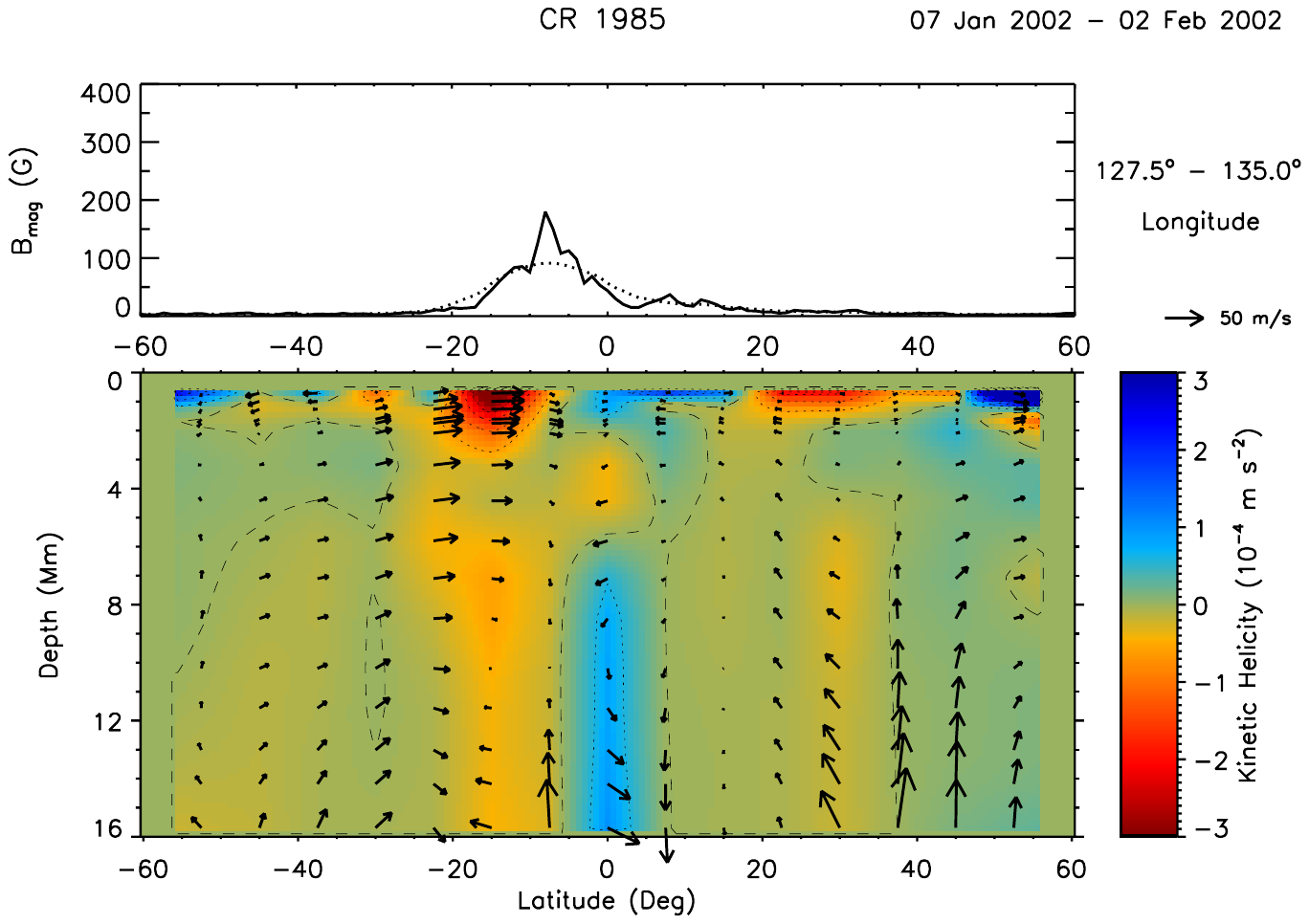}
\end{center}
\caption{Top: The unsigned magnetic flux (solid line) at
127.5\degree\todash135.0\degree\  longitude as a function of latitude
and binned over 15\degree\  (dotted line).
Bottom: The  kinetic helicity density at 127.5\degree\todash135.0\degree\
longitude as a function of latitude and depth. The kinetic helicity density
is the scalar product of the velocity and vorticity vector.
The arrows represent the meridional and vertical velocity components
with the vertical one increased by a factor of ten for visibility.
Average zonal and meridional flows have been subtracted.
Active region 9787 is noticeable as the location of strong helicity values
of opposite sign which coincides with the peak of the unsigned magnetic flux.
The region shows upflows at depths greater than about 10~Mm
and downflows at shallower depths at grid points of --15\degree\ to --7.5\degree\ latitude.
}
\label{fig:komm}
\end{figure}

Locations of strong active regions show, on average, extended divergent
horizontal flows at depths greater than about 10\todash12~Mm and convergent horizontal
flows closer to the surface (both with an amplitude of about 50 m s$^{-1}$).
AR 9787 shows this pattern as well (Figure~\ref{fig:komm}).
Using a mass conservation constraint, ring analysis infers vertical flows
associated with the horizontal flows. The convergent and divergent flow pattern,
is one of the most consistent characteristics of subsurface flows
associated with active regions.  It has been studied with several
helioseismic techniques \cite{gdl2001,bbl2004,haberetal2002,haberetal2004,zk2004,kommetal2005}.
These extended flows, however, should not be confused with the much more
localized moat flow (divergent, 250 m s$^{-1}$, see next section and \citeauthor{gizonetal2009},~\citeyear{gizonetal2009,gizonetal2009a}), which
cannot be resolved with 15\degree~aperture ring analysis.

Active regions are locations of flows with larger vorticity values;
they are locations of strong vertical gradients of the horizontal flows.
Strong active regions are identifiable by a dipolar pattern in zonal and
meridional vorticity \cite{masonetal2006} and AR 9787 appears to be no
exception (Figure~\ref{fig:komm}). The presence of active regions is barely noticeable
in vertical vorticity maps of this spatial resolution.  However, active regions
are, on average, characterized by cyclonic vorticity (counter-clockwise in the
northern hemisphere), which might be due to the Coriolis force acting on
 the flows \cite{spruit2003}. This agrees with observations with higher
spatial resolution \cite{dg2000,gd2003,zk2004}.

From direct surface measurements, it is well known that active regions rotate faster than quiet ones.
This has been confirmed with helioseismology \cite{gizon2004,zk2004}.
A ring-diagram analysis of about six years of GONG++ data shows that
the average zonal flow of active regions is about 4~m s$^{-1}$ larger than
 that of quiet regions  from the surface to a depth of 16~Mm \cite{kommetal2009}.
The difference is about one order of magnitude smaller than that derived
from surface measurements of active and quiet regions, which is most likely
a consequence of the rather large size of the dense-pack patches
and the resulting averaging over many different types of magnetic features.
Results from acoustic holography and time--distance analysis with higher
horizontal resolution support this interpretation \cite{bbl2004,gizon2004,zkd2004}.

A survey of 788 active regions observed with GONG++ makes it possible to
determine a signature of emerging magnetic flux in subsurface flows
associated with active regions \cite{khh2009}.  At depths greater than
about 10 Mm, upflows become stronger with time when new flux emerges.
At layers shallower than about 4 Mm, the flows might start to change from
downflows to upflows, when flux emerges, and then back to downflows after
the active regions are established. The flow response to emerging flux
agrees with numerical simulations of emerging flux tubes
\cite{fan2001,sr2005} where upflows indicate the beginning of flux emergence
and surface cooling due to adiabatic expansion leads to downflows along
the emerged loops.

\subsubsection{Sunspot Flows from Small Rings}\label{sec:haber}
\begin{figure}[ht]
\centering
\includegraphics[width=1.0\textwidth]{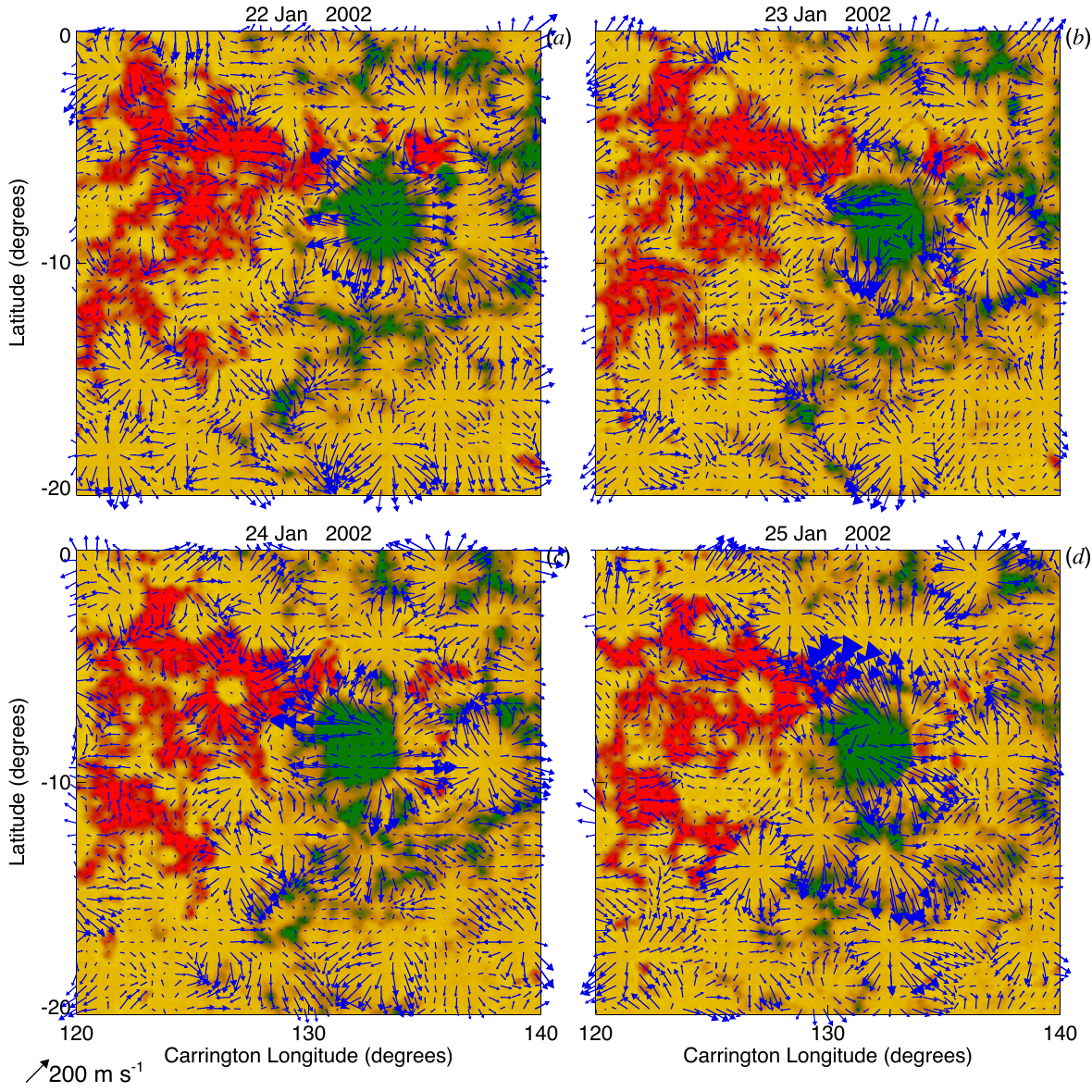}
\caption{Subsurface flows using high-resolution ring analysis of full-disk
MDI Dynamics Doppler data from 22\,--\,25  January 2002. MDI magnetograms
are underlaid with red and green specifying opposite polarities of the magnetic
field. The flows were determined from $f$-mode data and are thus representative
of flows within the top 2 Mm of the convection zone. The outflows from the sunspot
(shown in green) correspond to moat flows, while the cells of divergence seen
in the vicinity of the active region correspond to larger supergranular scales of about 50\,--\,60 Mm.
The magnetic network often appears at the edges of these larger cells.}
\label{fig:haber}
\end{figure}

Subsurface horizontal flows determined by high-resolution ring analysis
(HRRA) for the four days when AR 9787 was closest to the center of the
disk, show characteristic outflows from the lone sunspot corresponding
to moat flows (see Figure~\ref{fig:haber}).

The analysis was carried out on 2$^{\circ}$-diameter tiles whose centers were spaced  1$^{\circ}$ apart. Each flow arrow represents a spatial average over an entire tile and is itself the result of averaging the flows determined from all the fitted $f$ modes of the power spectrum for that tile since there are not enough modes to perform a true inversion. This means that the flows are characteristic of the gas from the surface down to a depth of to 2 Mm where the $f$ modes reside. The four-day panel (Figure \ref{fig:haber}) shows the evolution of large-scale zones of divergence around the sunspot corresponding to large supergranules as well as a seeming twist of the flows coming from the sunspot on 25 January. The flows are over-plotted on averaged MDI magnetograms for the given day where the green and red colors show magnetic fields of opposite polarity (the green circle corresponding to the sunspot).

\begin{figure}[ht]
\centering
\includegraphics[trim= 4mm 0mm 4mm 4mm, clip, width=1.0\textwidth]{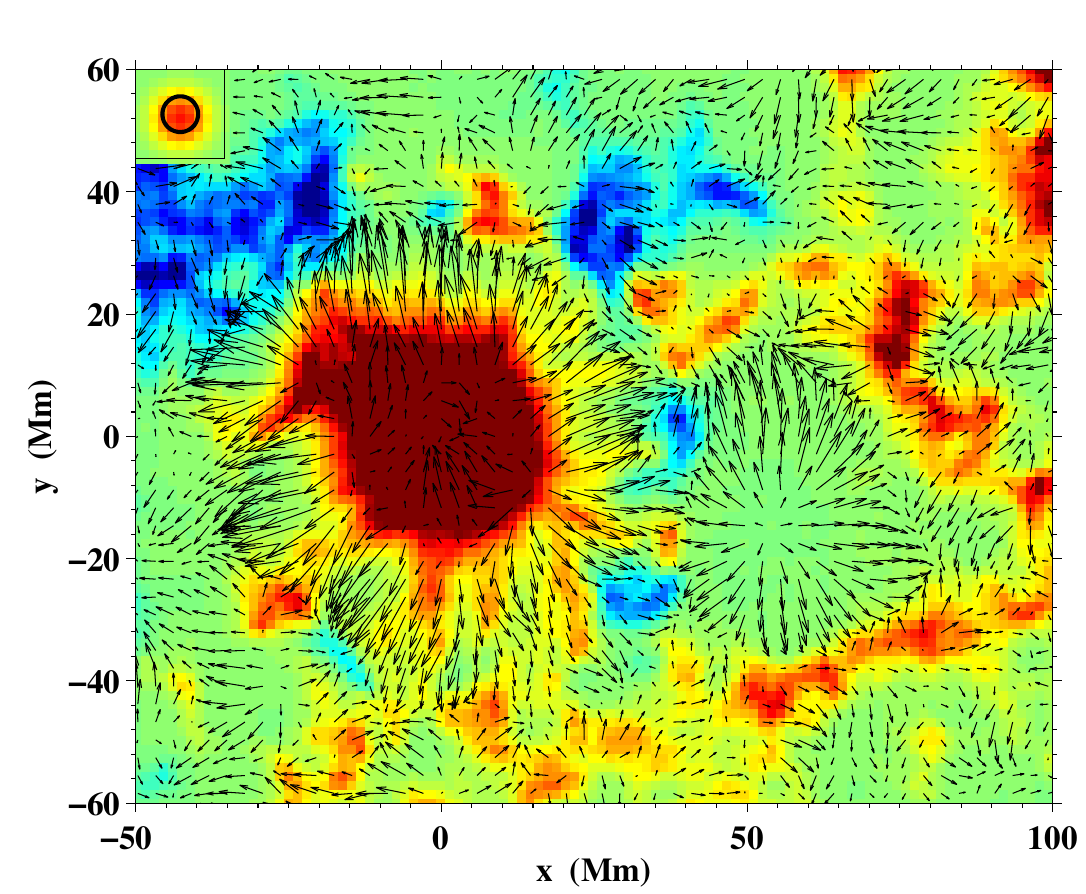}
\caption{Map of horizontal flows (arrows) at a depth of 1 Mm around the sunspot in AR
9787 using one day of MDI full-disk data and time--distance helioseismology on
January 24 (Gizon \etal,~2009). The spatial resolution is determined by the
width of the averaging kernel  (FWHM $7$~Mm) shown in the top-left corner.
The longest arrow corresponds to a flow of $450$~m s$^{-1}$. The surface
line-of-sight magnetic field is displayed in red and blue shades (saturated
at $\pm 350$~G). This inversion uses the modes $f$ and $p_1$ through $p_4$; it
is discussed by Jackiewicz, Gizon, and Birch (2008).}
\label{fig:gizon}
\end{figure}

Similar subsurface flow signatures were obtained by \citeauthor{gizonetal2009}~(\citeyear{gizonetal2009,gizonetal2009a}), who used $f$ to $p_4$ ridge-filtered time--distance travel times to produce linear inversions for flows (see \opencite{jgb2008}) around the sunspot in AR 9787. Figure \ref{fig:gizon} shows a summary plot of the subsurface horizontal flows around the sunspot, which appear very much consistent with the flows detected by HRAA, and direct observations of the moat flow (see Section\,\ref{mmf}). 

\subsection{Acoustic Halos}
\subsubsection{Observations}
The acoustic halo is an observed enhancement of high frequency (\textit{i.e.}, above the acoustic cut-off frequency at approximately $5.3$~mHz) acoustic power surrounding regions of strong magnetic field. \inlinecite{hb1998} demonstrate that this enhanced power at high frequencies tends to be prominent in intermediate magnetic-field strengths of 50-250 G and appear to be absent in the equivalent continuum intensity observations. In \citeauthor{gizonetal2009}~(\citeyear{gizonetal2009,gizonetal2009a}) it was demonstrated that AR 9787 has an extensive plage region, with the excess power at above 5.5~mHz being spatially correlated with these plage regions. It was also noticed that the strongest regions of acoustic halo occur between regions of opposite polarity.
\begin{figure}[ht]
\centering
\includegraphics[trim= 22mm 125mm 35mm 35mm, clip, width=1.0\textwidth]{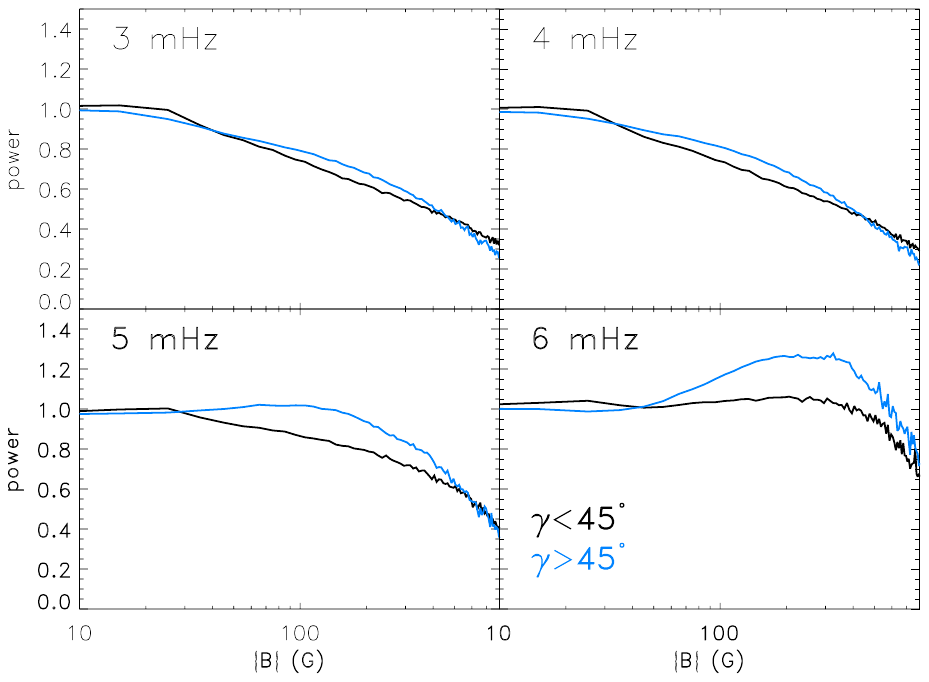}
\caption{Average normalized power against total magnetic-field strength (in bins of 10 G) for inclination $\gamma < 45^\circ$ (black) and  $\gamma > 45^\circ$ (blue) in frequency band passes as indicated. The importance of more horizontal field where $\gamma > 45^\circ$ for the acoustic halo is evident.}
\label{hs1}
\end{figure}

For our analysis, we calculate the vector magnetic field at the surface from the MDI magnetograms using a potential magnetic-field extrapolation. From this we define the inclination of the magnetic field from vertical [\,$\gamma$\,]. Figure~\ref{hs1} shows the average normalized power against total magnetic-field strength for $\gamma < 45^\circ$ (black) and $\gamma > 45^\circ$ (blue) for 1 mHz frequency band passes centered at 3, 4, 5 ,and 6 mHz as indicated. This highlights the importance of the field inclination for the acoustic halo.

In another recent analysis, \inlinecite{sb2009} further demonstrate that the acoustic halo is specifically confined to intermediate field strengths (between 100~G and 400~G) with $60^\circ < \gamma < 120^\circ$ from vertical. The power confined to this range of inclination has a peak which increases in frequency from 5.5~mHz to 6.5~mHz as the magnetic-field strength increases from 100~G to 400~G. In addition, they also find that the radial order ridges of the azimuthally summed power spectrum are shifted to higher wavenumbers than the quiet-Sun at constant frequency in the halo regions. The reason for this is, as yet, unexplained. 

\subsubsection{Theories}
A growing number of theories attempt to explain the phenomenon of wave velocity enhancements in the vicinity of active regions. The hypothesis of an altered wave excitation mechanism due to the presence of magnetic fields was proposed by \inlinecite{brownetal1992} and \inlinecite{braunetal1992}, and revisited more recently by \inlinecite{jacoutotetal2008}. Of course, it is also equally likely that the substantive differences in the wave propagation physics in active regions could also be leading to these enhancements, a possibility noted by a number of authors (\textit{e.g.} \opencite{hb1998}; \opencite{dbl1999}; \opencite{morettietal2007}).

From analyses of simulations of waves propagation through a model sunspot, \inlinecite{hanasoge2008} demonstrated that it was possible to obtain these halos without the requirement of enhanced sources. A wave scattering theory based on pure hydrodynamics was also put forth by \citeauthor{kuridzeetal2008}~(\citeyear{kuridzeetal2008,kuridzeetal2009}); their thesis was that bi-polar canopies create a secondary trapping region for the outward propagating high-frequency waves. Alternately, \inlinecite{kc2009} have suggested that upward propagating high frequency waves undergo conversion to fast modes, refract off the large Alfv\'{e}n speed gradient in the atmosphere, and return to the photosphere. This results in the same wave being observed twice, once while going upwards and the second time, on its way back down, thereby causing an enhancement. Finally, \inlinecite{hanasoge2009} has proposed that the presence of a large number of wave scatterers in the vicinity of active regions may cause preferential scattering 
into low mode-mass waves, \textit{i.e.} waves whose energy is focused in the near-surface layers. Relative to the original (quiet Sun) set of modes, this scattered configuration contains stronger surface velocity signatures (\textit{i.e.} lower net mode mass), thereby leading to the halos. Observationally, it can be confirmed that the velocity enhancements are primarily present at high-$l$, in the region of the power spectrum where the modes possess the lowest mode masses.

In reality, it may well be that there are multiple mechanisms at play, \textit{i.e.}, altered wave excitation, reflection/refraction of waves by magnetic fields, and preferential scattering into waves with low mode masses. Further studies are required before any solid conclusions can be drawn.

\subsection{Comparison of Observed Cross-Covariances with Simulations}\label{cameron_comp}
\begin{figure}[ht!]
\centering
\begin{tabular}{cc}
\includegraphics[trim= 30mm 0mm 30mm 10mm, clip, width=0.41\textwidth]{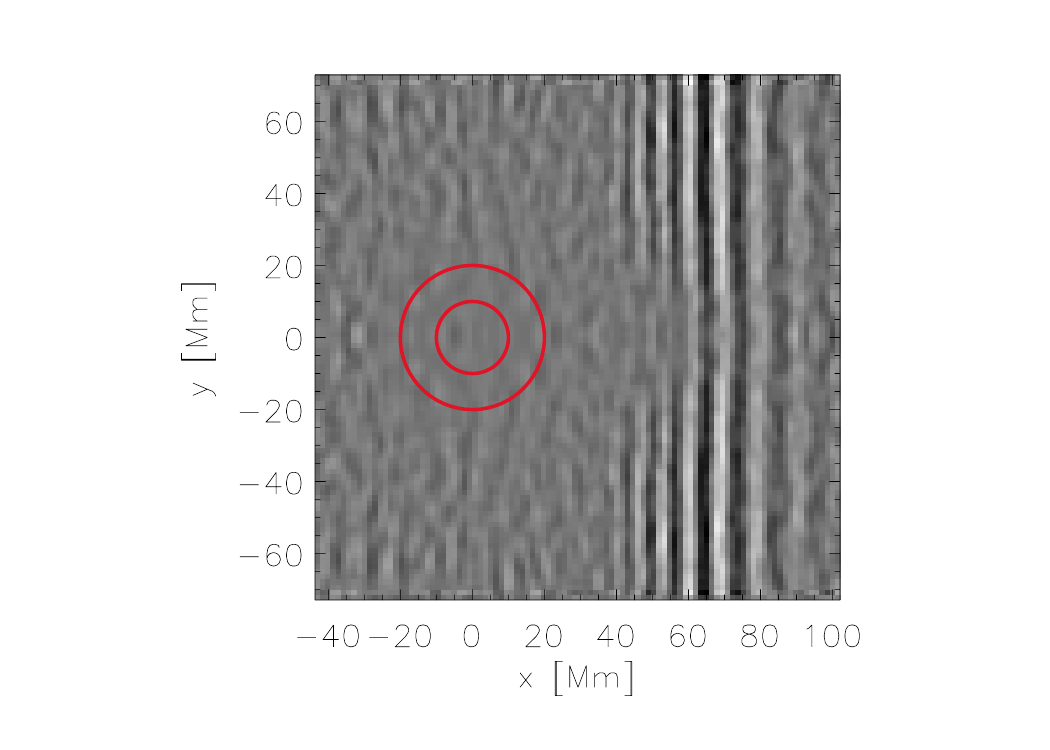}&
\includegraphics[trim= 30mm 0mm 30mm 10mm, clip, width=0.41\textwidth]{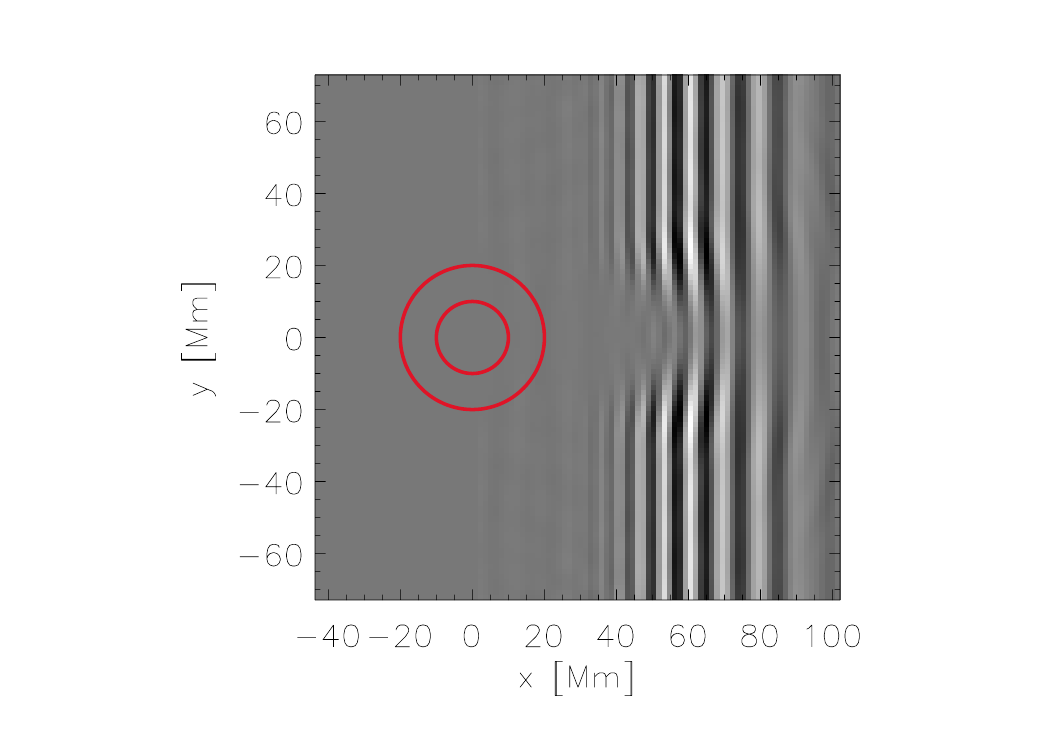}\\
\includegraphics[trim= 30mm 0mm 30mm 10mm, clip, width=0.41\textwidth]{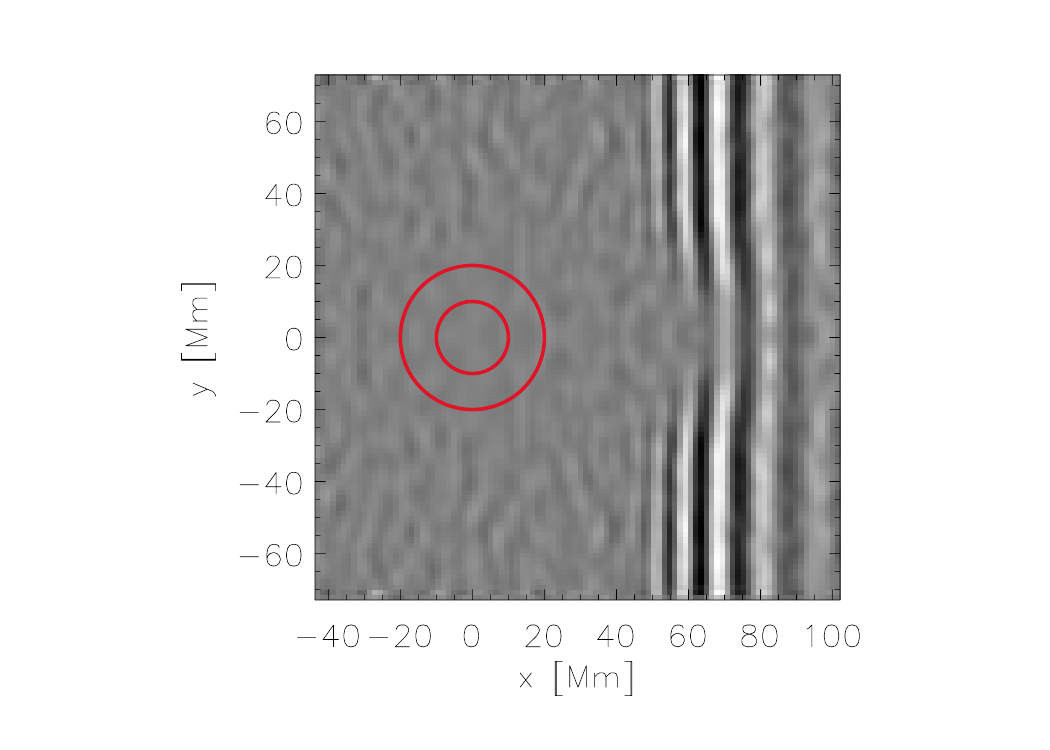}&
\includegraphics[trim= 30mm 0mm 30mm 10mm, clip, width=0.41\textwidth]{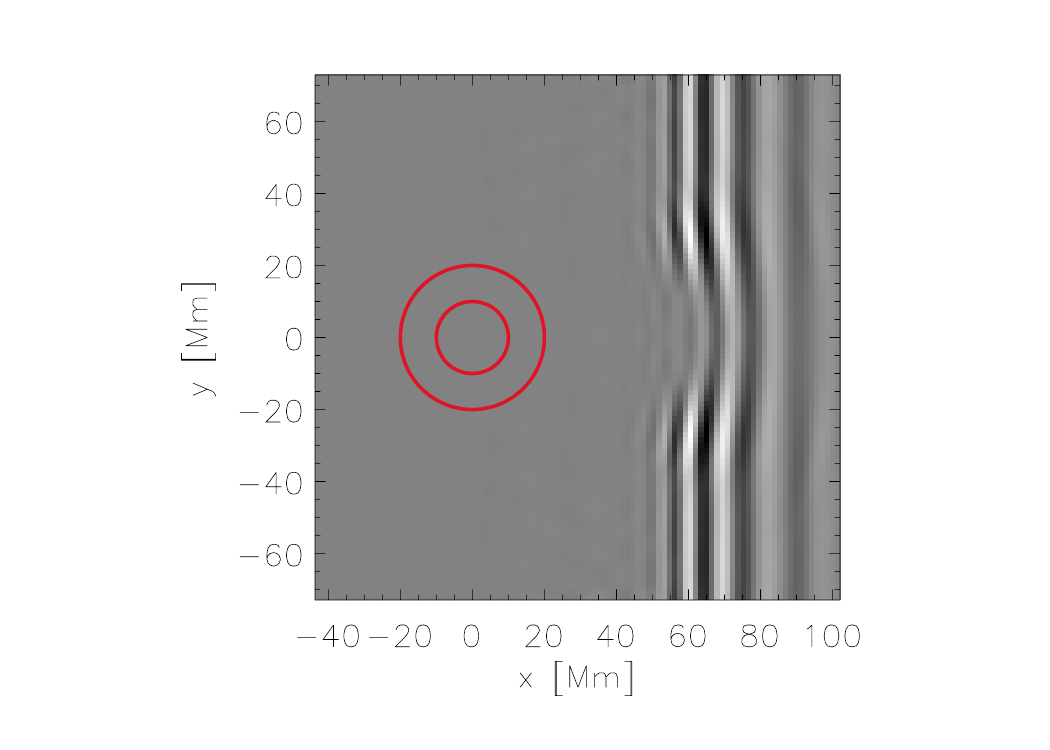}\\
\includegraphics[trim= 30mm 0mm 30mm 10mm, clip, width=0.41\textwidth]{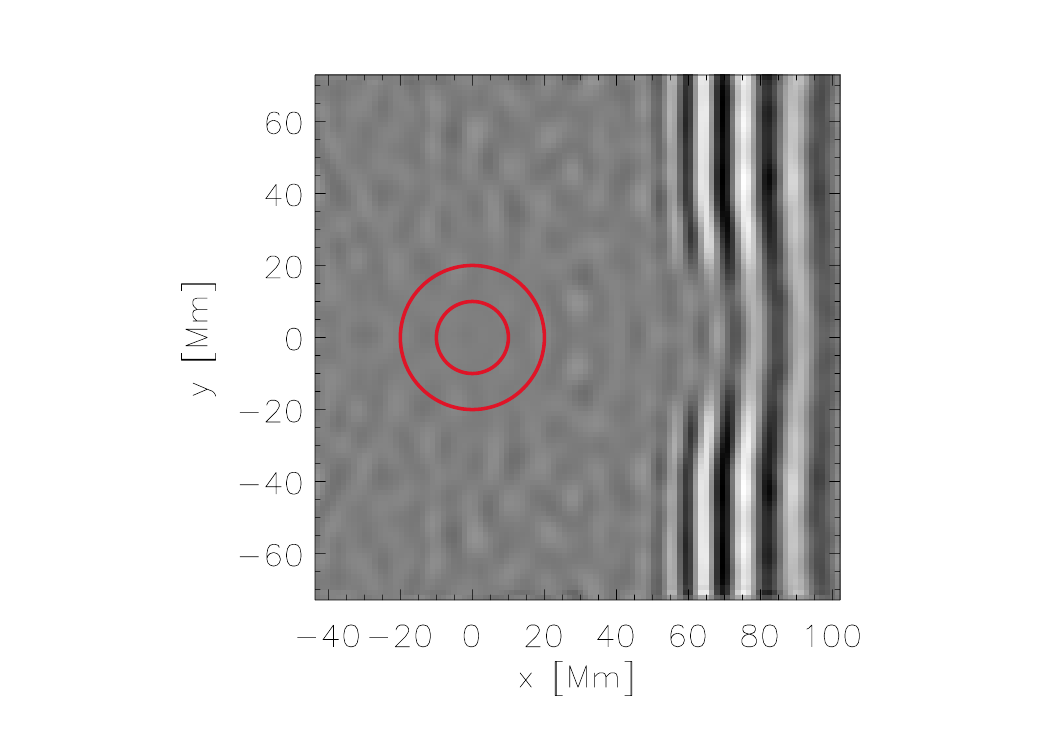}&
\includegraphics[trim= 30mm 0mm 30mm 10mm, clip, width=0.41\textwidth]{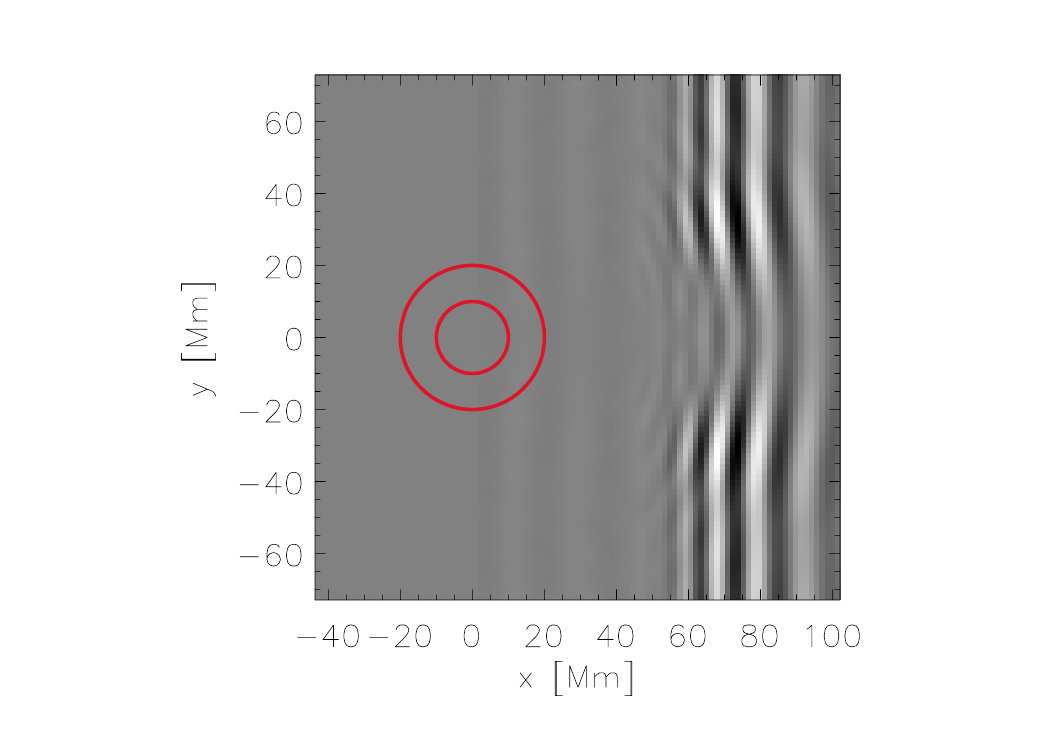}
\caption{Left panels: Observed MDI cross-covariance functions between the Doppler velocity averaged over the line x\,=\,-40\,Mm and the Doppler velocity at any other point $(x,y)$ in the map.
The sunspot AR 9787 is located at the origin and the circles show the boundaries of the umbra and penumbra. The cross-covariance is averaged over seven days. The top left panel shows the cross-covariance for the $f$-mode ridge at time-lag $t=185$~minutes,
the middle left panel for the $p_1$-mode ridge at time-lag $t=165$~minutes, and bottom panel for the $p_2$-mode ridge at time-lag $t=145$~minutes. The right panels show SLiM simulations of the propagation of $f$, $p_1$, and $p_2$ wave packets in the $+x$-direction through the semi-empirical sunspot model of Cameron \etal~(2010). The vertical velocity in the simulation is directly comparable to the observed cross-covariances.}
\label{fig:cameron}
\end{tabular}
\end{figure}
\inlinecite{cgd2008} and \citeauthor{gizonetal2009}~(\citeyear{gizonetal2009,gizonetal2009a}) showed that the cross-covariance of the MDI Doppler velocity is a very useful quantity to study the scattering of solar waves by the sunspot in AR 9787. The cross-covariance between two points is closely related to the Green's function between these two points (see, \eg \opencite{gbs2010}, and references therein).
By extension, the cross-covariance between the signal averaged over a line and any other point can be used to study the interaction of plane wave packets with the sunspot.

The left panels in Figure~\ref{fig:cameron} show such cross-covariances at particular values of the correlation time-lag, after the wave packets have traversed the sunspot. While \citeauthor{gizonetal2009}~(\citeyear{gizonetal2009,gizonetal2009a}) had studied only $f$-mode wave packets, here we also show the observed cross-covariances for $p_1$ and $p_2$ wave packets. The wave packets are constructed using standard ridge filters (wave power peaks near 3 mHz).  The observed cross-covariances show that, in all cases, solar waves speed up through the sunspot and that their amplitudes are reduced relative to the quiet Sun. The stochastic noise in the observations is reduced by exploiting all available symmetries, including the near-cylindrical symmetry of the sunspot.

The cross-covariances also provide very strong observational constraints to test the validity of sunspot models. Here we have used the SLiM code to test the semi-empirical model of AR 9787 by \inlinecite{cameronetal2010} (see Section~4.3.1 and Figure~4) by propagating $f$, $p_1$, and $p_2$ wave packets through the sunspot model. The initial conditions of the simulations are chosen such that the vertical component of velocity in the simulation is directly comparable with the cross-covariance function. As can be seen in Figure~\ref{fig:cameron}, the simulations reproduce the basic features of the observations. This indicates that the sunspot model has a seismic signature that is close to the real sunspot, which is very encouraging.

\section{Discussion and Perspective}

\subsection{Sunspot Structure: A Critical Assessment of Existing Models}
Recent years have seen an increased interest in questions relating to the structure of sunspots, fuelled by a fortunate coincidence of several factors. After decades of more gradual improvements, classical (visible light) observations have quickly improved in quality over the past few years. With the Swedish 1-m Solar Telescope, spatial resolution achieved has made a jump (with a resolution of 0.1" achieved regularly in the blue), while time coverage has vastly improved through the long time sequences from the \textit{Hinode} satellite. Secondly, helioseismic observations have now approached the point where they can be turned into powerful diagnostics of spot structure, as discussed elsewhere in this text. Finally, and perhaps most dramatically, the realism achieved by 3D radiative MHD simulations has now opened the perspective of a definitive physical interpretation of the phenomenology of sunspot structure, including umbral dots, light bridges, penumbral filaments, the Evershed flow, the moat flow, and their relations to each other. It is likely that understanding will soon replace the historically evolved patchwork of mutually inconsistent views and physically dubious ad-hoc models. The following gives a brief perspective on these developments. For more detail, see the reviews in \inlinecite{scharmer2009} and \inlinecite{ns2009}.

As discussed in Section\,\ref{spruit}, the structure of a sunspot as observed at the surface is kept together by forces deeper down. Assuming this anchoring at depth as given (but yet to be explained), one can seek a physical interpretation of the surface structure observed in umbra and penumbra. At the observed field strength, 1\,--\,3 kG, the magnetic field is strong enough to suppress convection. This explains why spots are dark but it does not explain the particular time dependent fine structure observed.

In early observational work (\textit{e.g.} \opencite{mamadazimov1972}) the mix of long dark and bright filaments has been interpreted as showing a magnetic field (dark, as in the umbra) on top of the normal photosphere (shining through in the bright filaments). This explained the general appearance of the penumbra and the nearly photospheric brightness of the bright structures, but required that the penumbral magnetic field only touches over the photosphere (a ``thin penumbra'', with optical depth only of order unity). The field would have to be essentially horizontal in the penumbra, field lines crossing the solar surface only in the umbra. This does not agree with the observed inclination of the penumbral field. In fact, most of the magnetic flux of a spot crosses the surface through the penumbra, not the umbra. The penumbral field, on average, is therefore not horizontal, and the observed field must continue to some depth below the surface.

This has led to interpretations in terms of convection in a magnetic field extending to a substantial depth below the surface (a ``thick penumbra''). An influential conceptual picture was the \inlinecite{danielson1961} model of convective ``rolls'': an overturning flow in a plane perpendicular to a horizontal magnetic field.  This idea has led to  a ``magnetoconvection'' view of the penumbra, which interprets the observed filamentary structures as turbulent fluctuations in a mean magnetic field inclined at some angle to the solar surface (see \textit{e.g.} \opencite{tiw2004} and references therein). It also plays a role as a conceptual picture underlying global models of sunspot structure, such as \inlinecite{js1994}.

A key aspect of any realistic model of magnetic surface structure is the transition from the gas pressure dominated regime below the photosphere to the magnetically dominated atmosphere. In the penumbra this takes place just at the photospheric level. The transition takes place over a couple of pressure scale heights. This is no more than the horizontal resolution of most telescopes, and comparable to the smallest horizontal structures seen in the penumbra. With photospheric observations one is thus looking at the {\em thin interface between two physically different regimes}, rather than a slice through some quasi-uniform turbulent medium. The consequences of this fact have not been realized in most of the currently popular views.

Observations of this interface show a mix of magnetic-field strengths and inclinations. This has led to the idea of inclusions (``flux tubes'') embedded in a background field of different direction \cite{sm1993}. This has become the dominant theme in interpretations of penumbral structure (\textit{e.g.} \opencite{mp2000}; \opencite{bellotrubioetal2003}; \opencite{bbc2004}; \opencite{borrero2007}).

In the atmosphere, however, where $B^2/8\pi\gg P$, forces other than magnetic are small. The magnetic field in such a case is nearly {\em force free}. This provides strong restrictions on the physically realizable field configurations. These constraints are actually violated in most proposed ideas about penumbral structure, especially the floating flux tubes in the embedding and uncombed type models. An exception is the proposal by \inlinecite{martensetal1996}, who propose a structure of magnetic sheets of different directions separated by force-free currents.

Force-free, but not current-free, configurations also cause problems however. The strength of the magnetic field decreases away from the spot. A force-free field has the property that a difference in field-line direction between neighboring magnetic surfaces {\em increases with decreasing field strength} along the field (this is related to the magnetic torque being conserved along field lines, see discussion in \opencite{ss2006a}). Force free models thus predict that differences in inclination should become more prominent with height above the photosphere and distance from the umbra. Observations in the chromosphere (the ``superpenumbra'') do not support this. Instead, magnetic-field line directions traced by chromospheric indicators appear to become more uniform with height. This is not consistent with force-free models like those of \inlinecite{martensetal1996}. Instead, these observations indicate that the field in the atmosphere is much closer to a {\em potential} (current free) configuration.

The ``gappy penumbra'' model \cite{ss2006a,ss2006}, based on the cluster model of a sunspot, explains how the pattern of strong irregularities decreasing with height above the surface comes about naturally in a potential field, as a result of gaps between field lines created below the surface by overturning convection\footnote{By ``overturning'' we mean here the convective pattern also observed in realistic simulations of stellar surface convection (see the review by \opencite{scharmer2009} for more detail). As opposed to traditional views approximating convection as closed cells or ``rolls'', almost nothing of the descending part of the flow returns to the surface but is replaced by upflows from larger depths. As opposed to the field-aligned rolls of the \inlinecite{danielson1961} kind, this pattern maintains a low field strength in the gaps through the process of {\em convective expulsion}. While these gaps may not be exactly field free (as none of the convection zone is), they are regions where the hydrodynamic forces dominate over magnetic forces, \textit{i.e.}, the field is ``below equipartition with convection''.}. At the same time this model addresses the long-standing heat flow problem: the question how the observed radiative energy flux is supplied to the penumbral surface. The low field strength in the gaps allows convection to supply the heat flux emitted by the surface unimpeded by magnetic forces. Convection in magnetic rolls, on the other hand, does not solve this problem, since a roll with the diameter of a penumbral filament does not contain enough thermal energy to maintain the radiative flux over the life of a filament. However, the gappy model has problems explaining the structure in the outer penumbra, where observations show nearly horizontal field (also confirmed in the numerical simulations of \opencite{rempeletal2009}), and the presence of returning magnetic flux (see Section\,\ref{field}).

Apart from the gappy model, a number of other physical models for the penumbral currently exist. The model of \inlinecite{sjs1998}~(see also \opencite{schlichenmaier2009} and references therein), which has the virtue of substantial physical detail, prescribes that magnetic flux tubes rising from below the surface carry heat and as well as an outward flow, interpreted as the Evershed flow. It has problems reproducing the observed heat flux from the penumbra, however, and the postulated rising flux tubes have not been identified in the numerical simulations. In the numerical simulations, both the heat flux and the Evershed flow are found to be driven by overturning convection, not in the form of magnetic rolls or flux tubes, but in regions low field strength (see \opencite{ns2009}).

Another proposal appeals to the process of ``turbulent flux-pumping'' \cite{thomasetal2002,weissetal2004}. In this process asymmetry between upflows and downflows in stratified convection effectively transports horizontal magnetic-field lines downward. Penumbral structure is then ascribed to this process. The source of the penumbral structure is said to be located outside the penumbra \cite{weissetal2004}. Alternatively, it has also been implied to be located in the penumbra itself (see \opencite{tiw2004}). The turbulent pumping mechanism appears to offer a plausible mechanism for producing the returning magnetic flux in the outer penumbra, as well as the apparent hysteresis observed in the transition between a pore and a sunspot \cite{weissetal2004}. However, there are also objections to this model that are based on the fact that the action of convection on a magnetic field overlying it, as in the outer penumbra, would be expected to have the opposite effect of turbulent pumping. Overturning convection would instead expel a magnetic field that is imposed from above, through the well-known process of convective expulsion \cite{zeldovich1956,weiss1966}. The process is seen in action in numerical simulations of horizontal magnetic fields overlying granulation by \citeauthor{steineretal2008}~(\citeyear{steineretal2008,steineretal2009}). 

As the quality of observations continues to improve, realistic 3D radiative numerical MHD simulations of sunspots \cite{sv2006,heinemannetal2007,snh2008,rsk2009,rempeletal2009} show a remarkable level of agreement with these observations. The agreement includes the properties of umbral dots, the inward propagating bright filaments, the dark cores overlying them, the varying aspect of penumbral structure with viewing angle, the varying field strengths and direction in penumbral filaments, the dependence of these on height, the moat flow and the Evershed flow. We expect that more advanced simulations in the near future will further improve our understanding of penumbrae, in particular with regards to their outermost parts. With this level of agreement with observations, there is little doubt that the simulations are reproducing the physics of sunspot structure as observed at the surface.

Next to the treatment of the atmospheric magnetic field, the physics of radiation is of equal importance for realism in numerical simulations. Cooling by radiation at the surface determines the thermal structure of the penumbra and drives the observed flows. On the other hand, it also determines the detailed appearance of penumbral structure at the optical depth unity surface. Any physically meaningful comparison with observations thus requires inclusion of radiation physics at a fairly well developed level. The fact that the level of realism needed for a meaningful interpretation of solar surface structure, magnetic or quiet, has now been achieved should be considered the most significant advance in theoretical solar physics of the past few decades.

\subsection{The Starspot Connection}
It is important to keep in mind that our Sun is not the only star that possesses spots. Moreover, it is more than likely that all late-type stars with
convective envelopes exhibit spots, or ``starspots'' (\textit{e.g.} see reviews by \opencite{strassmeier2009} and \opencite{berdyugina2005}). Spotted stars constitute roughly 90\% of all stars in the Milky Way, basically all GKM stars and a large fraction of F and L and T dwarfs are spotted, representing a mass range from the brown dwarf limit up to the $2.4$~M$_\odot$ of a G giant, and an age range from the pre-main sequence phase up to the asymptotic giant branch. Stars with planets can also be affected by magnetic processes and their magnetic environment may even affect close-in planets and back-react on to the star (\textit{e.g.} \opencite{swb2003}; \opencite{catalaetal2007}; \opencite{kds2008}; \opencite{lanza2008}). Therefore, we must understand sunspots first, before we can understand starspots.
\begin{figure}[ht]
\centering
\includegraphics[angle=0,width=0.96\textwidth, clip]{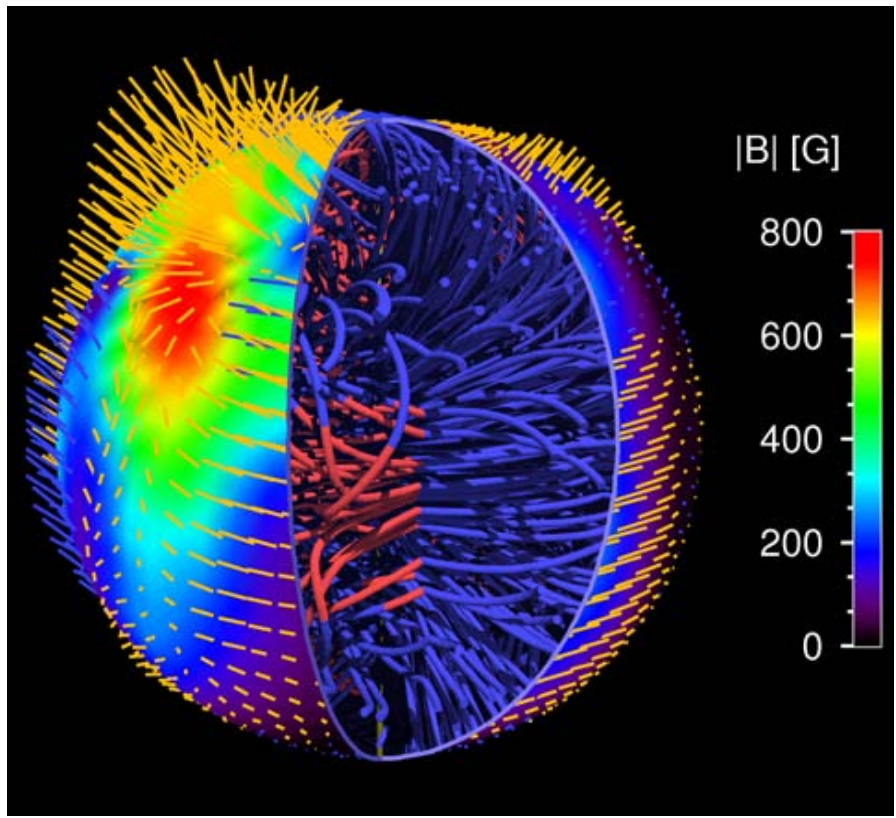}
\caption{A starspot on the K subgiant II~Pegasi. The surface temperature and the photospheric magnetic field is a Zeeman--Doppler observation while the interior shows a numerical simulations of a non-axisymmetric dynamo in a fully convective star. There, polarity in the western hemisphere mirrors
the polarity in the eastern hemisphere. On the surface, red means a field strength of up to 800~G according to the color bar while blue means basically no detection. In the interior, red means the field is pointing outwards, blue means the field points inwards.}
\label{iipeg}
\end{figure}

Models of starspots do not exist yet but the mere size difference
suggests that a scaling from a sunspot model is not appropriate (see \textit{e.g.}  \opencite{su2004}).
While sunspots typically cover 10$^{-4}$ to 10$^{-5}$ of the solar
surface and only during solar maximum reach about 10$^{-3}$, the
record holder among other stars is still the one big starspot seen
on the K0 giant XX~Tri in December 1997 \cite{strassmeier1999}. Based on
the star's Hipparcos distance of 197~pc, the spot's area covers
approximately $12\times20$~$R^2_{\odot}$,
\textit{i.e.} 22\% of the star's hemisphere or 10\,000 times the area of the largest sunspot group.
Clearly, its emergence, structure, and decay must be
addressed on a global scale. Figure~\ref{iipeg}
shows a Zeeman--Doppler image of the active spotted star II~Pegasi
\cite{carrolletal2007} combined with a numerical simulation of a
non-axisymmetric dynamo. Although just a demonstration example, it nonetheless
highlights the expected global starspot--dynamo connection.

\subsection{Conflicting Helioseismic Observations}	
\begin{figure}[ht]
\centering
\includegraphics[trim= 25mm 25mm 25mm 10mm, clip, width=0.95\textwidth]{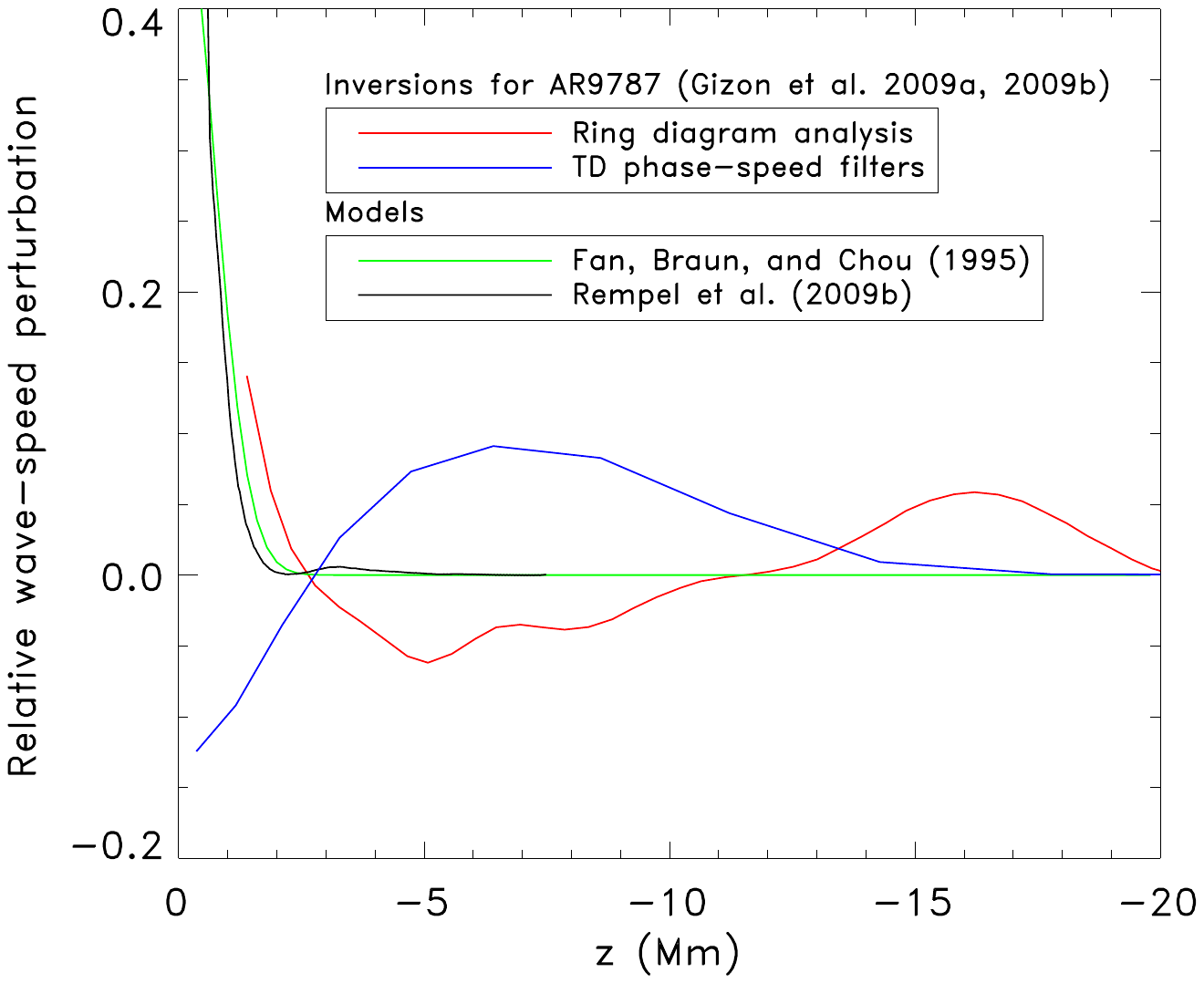}
\caption{Comparison of different helioseismic methods used to infer
wave speed perturbations below AR 9787 ($\delta c_w/c$). 
The red and blue curves show the ring-diagram and
phase-speed filtered time--distance results, respectively, from Gizon
\etal~(2009a, 2009b). The time--distance result is shown along the axis of the sunspot; the ring-diagram results has been scaled by a factor of 10. 
The black curve indicates the fast mode speed perturbation ($(c_{\rm f}-c)/c$) from
the radiative MHD simulations of Rempel \etal~(2009b), described in
Section\,\ref{rempel}; it approaches the value 75 at $z=0$~Mm. 
The green line represents the on-axis wave-speed perturbations deduced from the phenomenological model of Fan, Braun, and Chou (1995), based on the Hankel phase-shift measurements of Braun (1995).}
\label{fig:conflict}
\end{figure}

A prevalent, largely phenomenological, approach to modeling the subsurface structure of sunspots has been to treat the regions of magnetism as perturbations to the background wave speed. These types of models have been constructed
using a variety of local-helioseismic procedures. A comparison of structural
(wave-speed) inversions for AR 9787 using both ring-diagram analysis
and time--distance helioseismology with phase-speed filters was presented by \citeauthor{gizonetal2009}~(\citeyear{gizonetal2009,gizonetal2009a}), and is partially reproduced in Figure~\ref{fig:conflict}.

As is evident, these two inversions give subsurface wave-speed profiles with opposite signs and different amplitudes.
\citeauthor{gizonetal2009}~(\citeyear{gizonetal2009,gizonetal2009a}) discussed a number of factors which could contribute to
such a disagreement, including the observation that the ring-diagram inversions
include a treatment of near-surface effects absent from the time--distance analyses,
and the apparent sensitivity of the measurements to
details of the analysis which are not included in the models. A prominent
example is the sensitivity of time--distance measurements to parameters of the filters
(Section\,\ref{filters}). As suggested by \inlinecite{bb2008}, some evidence of a
strong near-surface contribution to helioseismic measurements in sunspots is
shown in the forward modeling of Fourier--Hankel measurements performed by \inlinecite{fbc1995}.
It is worthwhile, for comparative purposes, to therefore plot the results of the subsurface perturbation suggested by \inlinecite{fbc1995} in Figure~\ref{fig:conflict}. An important caveat in this comparison is that the Hankel phase-shift measurements
\cite{braun1995}, from which this wave speed result was inferred, were of different sunspots. However, \inlinecite{bb2008} show that these phase-shift measurements agree favorably with more recent ridge-filtered holography measurements of sunspots of similar size. The relative fast-wave speed perturbations of the sunspot model of \inlinecite{rempeletal2009} presented in Section\,\ref{rempel} are also included in Figure \ref{fig:conflict} for reference: this model also displays positive wave-speed perturbations in the first 2~Mm below the surface.

\subsection{Emergence of a New Paradigm in Sunspot Seismology}
Keeping in mind all of the above caveats, it is worth noting that three out of four
curves shown in Figure~\ref{fig:conflict} are consistent with a strong, positive wave-speed perturbation extending about 2.5 Mm below the surface. Below this depth, the helioseismic inversions show considerably stronger deep wave-speed perturbations (albeit, of opposite signs) than the other methods.
Braun and Birch (2006) have argued that shallow
wave-speed perturbations like those suggested by \inlinecite{fbc1995} are required to explain
the systematic frequency dependence in the mean travel times observed in sunspots. Of course,
it is well to keep in mind the warnings of \citeauthor{gizonetal2009}~(\citeyear{gizonetal2009,gizonetal2009a}), namely the possible naivety of modeling potentially complicated effects of the magnetic field in terms of an equivalent sound-speed perturbation. 

Many of these magnetic effects are more suitably explored through the numerical methods discussed in Section\,\ref{forward_models}. Perhaps the strongest argument in favour of a shallow, fast wave-speed model is provided by the linear simulations of MHD wave propagation (Section\,\ref{cameron_comp}). These forward numerical computations show that a simple semi-empirical sunspot model extending no deeper than 2 Mm (Figure \ref{fig:cameronss}) is capable of reproducing many features of the helioseismic measurements, in particular the cross-covariance signatures of the sunspot in AR 9787. While additional linear simulations will be needed to confirm this claim, the realistic radiative simulations of sunspot-like structure by \inlinecite{rempeletal2009} will provide the ultimate testbed to validate the forward and inverse methods of sunspot seismology.

Finally we note that sunspot seismology will benefit greatly from improved observations by the \textit{Solar Dynamics Observatory} (SDO), scheduled to be launched shortly. The \textit{Helioseismic and Magnetic Imager} (HMI) instrument onboard SDO will not only provide higher-resolution Doppler images of the solar disk, but it will also provide full vector magnetograms. Reliable measurements of the photospheric magnetic field will be key to constraining the near-surface layers of sunspot models, as detailed models of the surface layers are a necessity in order to probe the deeper structure of sunspots.

%
 \begin{acks}
This parametric study of sunspot models is supported by the European Research Council under the European Community's Seventh Framework Programme (FP7/2007-2013)/ERC grant agreement \#210949,  ``Seismic Imaging of the Solar Interior'',  to PI L. Gizon (Milestone \#3). The follow-up analysis of AR 9787 was carried out at the Third HELAS Local Helioseismology Workshop, which was held in Berlin on 12\,--\,15 May 2009 and supported by the European Commission under the Sixth Framework Program of the European Union. Portions of this work were also supported by the NASA
SDO Science Center and Heliophysics Guest Investigator programs through contracts NNH09CE41C and NNG07EI51C to NWRA under
PI D.C. Braun. The authors would also like to acknowledge Kaori Nagashima and Takashi Sekii for providing the \textit{Hinode} observation used in Figure~\ref{fig:hinode}, and the anonymous referee for useful comments that helped improve the paper.
\end{acks}

%
\newpage
\bibliographystyle{spr-mp-sola-cnd} 
\bibliography{SOLA1202R2_Moradi.bbl}

\end{article}
\end{document}